\documentclass{svmult}
\usepackage{makeidx}  
\usepackage{graphicx}  
\usepackage{subfigure}   
\usepackage{epstopdf} 
\usepackage{mathptmx,helvet,courier}  
\usepackage{multicol}        
\usepackage[bottom]{footmisc} 
\usepackage{cite}
\usepackage{amsmath}\usepackage{amssymb}

\def\rcgindex#1{\index{#1}}
\def\myidxeffect#1{{\bf\large #1}}

\makeindex             

\begin{document} 
\title*{Nonlinear vortex light beams supported and stabilized by dissipation}
\toctitle{Nonlinear vortex light beams supported and stabilized by dissipation}
 \titlerunning{Nonlinear vortex light beams supported and stabilized by dissipation}
 \author{Miguel A. Porras
\and Carlos Ruiz-Jim\'enez \and M\'arcio Carvalho}
\authorrunning{M. A. Porras, C. Ruiz-Jim\'enez, and M. Carvalho}
\tocauthor{Miguel A. Porras, Carlos Ruiz-Jim\'enez and M\'arcio Carvalho}

\institute{M.A.~Porras
 \at Grupo de Sistemas Complejos, Universidad Polit\'ecnica de Madrid, ETSI de Minas y Energ\'\i{}a, Rios Rosas 21, 28003 Madrid, Spain,
 \email{miguelangel.porras@upm.es}
  \and Carlos Ruiz-Jim\'enez and M\'arcio Carvalho \at Grupo de Sistemas Complejos, Universidad Polit\'ecnica de Madrid,
ETSI Agron\'omica, Alimentaria y de Biosistemas, Ciudad Universitaria s/n, 28040 Madrid, Spain}

 \maketitle

\abstract{We describe nonlinear Bessel vortex beams as localized and stationary solutions with embedded vorticity to the nonlinear Schr\"odinger equation with a dissipative term that accounts for the multi-photon absorption processes taking place at high enough powers in common optical media. In these beams, power and orbital angular momentum are permanently transferred to matter in the inner, nonlinear rings, at the same time that they are refueled by spiral inward currents of energy and angular momentum coming from the outer linear rings, acting as an intrinsic reservoir. Unlike vortex solitons and dissipative vortex solitons, the existence of these vortex beams does not critically depend on the precise form of the dispersive nonlinearities, as Kerr self-focusing or self-defocusing, and do not require a balancing gain. They have been shown to play a prominent role in  ``tubular" filamentation experiments with powerful, vortex-carrying Bessel beams, where they act as attractors in the beam propagation dynamics. Nonlinear Bessel vortex beams provide indeed a new solution to the problem of the stable propagation of ring-shaped vortex light beams in homogeneous self-focusing Kerr media. A stability analysis demonstrates that there exist nonlinear Bessel vortex beams with single or multiple vorticity that are stable against azimuthal breakup and collapse, and that the mechanism that renders these vortexes stable is dissipation. The stability properties of nonlinear Bessel vortex beams explain the experimental observations in the tubular filamentation experiments.
 }

\section{Introduction}
\label{sec:porras-intro}
  \rcgindex{\myidxeffect{A}!absortion (multi-photon)}
  \rcgindex{\myidxeffect{A}!attractor}
  \rcgindex{\myidxeffect{A}!angular momentum}
  \rcgindex{\myidxeffect{B}!Bessel beams}
  \rcgindex{\myidxeffect{C}!collapse}
  \rcgindex{\myidxeffect{C}!chaos}
  \rcgindex{\myidxeffect{D}!dissipation}
  \rcgindex{\myidxeffect{F}!filamentation}
  \rcgindex{\myidxeffect{K}!Kerr effect (optical)}
  \rcgindex{\myidxeffect{L}!linear-stability}
  \rcgindex{\myidxeffect{O}!optical Kerr effect}
  \rcgindex{\myidxeffect{M}!multi-photon absorption}
  \rcgindex{\myidxeffect{N}!nonlinear Optics}
  \rcgindex{\myidxeffect{N}!nonlinear absortion}
  \rcgindex{\myidxeffect{N}!nonlinear Schr\"odinger equation}
  \rcgindex{\myidxeffect{S}!self-action effects}
  \rcgindex{\myidxeffect{S}!self-focusing}
  \rcgindex{\myidxeffect{S}!symmetry breaking}
  \rcgindex{\myidxeffect{S}!stability (linear analysis)}
  \rcgindex{\myidxeffect{T}!topological charge}
  \rcgindex{\myidxeffect{V}!vortex beams}
  \rcgindex{\myidxeffect{V}!vortex solitons}
  \rcgindex{\myidxeffect{V}!vortex stability}


Self-trapping of optical beams in nonlinear media has been one of central topics in nonlinear optics \cite{porras-malomed2005,porras-mihalache2012}. There is a particularly strong and sustained interest in self-trapped beams with embedded vorticity, tubular beams or vortex solitons, carrying orbital angular momentum, \cite{porras-desyatnikov2005} first introduced theoretically in \cite{porras-kruglov1985,porras-kruglov1992}, and in the problem of achieving the stability of these vortex-carrying structures \cite{porras-kivshar2000,porras-desyatnikov2005}. Their applications have opened new perspectives in information encoding, quantum entanglement, all-optical data-processing \cite{porras-kivshar2003}, optical trapping \cite{porras-curtis2003,porras-daly2015}, and diverse forms of transference of optical angular momentum from light to matter, e. g., to micro- and nano-particles, Bose-Einstein condensates or atoms \cite{porras-allen1999,porras-tabosa1999,porras-masalov1997}.

A major issue with multidimensional solitons is their stability \cite{porras-kuznetsov2011}. The ubiquitous self-focusing cubic nonlinearity gives rise to critical collapse in two dimensions \cite{porras-berge1998,porras-sulem1999,porras-fibich2015}, which destabilizes the solitons families. Vortex solitons are particularly prone to the instability initiated by azimuthal perturbations, which breaks their cylindrical symmetry, splitting it into fragments \cite{porras-kivshar2000,porras-desyatnikov2005}. Stabilization of vortex solitons was shown to be possible with specifically designed or ``tailored" nonlinearities, such as cubic and quintic ones with opposite signs \cite{porras-quirogateixeiro1997,porras-pazalonso2005}, or nonlocal nonlinearities \cite{porras-yakimenko2005}. Their practical implementation, however, requires a careful search for materials, such as liquid CS$_2$ for the cubic-quintic nonlinearity \cite{porras-falcaofilho2013}, or lead-doped glass for the thermal nonlocal nonlinearity \cite{porras-rotschild2005}. A related topic is the stability of vortex solitons in dissipative systems, often modeled by complex Ginzburg-Landau equations with cubic and quintic nonlinearities, which support stable dissipative solitons and vortex solitons \cite{porras-malomed2002,porras-mihalache2006,porras-aranson2002,porras-akhmediev2007,porras-knobloch2015}.

In this Chapter we approach the problem of the achievement of stationary and stable propagation of vortex beams in homogeneous media with self-focusing Kerr nonlinearity from a different perspective, and report on the existence of stable vortex beams in standard optical materials \cite{porras-porras2014,porras-porras2016}. In the context of the research on the phenomenon of filamentation induced by powerful laser pulses \cite{porras-skupin2006,porras-couairon2007}, Ref. \cite{porras-porras2004} reported the existence of light beams that can propagate with unchanged transversal intensity pattern, including any attenuation, in self-focusing Kerr media while their power is being continuously dissipated into matter via multi-photon absorption, eventually ionizing the medium. Multi-photon absorption of different orders takes place in common dielectric media, such as silica, water or air. It is a collapse-arresting mechanism that plays an essential role in filamentation \cite{porras-couairon2007,porras-couairon2006,porras-polesana2008,porras-polesana2006,porras-gaizauskas2009,porras-schwarz2000}. These light beams, originally called ``nonlinear unbalanced Bessel beams", belong to the family of conical waves and as such they transport, ideally, infinite power. They then posses an intrinsic power reservoir that in these beams flows permanently from the linear outer Bessel-like rings towards the nonlinear central peak of intensity, where most of nonlinear power losses due to multi-photon absorption take place, replenishing them. A fundamental difference from complex Ginzburg-Landau-based models is that the propagation with nonlinear losses persists without any balancing gain. Physically realizable versions of these nonlinear Bessel beams with finite amount of power have been shown to play a crucial role in various experiments \cite{porras-polesana2005,porras-polesana2006,porras-polesana2007}, particularly in the filamentation induced by ultrashort pulsed (linear) Bessel beams \cite{porras-polesana2008,porras-polesana2007}, where the nonlinear Bessel beams act as attractors in the nonlinear propagation \cite{porras-porras2015}.

It has recently been shown \cite{porras-porras2014,porras-jukna2014} that the above nonlinear Bessel beam is only the fundamental member of an infinite family of vorticity-carrying nonlinear Bessel beams with arbitrary integer topological charge $s$. Throughout this Chapter we will refer to these beams as nonlinear Bessel vortex beams (nonlinear BVBs). In them, power and angular momentum spiral inward permanently from the power and angular momentum reservoirs towards the inner, more intense rings surrounding the vortex, where they are transferred to matter. Simultaneous absorption of $M$ photons, each photon carrying (arbitrarily high) orbital angular momentum $\hbar s$ ($s=\pm 1, \pm 2,\dots $) could be used as an efficient method of optical pumping of angular momentum \cite{porras-masalov1997,porras-tabosa1999}. Differently from other vortex-carrying beams, as standard vortex solitons, they may exist in common transparent dielectric as optical glasses, liquid and gases at high enough powers. Nonlinear BVBs have subsequently been realized in experiments, and have been employed for laser-powered material processing \cite{porras-xie2015}.

In the first part of this Chapter we review the properties of nonlinear BVBs, their structure and the refilling mechanism that allows the stationary propagation with dissipation. By means of numerical simulations, we also show that nonlinear BVBs are attractors of the dynamics of linear Bessel beams. For each ideal (infinite power), linear BVB introduced in the nonlinear midium, there is a specific, ideal, nonlinear BVB that governs the dynamics. It is the one preserving three the cone angle, the topological charge and the inward power flux of the linear beam \cite{porras-porras2014}.

In the experiments of filamentation with vortex-carrying Bessel beams, three different propagation regimes in the so-called Bessel zone after the generating axicon have been reported to exist \cite{porras-xie2015}. Under certain conditions, usually associated with large cone angles, a steady or tubular filamentation regime is observed, which is identified as a nonlinear BVB. Under other conditions, however, azimuthal breakup takes place, and rotating filaments, or non-rotating, disordered filaments have been observed \cite{porras-jukna2014,porras-xie2015}.

These observations suggest that stable nonlinear BVB may exist. In the second part of this Chapter we review our research on the stability properties of nonlinear BVBs and demonstrate that there indeed exist nonlinear BVBs of any topological charge that are completely stable against the azimuthal breakup and collapse alike in self-focusing Kerr media with nonlinear absorption \cite{porras-porras2016}. We predict the stability by the analysis of the linearized equations for small perturbations, and verify it in direct numerical simulations of the propagation of randomly perturbed nonlinear BVBs. The linear-stability analysis is also used to demonstrate that the stability is imposed by the nonlinear absorption effect. Its stabilizing action  was previously pointed out in other contexts, such as the stabilization against radial and temporal perturbations of zero-vorticity beams \cite{porras-polesana2007,porras-porras2015} and light bullets \cite{porras-porras2013}, but not against azimuthal breaking of nonlinear vortex beams. In the stability analysis we focus on the most destabilizing cubic self-focusing, which, by it self, cannot support any stable 2D patterns, but similar results can be readily obtained for more general nonlinearities, since the existence of nonlinear BVBs does not critically depend on the particular nonlinearities, such as the order of the multi-photon absorption or the form of the Kerr nonlinearity \cite{porras-porras2004,porras-porras2014,porras-jukna2014}.

In relation to the experiments \cite{porras-xie2015}, our linear-stability analysis and diagnostic numerical simulations of the nonlinear propagation allow us to propose a common explanation of the observed tubular, rotary and speckle-like dynamical regimes in the Bessel region after the axicon \cite{porras-xie2015}. In all the three cases, there exists an attracting nonlinear BVBs that determines the dynamics: as in the ideal case, it is the nonlinear BVB  with the same cone angle, topological charge, and inward power flux as the linear Bessel beam that would be formed about focus of the axicon (middle of the Bessel zone) under linear conditions of propagation. This fact is confirmed by the numerical observation that the dynamics in the Bessel zone corresponds to that expected from the development of instability, if any, of that specific nonlinear BVB. The tubular regime is observed if the attracting BVB is stable according to the linear-stability analysis, or its instability is weak to develop over the Bessel distance. Otherwise, the azimuthal symmetry breaking leading to rotary or speckle-like regimes observed in the Bessel zone closely mimics that observed in the development of the instability of the nonlinear BVB, which acts in this case as an unstable, chaotic attractor.

\section{Propagation of Bessel vortex beams in nonlinear media with nonlinear losses}
\label{sec:porras-prop}

To illustrate how nonlinear BVBs arise spontaneously in optical media, we first analyze the propagation of high-order (vortex-carrying) Bessel beams \cite{porras-durnin1987,porras-durninmiceli1987}, or linear Bessel vortex beams (linear BVBs for short), at sufficiently high intensities, typically of the order of TW/cm$^2$, in a (linearly) transparent dielectric. We assume that the propagation is suitably modeled by the nonlinear Schr\"odinger equation (NLSE) with cubic, and possibly quintic, dispersive nonlinearities, and a dissipative term that accounts for nonlinear power losses associated with multi-photon absorption processes at these high intensities. For the light beam $E =A\exp[-i(\omega t - kz)]$ of angular frequency $\omega$ and linear propagation constant $k = (\omega/c)n$, where $n$ is the linear refractive index and $c$ is the speed of light in vacuum, and of complex envelope $A$, this NLSE reads as
\begin{equation}
\frac{\partial A}{\partial z}=\frac{i}{2k}\Delta_\perp A+if(|A|^2)A-\frac{\beta^{(M)}}{2}|A|^{2M-2}A \, ,
\label{NLSE2}
\end{equation}
where $\Delta_\perp=\partial^2_r+(1/r)\partial_r+(1/r^2)\partial^2_\varphi$ is the transversal Laplace operator, $(r,\varphi)$ are polar coordinates in the transversal plane, $\beta^{(M)}>0$ is the $M$-photon absorption coefficient, and dispersive cubic and quintic nonlinearities are included in the term with $f(u)\equiv k (n_2u+n_4u^2)/n$, $n_2$ and $n_4$ being nonlinear refractive indexes. The multi-photon order $M$ is determined by the medium and the light wavelength \cite{porras-couairon2007}. In air, for instance, $M$ ranges from $3$ to $8$ in the wavelength range $248$--$800$ nm \cite{porras-polesana2008,porras-schwarz2000}.

Linear Bessel beams are, in the paraxial form implicit in Eq. \eqref{NLSE2}, the solutions $A(r,\varphi,z) \propto J_s(\sqrt{2k|\delta|} r)e^{is\varphi}e^{i\delta z}$ of Eq. \eqref{NLSE2} when all nonlinear terms are neglected, where $\delta =-k\theta^2/2<0$ is the shortening of the axial wave vector due to the conical structure, $\theta= \sqrt{2|\delta|/k}$ is the cone angle of the Bessel beam, and $s=\pm1, \pm 2 \dots$ is the topological charge of the vortex at $r=0$. For a more comprehensive analysis of the nonlinear propagation of a linear BVB of given cone angle, we introduce dimensionless variables
 \begin{equation}  \label{SCALING}
\rho \equiv k\theta r=\sqrt{2k|\delta |}r, \quad \zeta \equiv |\delta |z,\quad \tilde{A}\equiv \left(\frac{\beta^{(M)}}{2|\delta |}\right)^{\frac{1}{2M-2}}A\, ,
\end{equation}
with which Eq. \eqref{NLSE2} rewrites as
\begin{equation}
\frac{\partial \tilde{A}}{\partial \zeta}=i\Delta_\perp\tilde{A}+i\tilde{f}(|\tilde{A}|^2)\tilde{A}-|\tilde{A}|^{2M-2}\tilde{A} \, ,
\label{NLSE3}
\end{equation}
where the transversal Laplace operator is now $\Delta_\perp\equiv \partial^2_\rho+(1/\rho)\partial_\rho+(1/\rho^2)\partial^2_\varphi$, and $\tilde{f}(u)\equiv \alpha_2 u + \alpha_4 u^2$, with
\begin{equation}
\alpha_2\equiv \frac{kn_2}{n|\delta|}\left(\frac{2|\delta|}{\beta^{(M)}}\right)^{\frac{1}{M-1}} \, ,\quad
\alpha_4\equiv \frac{kn_4}{n|\delta|}\left(\frac{2|\delta|}{\beta^{(M)}}\right)^{\frac{2}{M-1}} \, ,
\end{equation}
are determined by the medium properties at the light wavelength and the cone angle. In water at $527$ nm, for example, with the values $n=1.33$, $n_2=2.7\times 10^{-16}$ cm$^2/$W, $n_4\simeq 0$, $M=4$ and $\beta^{(4)}=2.4\times 10^{-36}$ cm$^5/$W$^3$ \cite{porras-polesana2007}, and with typical cone angles $\theta=3^\circ,2^\circ,1^\circ,0.5^\circ$, we obtain the respective values $\alpha_2\simeq 0.76,1.31,3.30,8.31$, and we can assume $\alpha_4\simeq 0$. Similar values are obtained for other media at other wavelengths, and are therefore considered below. In these dimensionless variables, the linear BVB solution of Eq. \eqref{NLSE3} without nonlinear terms is written as $b_0J_s(\rho) \exp (is \varphi) \exp (-i \zeta)$, where $b_0$ determines its amplitude.

The existence of nonlinear BVBs is suggested by the following numerical simulations. Solving the NLSE in Eq. \eqref{NLSE3} with the linear BVB $\tilde A(\rho,\varphi,0)= b_0 J_s(\rho)e^{is\varphi}$ as initial condition at the entrance plane $\zeta=0$ of the medium, we observe that it is not completely absorbed, as would happen to a plane wave or to a Gaussian beam; on the contrary, it transforms spontaneously into a nondiffracting and non-attenuating beam. This fact was originally predicted and observed for the fundamental (vortex-less) Bessel beams a decade ago \cite{porras-porras2004,porras-polesana2007}, but this phenomenon is seen here to occur also with BVBs \cite{porras-porras2014,porras-jukna2014}.
\begin{figure}[tb]
\begin{center}
\includegraphics[width=4cm]{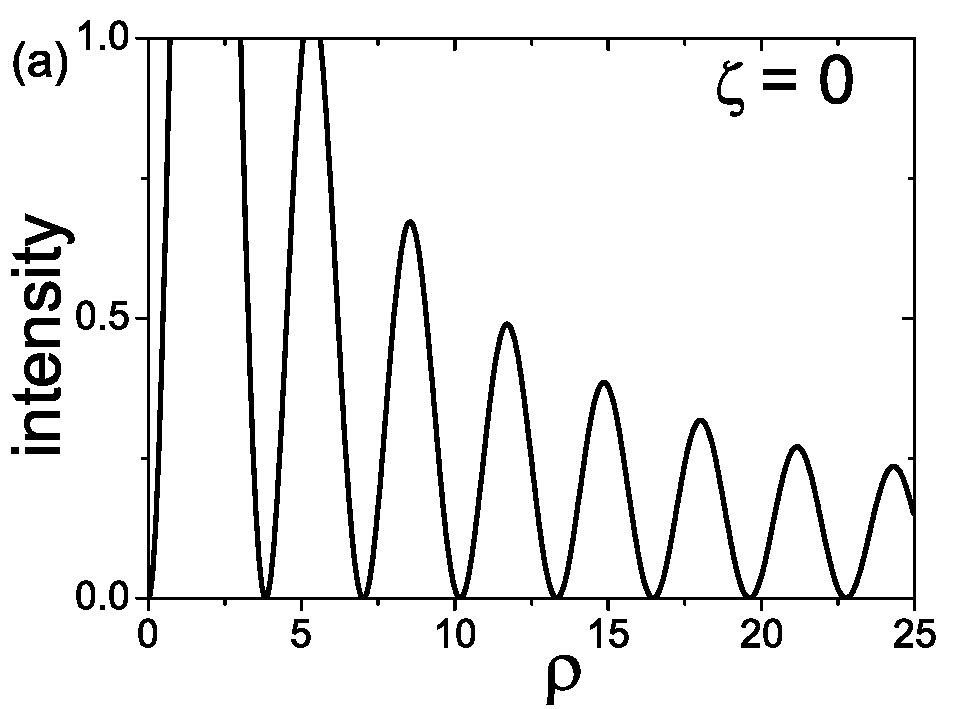}\includegraphics[width=4cm]{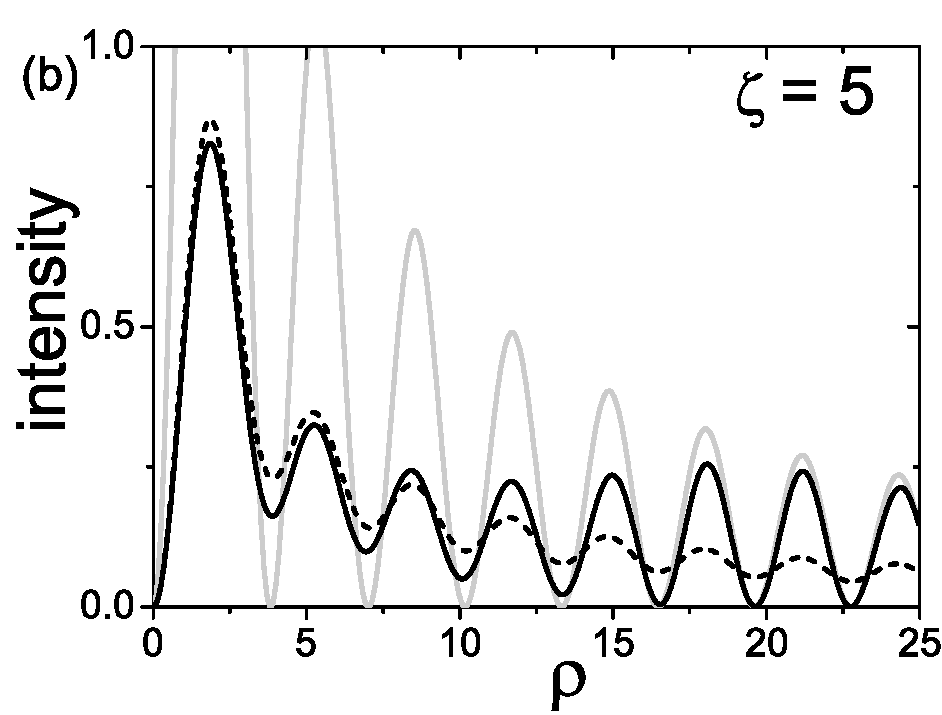}\\
\includegraphics[width=4cm]{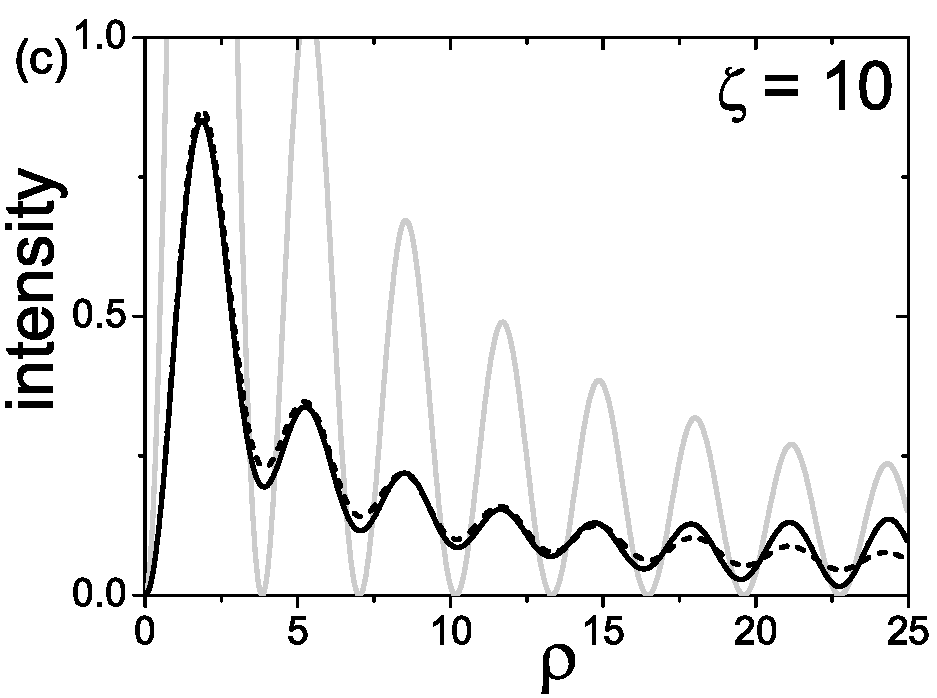}\includegraphics[width=4cm]{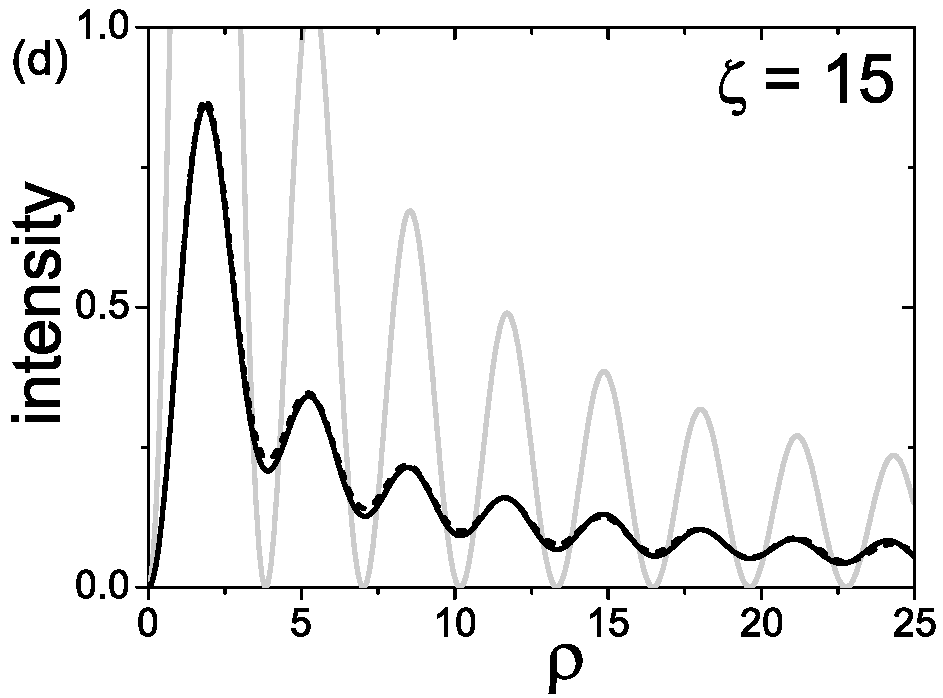}
\includegraphics[width=8cm]{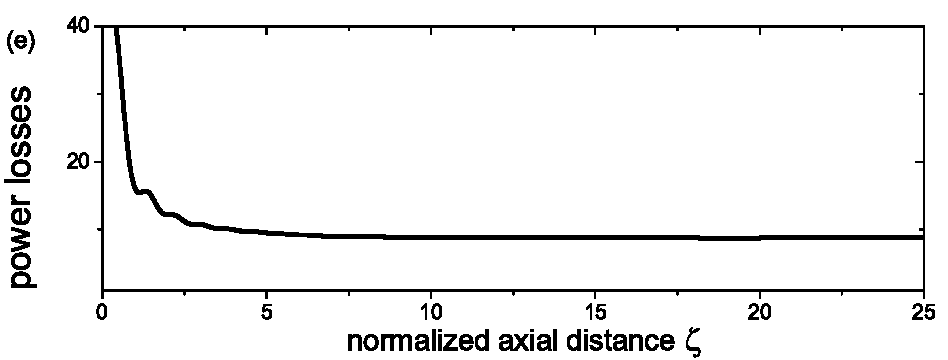}
    \caption{\label{Fig1}(a-d) Radial intensity profiles at different propagation distances $\zeta$ when a linear BVB with topological charge $s=1$ and amplitude parameter $b_0=3$ is introduced into a medium with four-photon absorption ($M=4$) with negligible Kerr nonlinearity ($\alpha_2=0,\alpha_4=0$). The gray lines represent the initial linear BVB and the dashed lines the final nonlinear BVB with the same cone angle, topological charge and $b_0=1.6$. (e) Nonlinear power losses per unit length in Eq. \eqref{NLL} as a function of distance $\zeta$.}
\end{center}
 \end{figure}
In Figs. \ref{Fig1} (a-d), the radial profiles of intensity are represented at different propagation distances. For simplicity, a medium with negligible Kerr nonlinearity ($\alpha_j = 0$) is first considered. Once the propagating beam passed through a stage of strong absorption, a final steady state with a vortex is reached. This non-diffracting and non-attenuating final beam does not require Kerr non-linearities for its stationarity, as vortex solitons. The beam maintains its radial profile despite it is experiencing nonlinear losses per unit length given by
\begin{equation}
N(\infty)=2\pi\int_0^\infty d\rho \rho |\tilde{A}|^{2M} \, .
\label{NLL}
\end{equation}
As seen in Fig. \ref{Fig1}(e), these losses reach a positive constant value in the final steady regime. In Figs. \ref{Fig1} (a-d), we observe that the transofrmation into the final state starts in the central lobes and spreads conically along the cone $z=r/\theta$ (or $\zeta=\rho/2$ in our dimensionless variables) on propagation. The formation of a stationary state with similar characteristics takes also place in self-focusing and self-defocusing media, $\alpha_2>0$ and $\alpha_2<0$, as seen more concisely in Figs. \ref{Fig2} (a) and (b), with the only significant difference that central lobes of the final beam state are compressed or widened by the action for the respective Kerr nonlinearities. In all cases, the final beam state is a nonlinear BVB, whose intensity profile is depicted in the example of Fig. \ref{Fig1} by dashed lines, and whose properties will be studied in the next section.
\begin{figure}[tb]
\center
\includegraphics*[width=5.5cm]{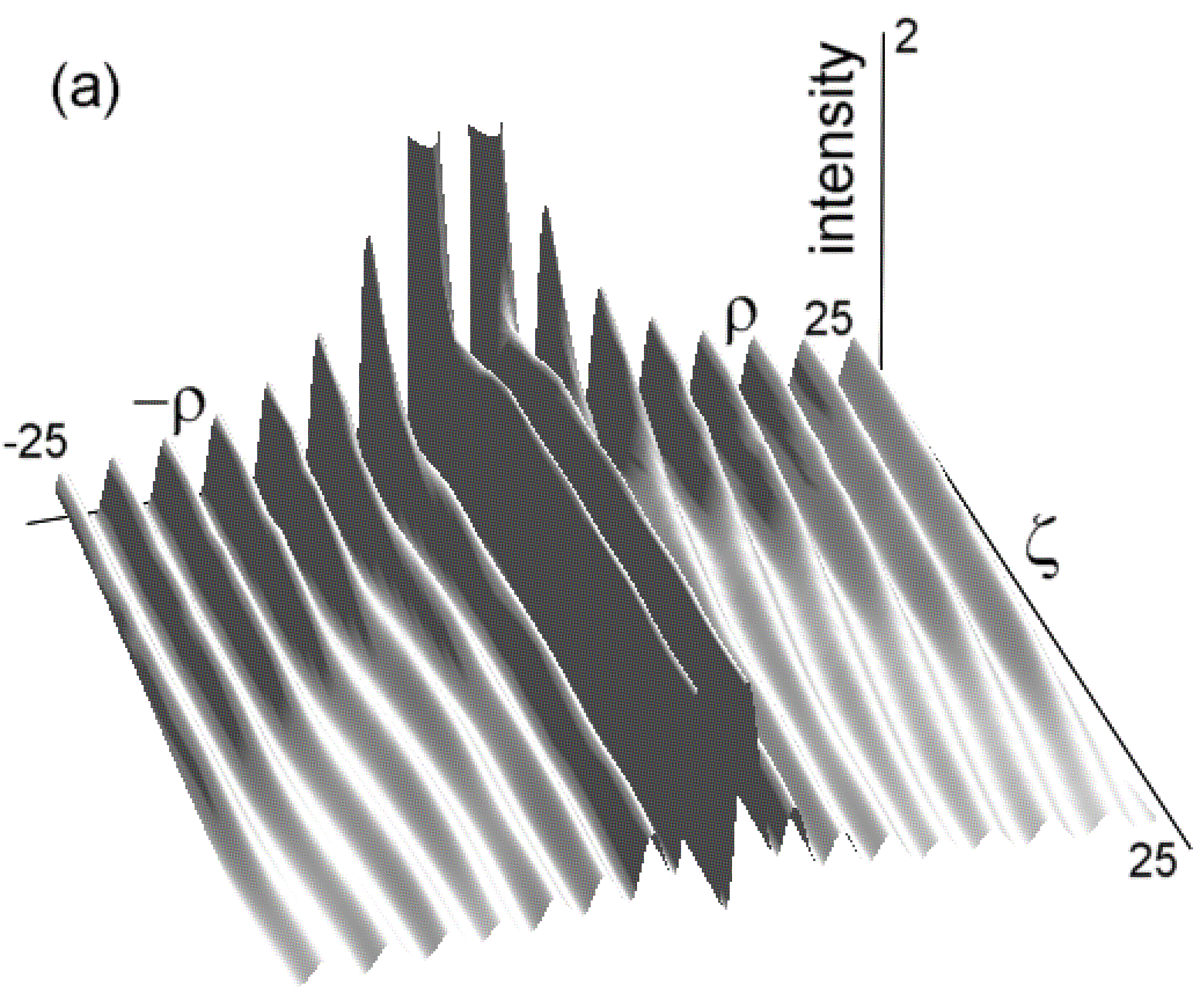}\includegraphics*[width=5.5cm]{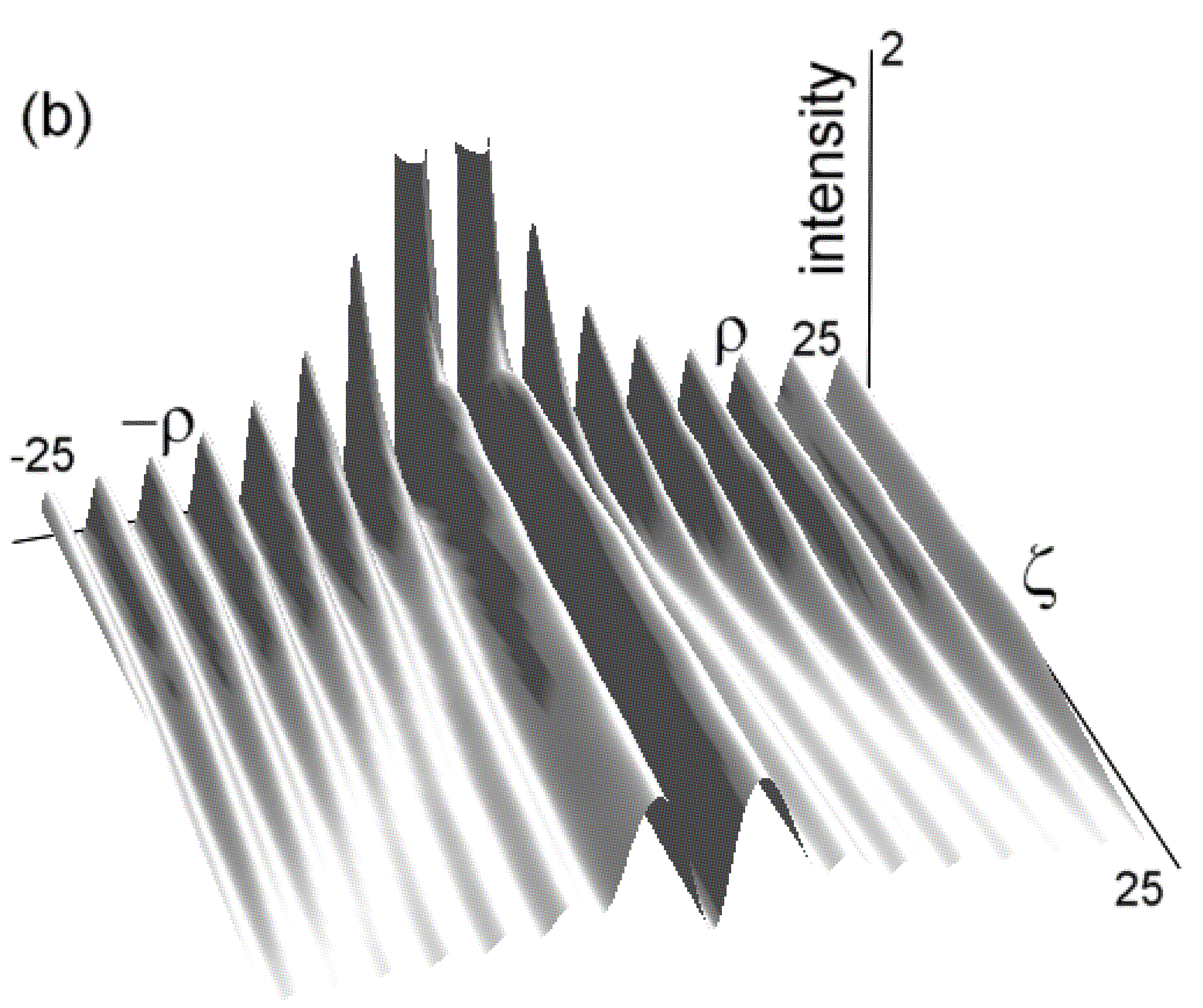}
\caption{\label{Fig2} Nonlinear dynamics of the intensity profile of the same linear BVB as in Fig. \ref{Fig1} introduced in (a) a saturable self-focusing medium with $\alpha_2 = 2$, $\alpha_4 = -1$ and in (b) a saturable self-defocusing medium with $\alpha_2 = -2$, $\alpha_4 = 1$ with four photon absorption.}
\end{figure}

\section{Nonlinear BVBs supported by nonlinear losses}
\label{sec:porras-bvbs}

As described in Refs. \cite{porras-porras2014,porras-jukna2014}, we search for nonlinear stationary solutions of the NLSE \eqref{NLSE2} whose intensity profile does not depend on the propagation coordinate $z$ of the form
\begin{equation}
A(r,\varphi,z)=a(r){\rm e}^{i\phi(r)}{\rm e}^{is\varphi}{\rm e}^{i\delta z} \, ,
\label{ANSATZ}
\end{equation}
where $\phi(r)$ and $a(r)>0$ are the radial phase and amplitude profiles to be determined, and $\delta$ is a constant. We do not initially assume that this constant is negative. In dimensionless variables the above Ansatz is
\begin{equation}
\tilde{A}(\rho,\phi,\zeta)=\tilde{a}(\rho){\rm e}^{i\phi(\rho)}{\rm e}^{is\varphi}{\rm e}^{i\,\text{sgn}(\delta)\zeta} \, ,
\label{ANSATZ2}
\end{equation}
where $\text{sgn}(\delta)$ is the sign of $\delta$. Substituting Eq. \eqref{ANSATZ2} in the NLSE \eqref{NLSE3} and separating real and imaginary parts we obtain the following ordinary differential equations for the amplitude and phase:
\begin{eqnarray}
\frac{d^2\tilde a}{d\rho^2}+\frac{1}{\rho}\frac{d\tilde a}{d\rho}-\left(\frac{d\phi}{d\rho}\right)^2 \tilde a +\tilde f(\tilde a^2) \tilde a - \text{sgn}(\delta)\tilde a -\frac{s^2}{\rho^2}\tilde a &=&0 \, , \label{AMP}\\
\frac{d^2\phi}{d\rho^2}+ \frac{1}{\rho}\frac{d\phi}{d\rho} + 2\frac{d\phi}{d\rho}\frac{d\tilde a}{d\rho}\frac{1}{\tilde a} + \tilde a^{2M-2}&=&0\,.
\label{PHAS}
\end{eqnarray}
Physically valid solutions must obey the boundary condition of localization, i.e., $\tilde{a}(\rho) \rightarrow 0$ as $\rho\rightarrow \infty$. In addition, around the vortex the amplitude is very small, and therefore the nonlinear effects are negligible; we then will demand the boundary condition $a(\rho)\simeq b_0 J_s(\rho)$ about the vortex, or, in other words, that the beam behaves as a linear BVB of a certain amplitude $b_0$. Since $J_s(\rho)\simeq \rho^{|s|}/(2^{|s|}|s|!)$ for $\rho\rightarrow 0$ \cite{porras-olver2010}, we can also write this boundary condition as
\begin{equation}
\tilde{a}(\rho)\simeq \frac{b_0}{2^{|s|}|s|!}\rho^{|s|} \quad \mbox{for} \quad \rho\rightarrow 0.
\label{INICIAL_HBLN}
\end{equation}
In absence of dissipation, this problem has solution with $\text{sgn}(\delta) = 1 $ for a discrete spectrum of values ​​of $b_0$, which constitute the family of vortex solitons \cite{porras-desyatnikov2005}, characterized by strongly localized transversal profiles around the vortex, and with $\text{sgn}(\delta) = -1$ for continuous values of $b_0$ from $0$ to $\infty$, which are called nonlinear Bessel beams in transparent media. In this case, only the amplitude equation (\ref{AMP}) is usually written because a constant phase satisfies Eq. \eqref{PHAS} \cite{porras-desyatnikov2005}. When absorption is included, there exists no solution with $\text{sgn}(\delta) = +1$, i.e., the vortex solitons family does not survive dissipation effects \cite{porras-silberberg1990}. However, with $\text{sgn}(\delta) = -1$, solutions to this problem still exist with a continuous spectrum of values ​​of $b_0$ up to a maximum value, $b_{0,\text{max}}$, whci hconstitute the family of nonlinear BVBs in Kerr nonlinear media with nonlinear absorption \cite{porras-jukna2014,porras-porras2014}.
\begin{figure}[tb]
\begin{center}
\includegraphics[width=5cm]{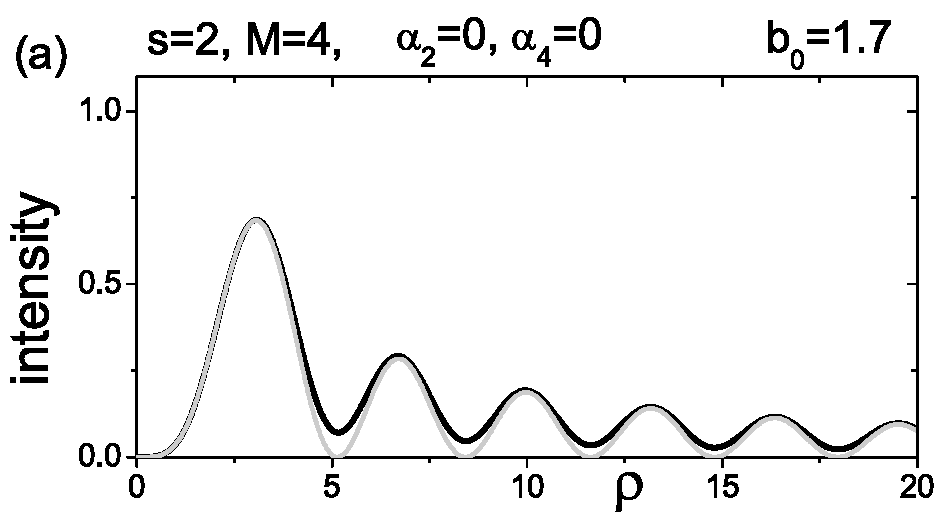}\includegraphics[width=5cm]{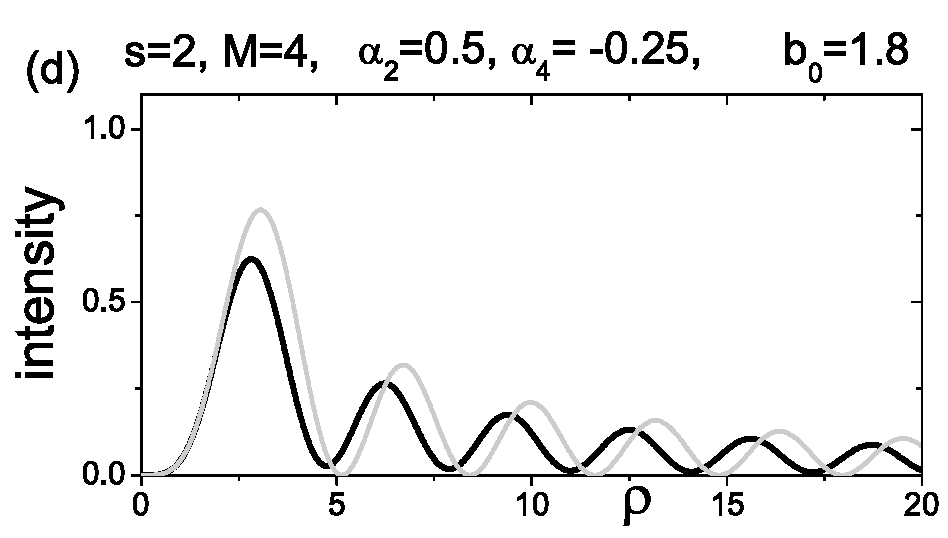}
\includegraphics[width=5cm]{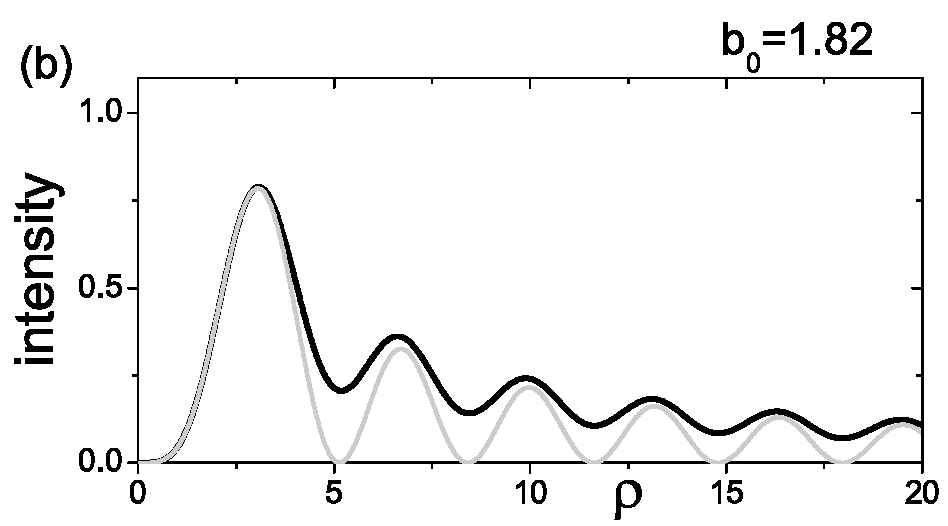}\includegraphics[width=5cm]{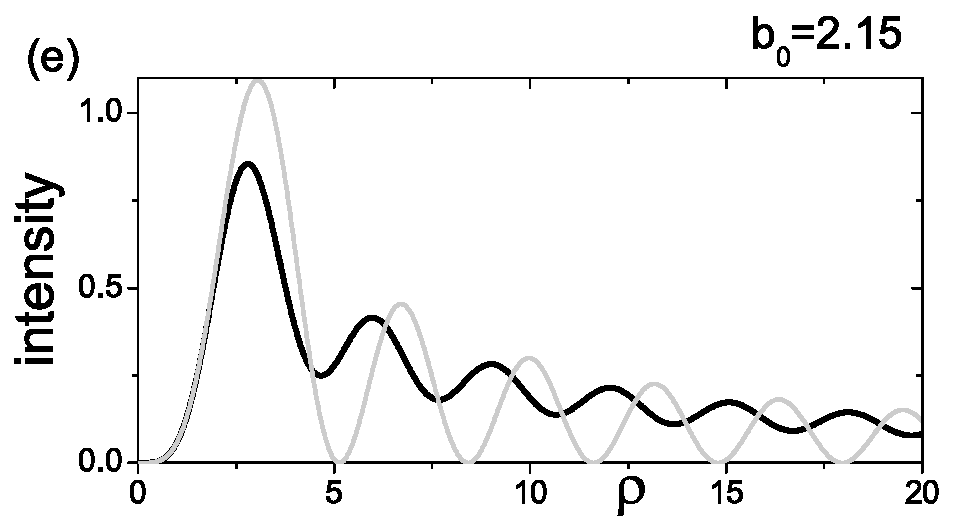}
\includegraphics[width=5cm]{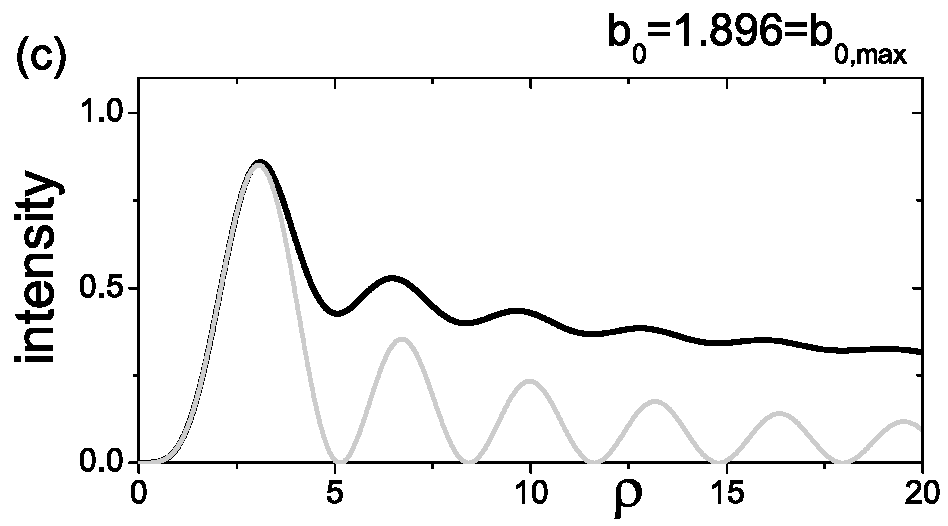}\includegraphics[width=5cm]{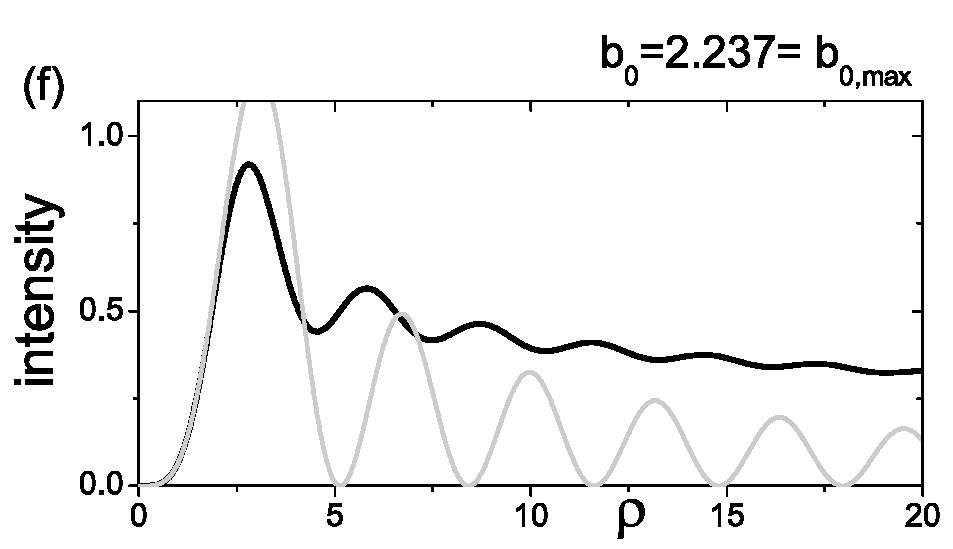}
\end{center}
\caption{\label{Fig3} For $\alpha_2=\alpha_4=0$, and $M=4$, intensity profiles of nonlinear BVBs with $s=2$ and (a) $b_0=1.7$, (b) $b_0=1.82$ and (c) $b_0=1.896$. For $\alpha_2=0.5$, $\alpha_4=-0.25$ and $M=4$, intensity profiles of nonlinear BVBs with $s=2$ and (d) $b_0=1.8$, (e) $b_0=2.15$ and (f) $b_0=2.237$. The dotted curves represent linear BVBs with $s=2$ and the same $b_0$ as the nonlinear BVBs.}
\end{figure}

A few examples of their intensity radial profiles $\tilde a^2(\rho)$ are shown in Fig. \ref{Fig3} (solid curves) and are compared to the linear BVB with the same vortex core (same $b_0$). At low amplitude (small $b_0$), the nonlinear BVB has a linear behaviour not only at the vortex core but at any radial distance. When $b_0$ increases up to $b_{0,\rm max}$, the contrast of the rings is gradually lost, and the inner rings become compressed (enlarged) in self-focusing (self-defocusing) media. In all cases, the outer rings at $\rho\rightarrow\infty$ decay in amplitude as $\rho^{-1/2}$ and oscillate at the same radial frequency as those of the linear BVB. In self-focusing or self-defocusing media, these outer rings are therefore radially shifted with respect to those of the linear BVB in order to match the compressed or enlarged inner rings.

These examples also illustrate that typical values of $b_0$ corresponding to intensities where the interplay between self-focusing and nonlinear absorption plays an essential role are of the order of unity. For example, with $M=4$, $s=1$, $\alpha_2=3.30$ and $\alpha_4=0$, values $b_0=1.2$ and $1.6$ pertain, respectively, to nonlinear BVBs with cone angle $\theta=1^\circ$ and peak intensities $0.77$ TW/cm$^2$ and $1.16$ TW/cm$^2$ in water at $527$ nm \cite{porras-polesana2007}.

Therefore, these nonlinear BVBs can propagate without any diffraction and any attenuation while their energy and angular momentum are continuously transferred to matter nonlinearly via multiphoton absorption. As pointed out above, the existence of these beams does not critically depend on the specific dispersive nonlinearities or on the order of multiphoton absorption. Only the details of the transversal structure and the specific maximum value $b_{0,\rm max}$ for existence depend on the choices of $s$, $\alpha_2$, $\alpha_4$ (or other dispersive nonlinearities), and multi-photon order $M$.

\section{Stationary propagation of nonlinear BVBs with nonlinear absorption: energy and angular momentum transfer to matter}
\label{sec:porras-stat}

\begin{figure}[tb]
\center
\includegraphics*[width=\textwidth]{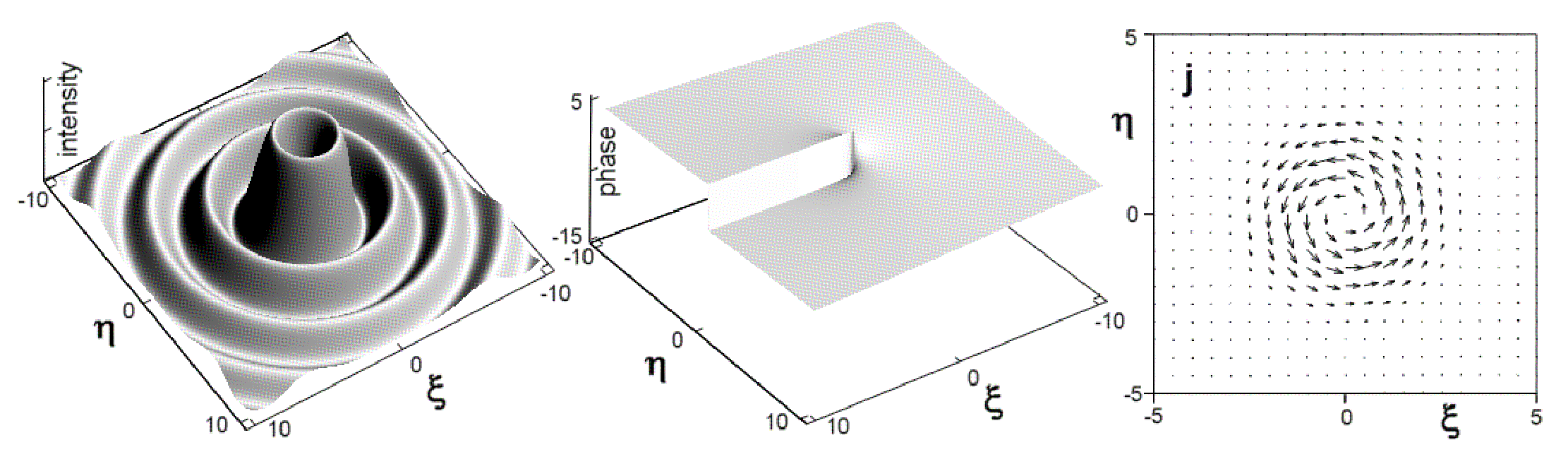}
\includegraphics*[width=\textwidth]{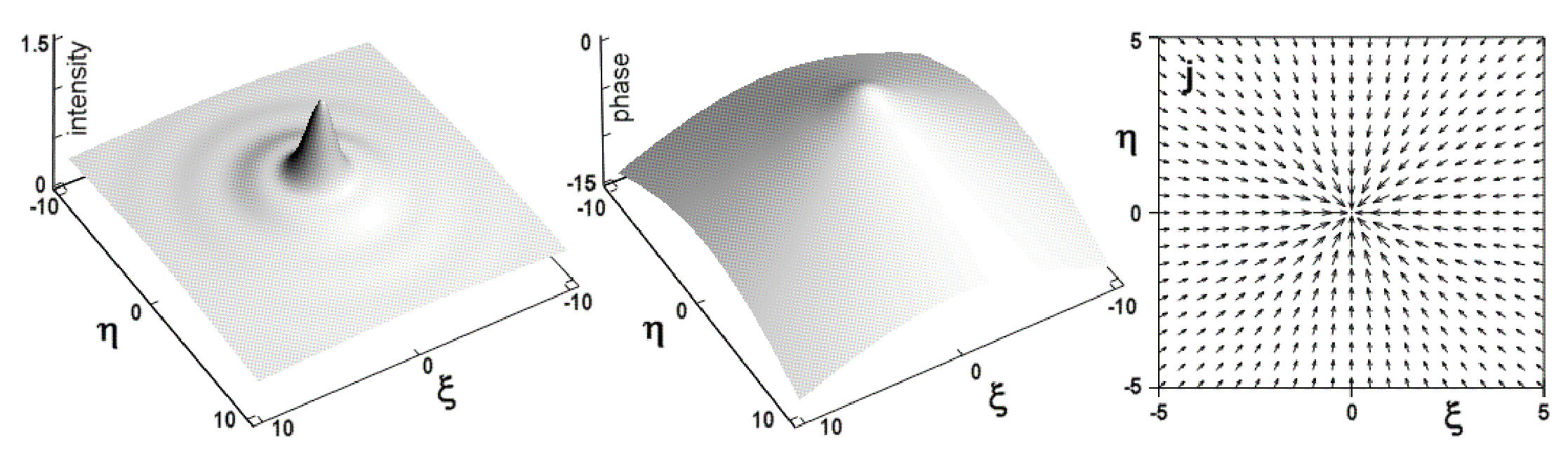}
\includegraphics*[width=\textwidth]{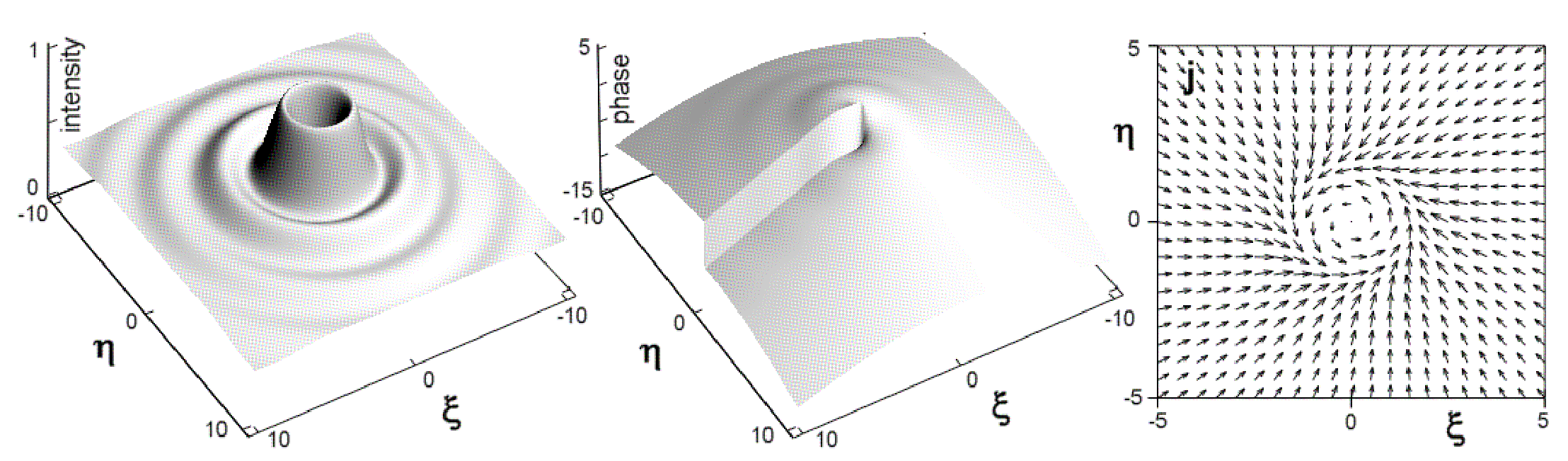}
\caption{\label{Fig4} Intensity profiles (first column), phase profiles (second column) and intensity current (third column) of the linear BVB with $s=1$, and $b_0=1.666$ (first row), of the vortex-less ($s=0$) nonlinear Bessel beam with $M=4$, $\alpha_2=\alpha_4=0$ and $b_0=1.174$ (second row) and of the nonlinear BVB with $s=1$, $M=4$, $\alpha_2=\alpha_4=0$ and $b_0=1.666$ (third row). The normalized cartesian coordinates in the transversal plane are $\xi=\sqrt{2k|\delta|}x$ and $\eta=\sqrt{2k|\delta|}y$ in the transversal plane.}
\end{figure}

The energy and angular momentum transferred continuously to matter is continuously replenished from an intrinsic reservoir of energy and angular momentum, which sustains the stationary propagation. To study this mechanism, we note that integration of Eq. \eqref{PHAS} in $\rho$ leads to
\begin{equation}\label{REFILLINGE}
-2\pi\rho \frac{d\phi}{d\rho}\tilde a^2 = 2\pi \int_0^{\rho} d\rho\rho \tilde a^{2M} \, ,
\end{equation}
or $-F(\rho)=N(\rho)$ for short. This relation expresses that the nonlinear power losses in any circle of radius $\rho$, $N(\rho)$, are replenished  by an inward power flux $F(\rho)$ crossing its circumference and coming from a power reservoir at large radial distances, as is illustrated in Fig. \ref{Fig4}. This is substantially the mechanism of stationarity of nonlinear BVB, which is only possible in beams carrying infinite power ---the reservoir--- as conical beams, and as originally proposed in Ref. \cite{porras-porras2004}. This mechanism has some peculiarities with nonlinear BVBs \cite{porras-porras2014}. If we write the complex envelope of a light beam (not necessarily stationary) in the form $\tilde A=\tilde a e^{i\Phi}$, the NLSE in Eq. \eqref{NLSE3} gives the continuity equation
\begin{equation}
\frac{1}{2}\partial_\zeta \tilde a^2 + \nabla_\perp \cdot \mathbf{j} = -\tilde a^{2M}
\end{equation}
for the intensity $\tilde a^2$, where the transversal current of the intensity is given by $\mathbf{j}=\tilde a^2 \nabla_\perp \Phi$, and $\nabla_\perp$ is the transversal gradient. The general condition for stationarity of the intensity pattern is then
\begin{equation}\label{REFILLINGDIF}
-\nabla_\perp\cdot \mathbf{j}=\tilde a^{2M}\, ,
\end{equation}
expressing in differential form, and by virtue of the divergence theorem, that the nonlinear power losses in any finite region of the transversal plane are refuelled by an inward power flux through its contour. Equation \eqref{PHAS} is the same as Eq. \eqref{REFILLINGDIF} for beams with radially symmetric intensity $\tilde a^2$ and phase $\Phi=\phi(\rho)+s\varphi - \zeta$. For these beams $\mathbf{j} = \tilde a^2 (d\phi/d\rho\mathbf{u}_\rho + s/\rho\mathbf{u}_\varphi)$, and Eq. \eqref{REFILLINGE} is obtained by integrating Eq. \eqref{REFILLINGDIF} over a circle of radius $\rho$.

Figure \ref{Fig4} shows the intensity (first column), phase (second column) and the intensity current (third column) for previously known stationary beams and for a nonlinear BVB. For linear BVBs (and also for vortex solitons) in transparent media (first row), the intensity current $\mathbf{j}=(\tilde a^2 s/\rho)\mathbf{u}_\varphi$ is azimuthal and solenoidal ($\nabla_\perp \cdot \mathbf{j}=0$). For the fundamental (vortex-less) nonlinear Bessel beam (second row), the intensity current in the transversal plane $\mathbf{j}=(\tilde a^2d\phi/d\rho)\mathbf{u}_\rho$ is radial inwards with a divergence that equals the nonlinear losses density $\tilde a^{2M}$. For the present nonlinear BVBs (third row), the intensity current has both azimuthal and radial inwards components, so that the power spirals inwards permanently from the intrinsic power reservoir towards the inner rings, where most of dissipation of energy into matter takes place.

A similar situation occurs with the angular momentum. The nonlinear BVB carries axial orbital angular momentum of density given by \cite{porras-desyatnikov2005} $L=\tilde a^2\partial\Phi/\partial\varphi=s\tilde a^2$ that is proportional to the intensity and to the topological charge $s$. Thus, as the intensity, the stationary angular momentum density is permanently flowing spirally with a current
\begin{equation}
s\mathbf{j} = s\tilde a^2 \left(\frac{d\phi}{d\rho}\mathbf{u}_\rho + \frac{s}{\rho}\mathbf{u}_\varphi \right)
\end{equation}
proportional to the current of the intensity, and this angular momentum current replenish their losses $s\tilde a^{2M}$ associated to the nonlinear absorption process according to the continuity equation $-\nabla_\perp \cdot (s \mathbf{j})= s\tilde a^{2M}$ for the angular momentum. Also, integration over a circle of radius $\rho$ leads to
\begin{equation}
-2\pi\rho s \frac{d\phi}{d\rho}\tilde a^2 = 2\pi \int_0^{\rho} d\rho\rho s \tilde a^{2M}\, ,
\end{equation}
which is analogous to Eq. (\ref{REFILLINGE}) for the energy, and expressing that the radial flux of angular momentum through any circle of radius $\rho$ replenishes their nonlinear losses within that circle when transferred to the material medium.

\section{Asymptotic behavior at large radius}
\label{sec:porras-asym}

\begin{figure}[t]
\sidecaption
\includegraphics*[width=3.5cm]{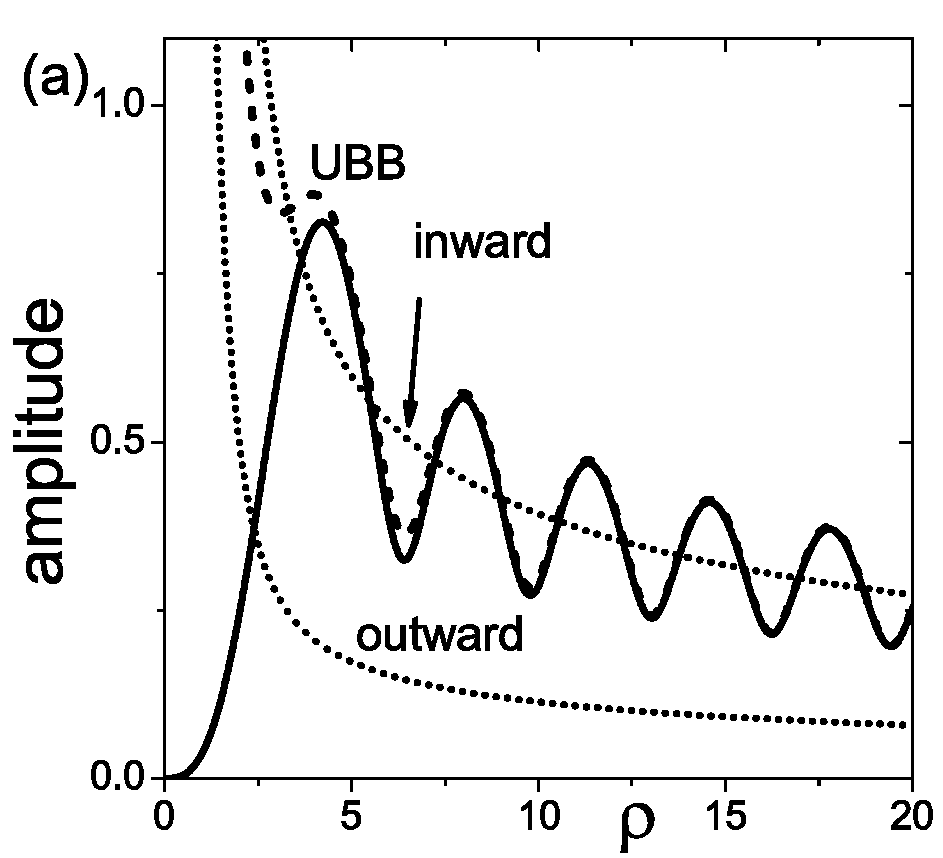}
\includegraphics*[width=4cm]{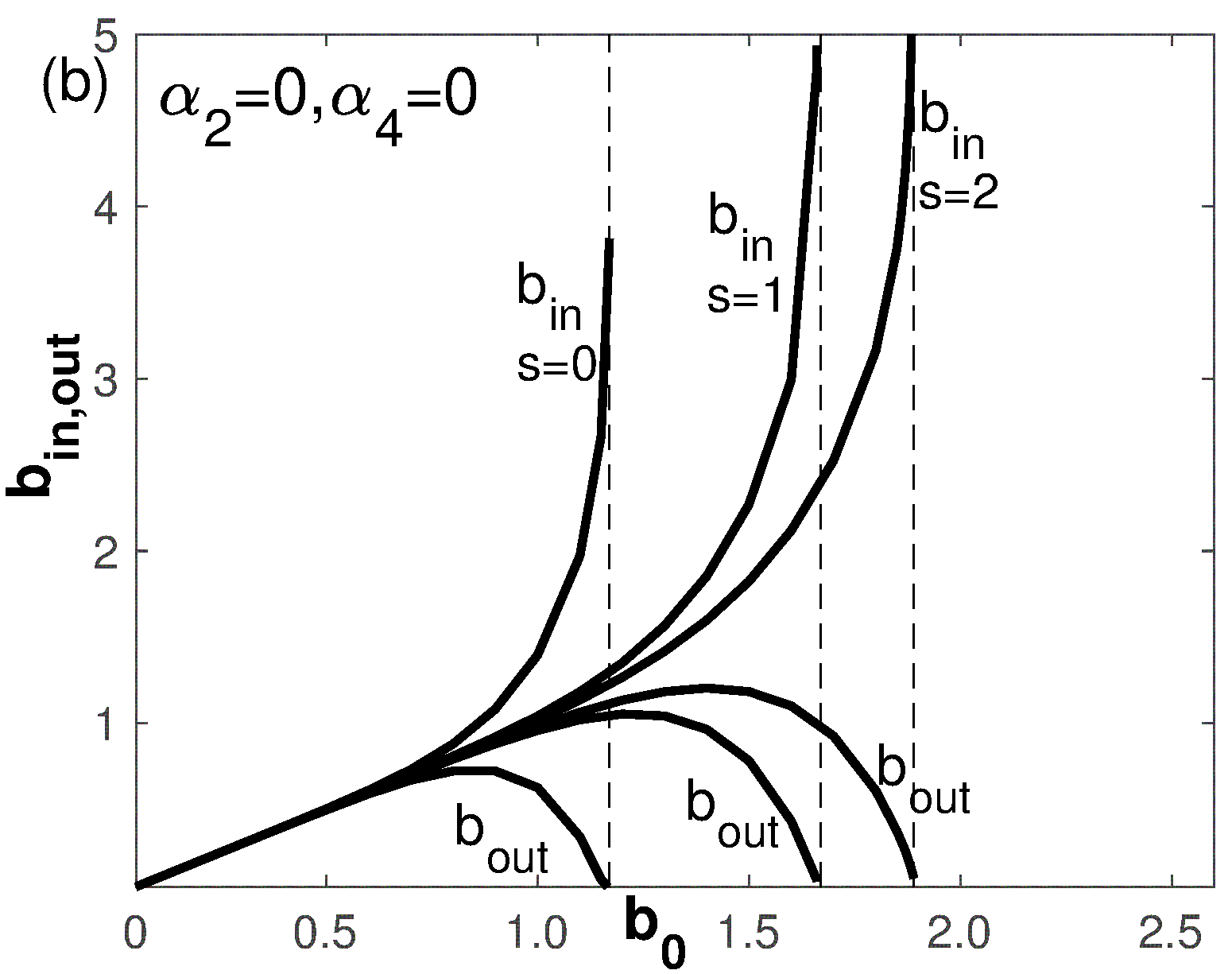}\includegraphics*[width=4cm]{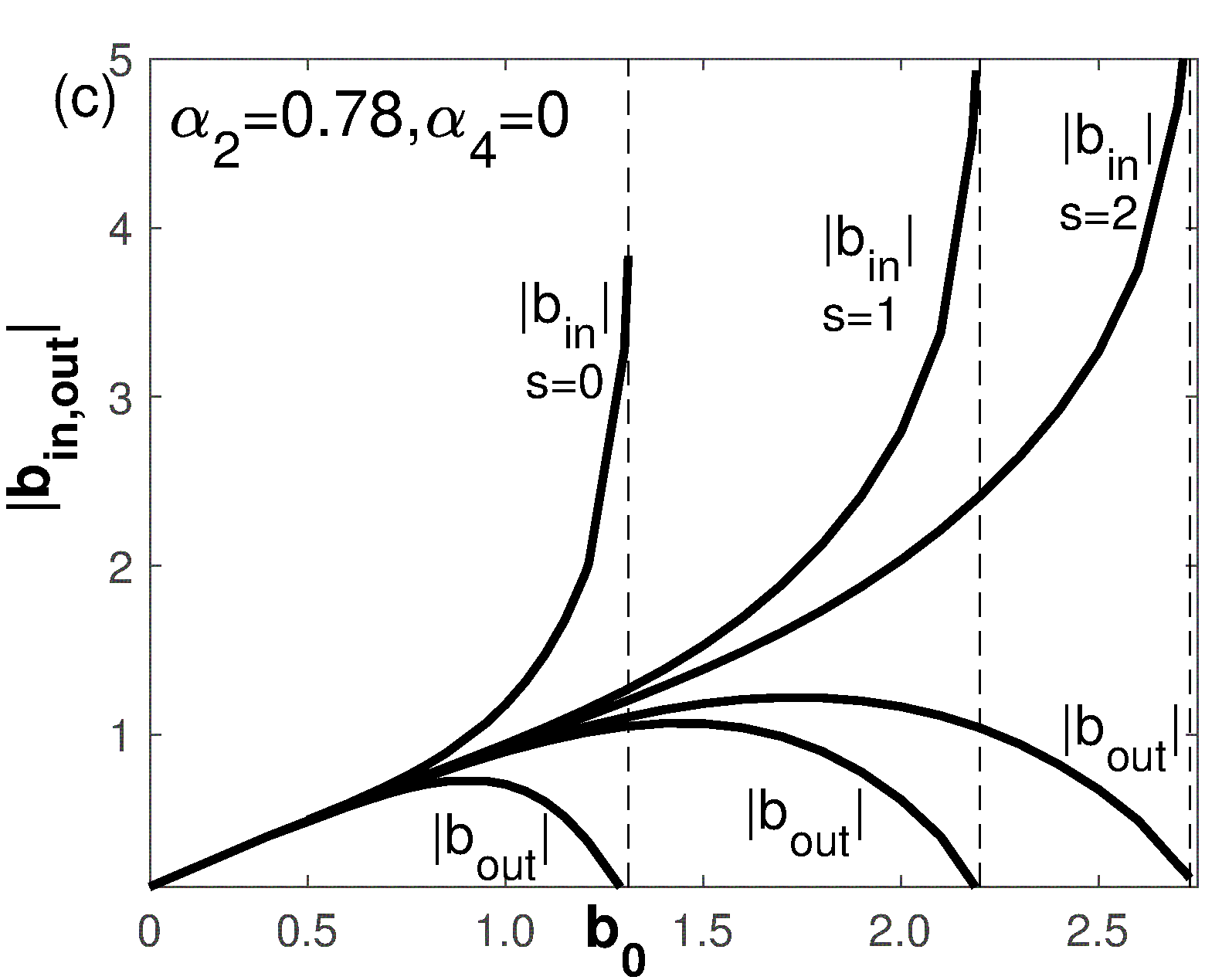}
\caption{\label{Fig5} (a) Amplitude of a nonlinear BVB ($M=4$, $\alpha_2=\alpha_4=0$, $l=3$ and $b_0=1.9$) (solid curve), its asymptotic linear unbalanced Bessel beam (UBB) (dashed curve), and its H\"ankel inward and outward components (dotted curves). (b) and (c) Values of $|b_{\rm out}|$ and $|b_{\rm in}|$ as functions of the amplitude $b_0$ of nonlinear BVBs with the indicated values of $s$ in media with the indicated values of $\alpha_2$, $\alpha_4$, and $M=4$. The curves end at the value $b_{0,\rm max}$ for each charge $s$ (vertical dashed lines).}
\end{figure}

The asymptotic behavior of nonlinear BVBs at large radii is also determined by their non-diffracting and non-attenuating properties in media with nonlinear absorption. As will be clear in Sect. \ref{sec:porras-selec}, this asymptotic behavior is of fundamental importance to understand the nonlinear dynamics of linear BVBs introduced in the medium. At the vortex core the nonlinear BVB behaves as the linear BVB $\tilde A\simeq b_0J_s(\rho)e^{is\varphi}e^{-i\zeta}$, or what is the same \cite{porras-olver2010},
\begin{equation}
\tilde A\simeq \frac{1}{2}\left[b_0 H^{(1)}_s(\rho)+ b_0H^{(2)}_s(\rho)\right]e^{is\varphi}e^{-i\zeta}\, .
\end{equation}
The H\"ankel beam $H^{(1)}_s(\rho)e^{is\varphi}e^{-i\zeta}$ carries power spirally outwards and the H\"ankel beam $H^{(2)}_s(\rho)e^{is\varphi}e^{-i\zeta}$ spirally inwards. Both have the same amplitude in the linear BVB, so there is no net transport of power in the cross section \cite{porras-salo2000}. In the nonlinear BVB, however, these components do not have equal amplitudes \cite{porras-porras2004}. Asymptotically at large $\rho$ the complex amplitude behaves as the ``unbalanced" Bessel beam
\begin{equation}
\tilde{A}(\rho,\phi,\zeta)\simeq \frac{1}{2}\left[b_\text{out}H_s^{(1)}(\rho)+b_\text{in}H_s^{(2)}(\rho)\right]{\rm e}^{is\varphi}{\rm e}^{-i\zeta},
\label{HBNL_HANKEL}
\end{equation}
as illustrated in the example of Fig. \ref{Fig5}(a), where the two interfering high-order H\"ankel beams have different amplitudes $|b_{\rm out}|$ and $|b_{\rm in}|$. From the equivalent asymptotic expressions $H^{(1,2)}_s(\rho)\simeq\sqrt{2/(\pi z)}\, e^{\pm i[\rho-(\pi/2)(s-1/2)]}$ at large $\rho$ \cite{porras-olver2010}, the condition that the inward radial flux $-F(\rho)=-2\pi\rho \tilde a^2 d\phi/d\rho=2\pi\rho \mbox{Im}[\tilde A(\partial \tilde A^\star/\partial\rho)]$ equals the total nonlinear power losses $N(\infty)$ leads to the relation
\begin{equation}\label{REFILLING2}
|b_{\rm in}|^2-|b_{\rm out}|^2=N(\infty)\, ,
\end{equation}
between $b_{\rm in}$ and $b_{\rm out}$. It then follows that $|b_{\rm in}|>|b_{\rm out}|$ for any nonlinear BVB in a medium with dissipation. The values of these amplitudes can easily be extracted from the numerically evaluated radial profiles of intensity of nonlinear BVBs as follows. Using the same asymptotic expressions, the radial intensity at large $\rho$ behaves as
\begin{equation}\label{MEAN}
\tilde a^2 \simeq \frac{1}{2\pi\rho}\left\{|b_{\rm out}|^2+|b_{\rm in}|^2 +
            2|b_{\rm out}| |b_{\rm in}|\cos\left[2\rho+\kappa \right]\right\}\, ,
\end{equation}
with $\kappa=-\pi s-\pi/2+\mbox{arg}(b_{\rm out}/b_{\rm in})$. Thus, $2\pi\rho \tilde a^2$ represents harmonic oscillations of contrast $C=2|b_{\rm in}| |b_{\rm out}|/(|b_{\rm in}|^2+|b_{\rm out}|^2)$ about the central value $R=|b_{\rm in}|^2+|b_{\rm out}|^2$. These properties are readily obtained from the numerically evaluated radial profiles of nonlinear BVBs, and from $C$ and $R$, the amplitudes $|b_{\rm in}|$ and $|b_{\rm out}|$ are obtained as
\begin{equation}
|b_{\rm in}|^2 = \frac{R}{2}\left(1+ \sqrt{1-C^2}\right), \quad |b_{\rm out}|^2 = \frac{R}{2}\left(1-\sqrt{1-C^2}\right) .
\end{equation}
They are represented as functions of $b_0$ in Figs. \ref{Fig5}(b) and (c) for nonlinear BVBs with different charges in different media. It turns out that $b_{\rm in}$ and $b_{\rm out}$ are real in absence of dispersive nonlinearities, and $\mbox{arg}\, b_{\rm in}=-\mbox{arg}\, b_{\rm out}$ with dispersive nonlinearities as self-focusing or self-defocusing.

\section{Propagation of Bessel vortex beams in nonlinear media revisited: the ``selection problem"}
\label{sec:porras-selec}

In the numerical simulations of propagation of linear BVBs, as in Fig. \ref{Fig1}, we observe that one of the nonlinear BVBs studied previously is formed spontaneously, acting thus as an attractor of the nonlinear dynamics. The nonlinear BVB has always the same cone angle and topological charge as the input linear BVB. The so-called ``selection problem" is the determination of the precise amplitude or intensity of the final BVB, i. e., the determination of the parameter $b_0$ of the final nonlinear BVB, say $b_0(\infty)$. This problem arose for the first time in Ref. \cite{porras-polesana2006} and remained unsolved for the vortex-less case. In the numerical simulation of Fig. \ref{Fig1}, for example, the linear BVB with $b_0(0) = 3$ is attracted by the nonlinear BVB with $b_0(\infty) = 1.60$.

\begin{figure}[tb]
\begin{center}
\includegraphics[width=3.9cm]{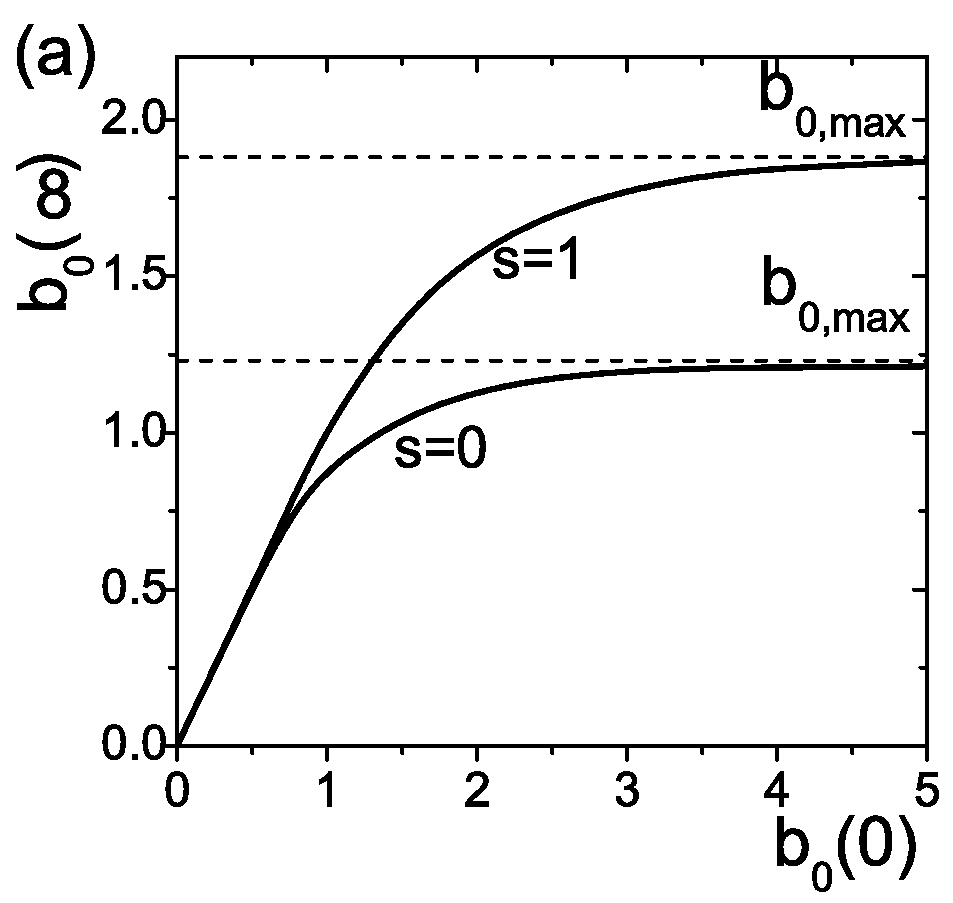}\includegraphics[width=3.9cm]{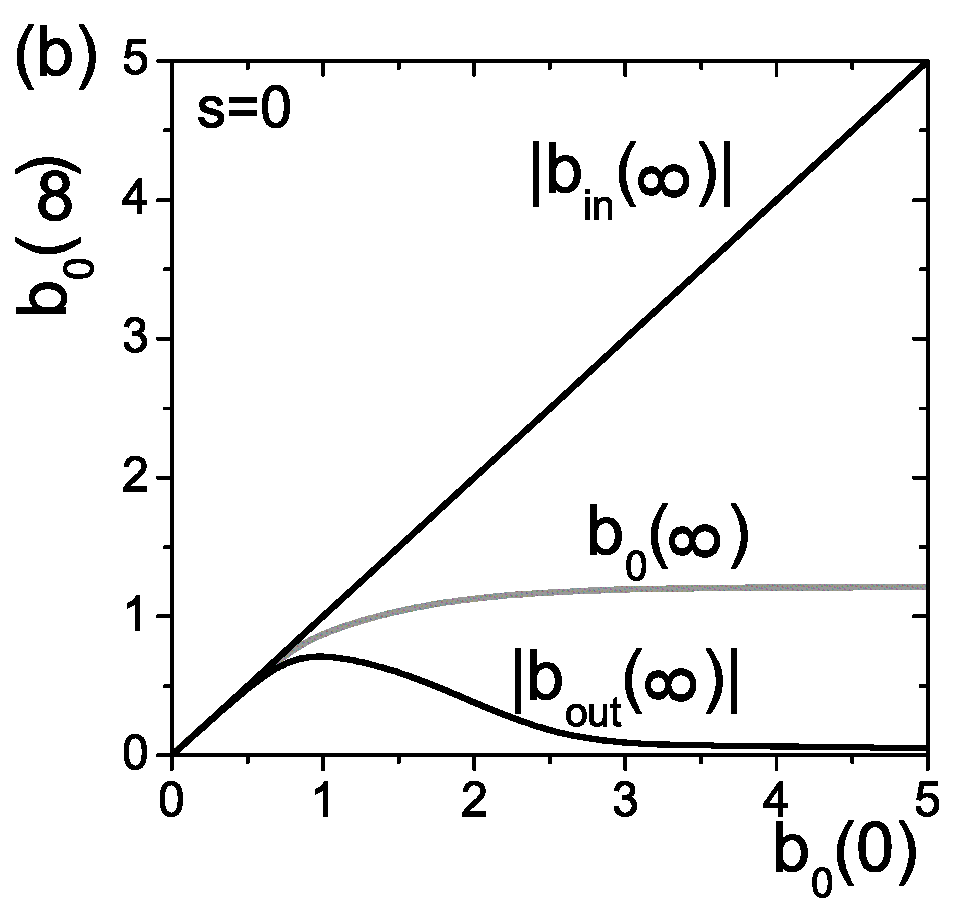}\includegraphics[width=3.9cm]{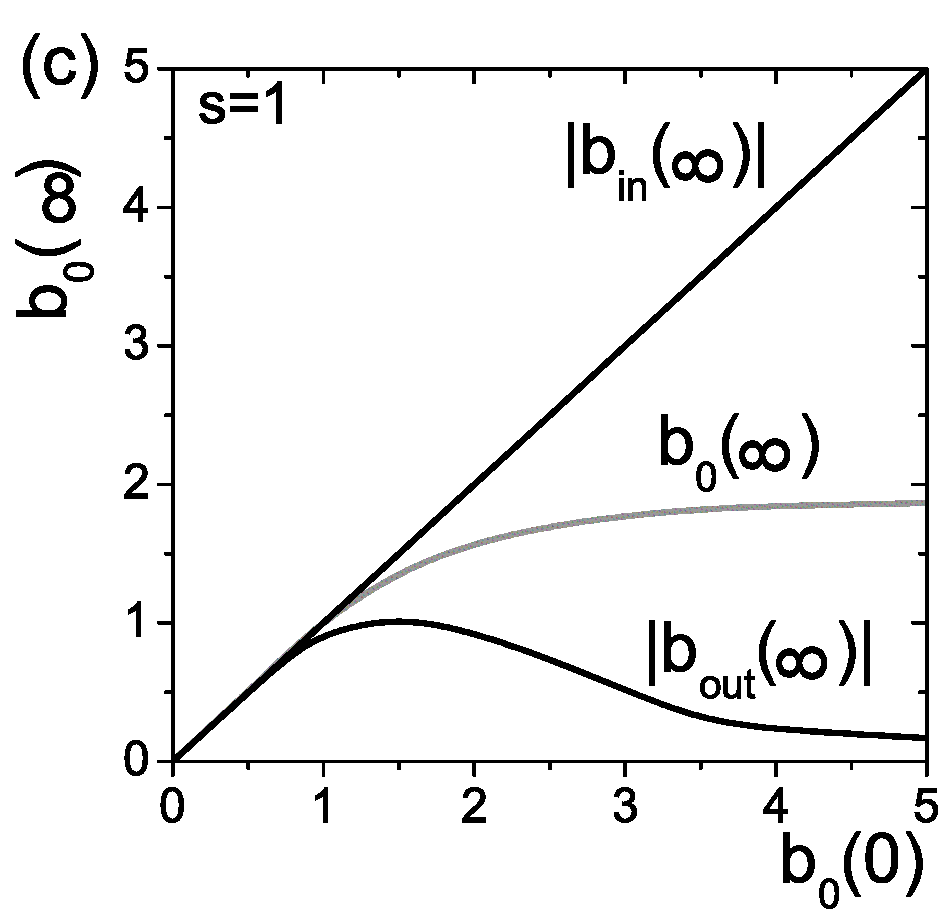}
\end{center}
\caption{\label{Fig6} (a) Values of the amplitude $b_0(\infty)$ of the attracting nonlinear BVB as a function of the amplitude $b_0(0)$ of the input linear BVB for $s=0$ and $s=1$ in a medium and cone angle such that $\alpha_2=0.5$ and $\alpha_4=-0.25$ and $M=4$, obtained by solving numerically the NLSE in Eq. (\ref{NLSE3}). (b) and (c) For the respective cases with $s=0$ and $s=1$, the black curves represent the amplitudes $|b_{\rm in}(\infty)|$ and $|b_{\rm out}(\infty)|$ of the inward and outward H\"ankel components of the attracting nonlinear BVB with the amplitude $b_0(\infty)$ (gray curves). As a function of $b_0(0)$, $|b_{\rm in}(\infty)|$ is then found to be the identity function, i. e., $|b_{\rm in}(\infty)|=b_0(0)$.}
\end{figure}

Figure \ref{Fig6} summarizes the results of a large number of similar numerical simulations, and unveils the law underlying the selection of the final nonlinear BVB, i. e., the value of $b_0(\infty)$. Figure \ref{Fig6}(a) depicts the pairs $[b_0(0),b_0(\infty)]$ of the input linear BVB and final nonlinear BVB extracted from these simulations for $s=0,1$ with a particular cone angle and medium (particular values of $\alpha$ and $M$). The attracting nonlinear BVB approaches that of maximum amplitude supported by the medium at very high intensities of the input linear BVB. On the other side, the selected nonlinear BVB does not differ substantially from the launched beam at low enough intensities.

The solution of the selection problem arises from the evaluation of the inward and outward amplitudes, $b_\text{in}(\infty)$ and $b_\text{out}(\infty)$ of the nonlinear BVB attractor with $b_0(\infty)$, represented in Figs. \ref{Fig6}(b) and (c) for the respective cases with $s=0$ and $s=1$. As seen, $|b_\text{in}(\infty)|$ is given by the identity function as a function of $b_0(0)$. Taking into account that for the linear BVB introduced in the medium $|b_\text{in}(0)| = b_0(0)$, we conclude that {\em the amplitude of the asymptotic inward H\"ankel component is conserved in the nonlinear propagation,} and only the amplitude of the outward H\"ankel component diminishes. The inward H\"ankel beam is indeed a linear beam that continuously brings power from the energy and angular momentum reservoir at arbitrarily large radial distances, and therefore is not affected by the nonlinear effects that take place in the inner rings. Thus, given the amplitude of the input linear BVB determined by $b_0(0)$, the nonlinear BVB selected as the final stage of the dynamics is that with $b_0(\infty)$ whose inward component equals $b_0(0)$, that is,
\begin{equation}
|b_{\rm in}(\infty)|=b_0(0)\,.
\end{equation}
Therefore, given a material medium and cone angle ($M$, $\alpha_2$ and $\alpha_4$) and a topological vorticity or charge $s$, graphs such as those in Figs. \ref{Fig5} (b) and (c) for the amplitude of the inward H\"ankel component can be obtained from the radial profiles of the nonlinear BVBs of different amplitudes supported by the medium. The nonlinear BVB attractor can be obtained from these graphs as one whose amplitude of the inward H\"ankel component coincides with the amplitudes $b_0$ of the linear BVB that is launched in the medium. These conclusions are drawn here with {\em ideal}, linear or nonlinear BVBs, but they will be seen in Sect. \ref{sec:porras-exp} to hold the same in actual settings with finite-power versions of these beams.

\section{Stability of nonlinear BVBs}
\label{sec:porras-stab}

It is well known that vortex solitons are prone to azimuthal breaking instability in self-focusing Kerr media \cite{porras-desyatnikov2005}. Only in specifically tailored media, as specific saturable, cubic-quintic nonlinearities \cite{porras-quirogateixeiro1997,porras-pazalonso2005}, or nonlocal Kerr nonlinearities \cite{porras-yakimenko2005}, stable vortex solitons have been reported to exist. Stability of some nonlinear BVBs is suggested by the above simulations of the propagation of linear BVBs attracted by a specific nonlinear BVB. Also, nonlinear BVBs have been observed to be spontaneously formed in experiments under specific conditions associated with large cone angles \cite{porras-xie2015}, which further supports the existence of stable nonlinear BVB in media such as glasses.

In this section we perform a linear-stability analysis of the nonlinear BVBs and find there exists a subset of them that is stable against radial and azimuthal perturbations. The results from the stability analysis are corroborated by direct numerical simulations of the propagation of perturbed nonlinear BVBs. For simplicity, in this section we only consider the more destabilizing self-focusing (positive) cubic nonlinearity and dissipation, but, as the existence of nonlinear BVBs, similar results regarding stability or instability are readily seen to hold with more general dispersive nonlinearities. We will then write $\alpha\equiv \alpha_2$, and consider $\alpha_4=0$.

The linear-stability analysis applied to our case will be the usual one \cite{porras-desyatnikov2005}, where we take a perturbed nonlinear BVB as
\begin{equation} \label{PERT}
\tilde{A}=\tilde{A}_{s}+\varepsilon \left[ u_{m}(\rho )e^{i\kappa \zeta +im\varphi }+v_{m}^{\star}(\rho )e^{-i\kappa ^{\star }\zeta -im\varphi }\right]e^{is\varphi -i\zeta }\, ,
\end{equation}
being $\tilde A_s=\tilde a e^{i\phi}e^{is\varphi-i\zeta}$ a nonlinear BVB, $\varepsilon$ an infinitesimal amplitude of perturbations with eigenmodes [$u_m$($\rho$), $v_m$($\rho$)] and integer winding number $m$. Valid solutions $u_m(\rho$), $v_m(\rho$) to our system must comply with the usual boundary conditions $u_m\sim\rho^{|s+m|}$ and $v_m \sim \rho^{|s-m|}$ for $\rho\rightarrow 0$, as well as vanishing for $\rho\rightarrow \infty$. Substituting the above Ansatz in Eq. \eqref{NLSE3} gives the following linearized equations:
\begin{equation}
\left(
\begin{array}{cc}
H_{+} & f \\
-f^{\star } & -H_{-}^{\star }%
\end{array}
\right) \left(
\begin{array}{c}
u_{m} \\
v_{m}%
\end{array}
\right) =\kappa \left(
\begin{array}{c}
u_{m} \\
v_{m}%
\end{array}
\right) ,  \label{EIGEN}
\end{equation}
where
\begin{equation}
f\equiv [\alpha \tilde{a}^{2}+i(M-1)\tilde{a}^{2M-2}]e^{2i\phi}\, ,
\end{equation}
and
\begin{equation}
H_{\pm }\equiv \frac{d^{2}}{d\rho ^{2}}+\frac{1}{\rho }\frac{d}{d\rho }-
\frac{(s\pm m)^{2}}{\rho ^{2}}+1+(2\alpha \tilde{a}^{2}+iM\tilde{a}^{2M-2}).
\label{H}
\end{equation}
\begin{figure}[tb]
\begin{center}
\includegraphics[width=3.9cm]{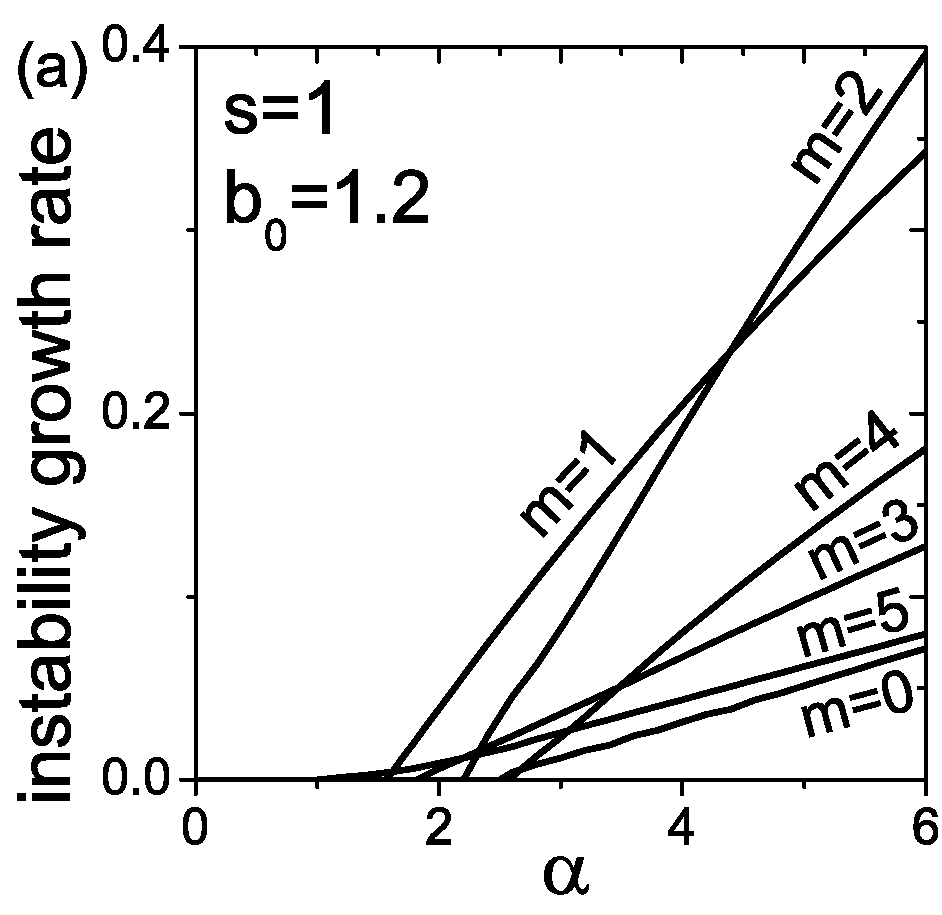}\includegraphics[width=3.9cm]{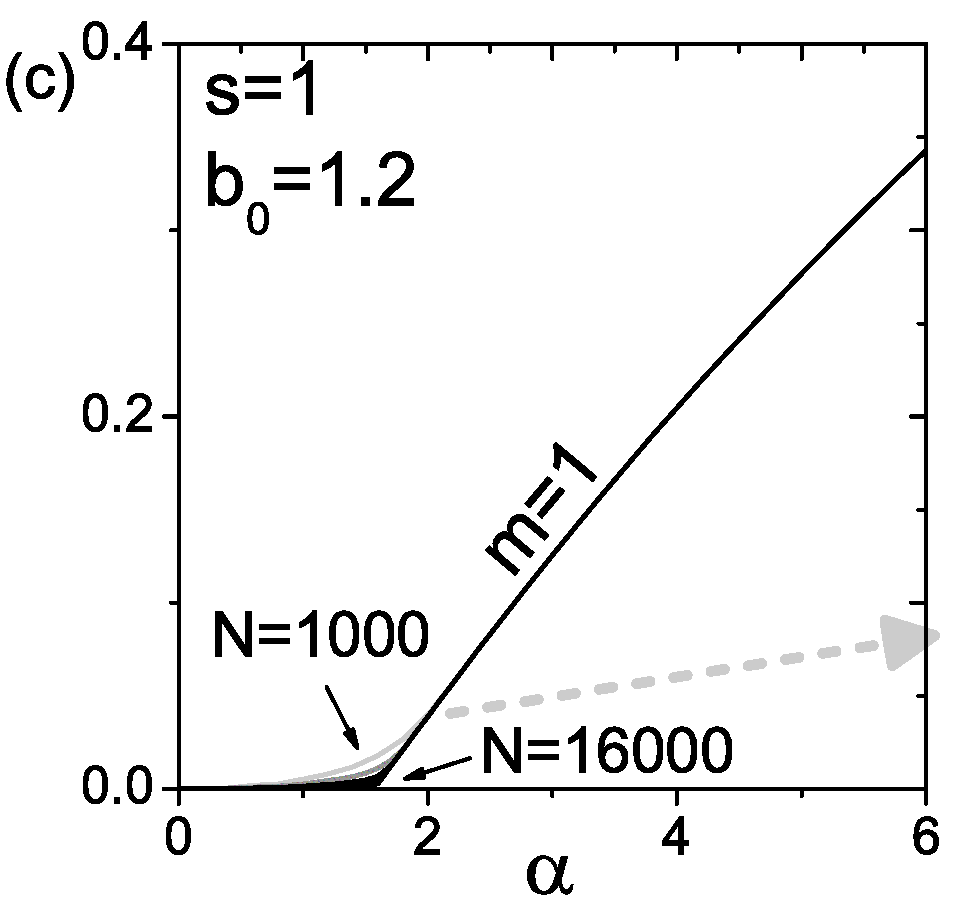}\includegraphics[width=3.9cm]{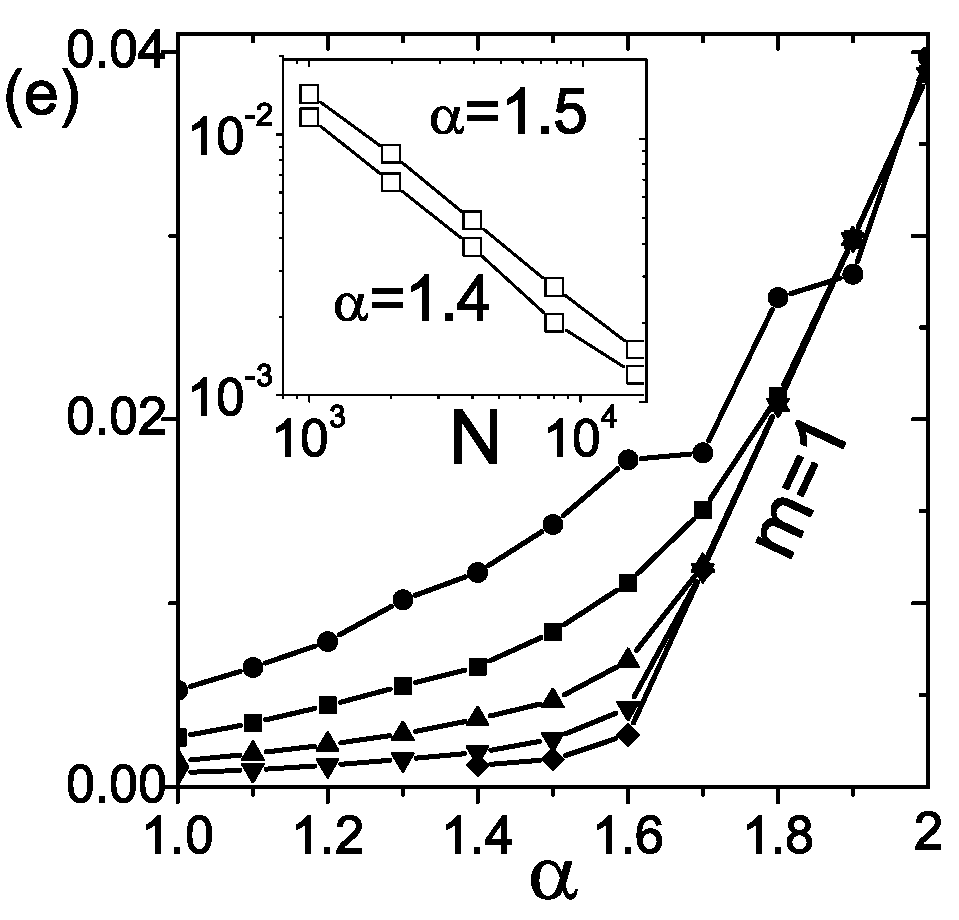}
\includegraphics[width=3.9cm]{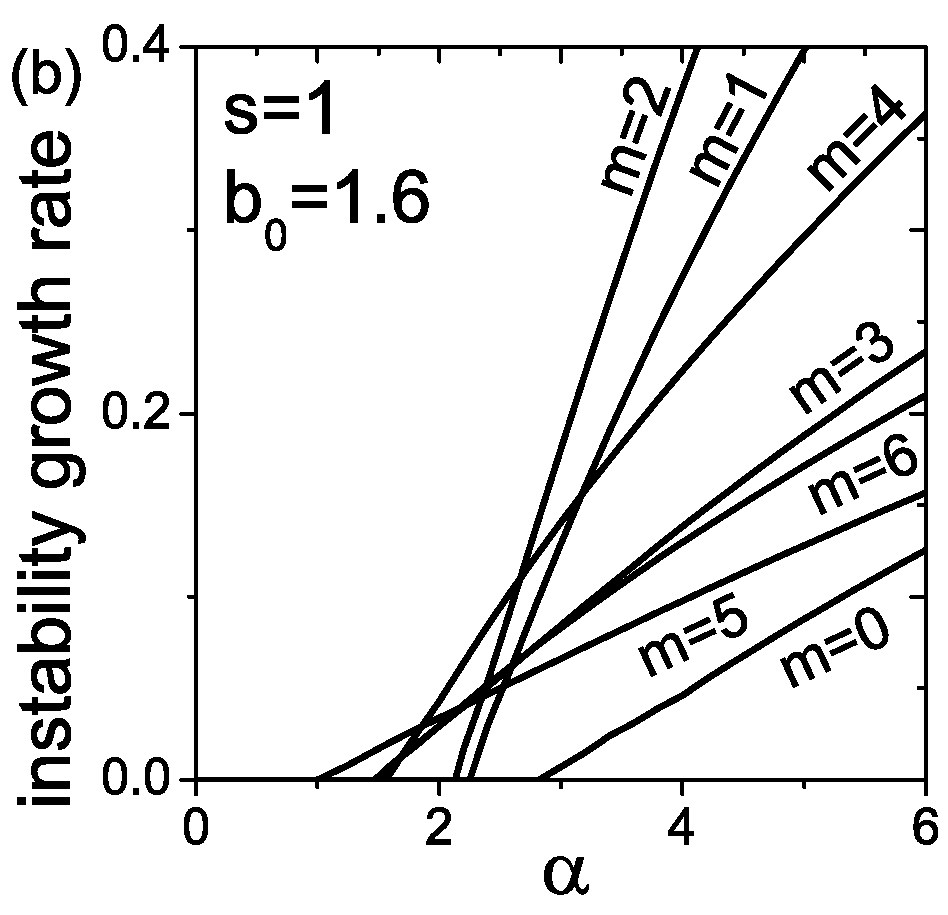}\includegraphics[width=3.9cm]{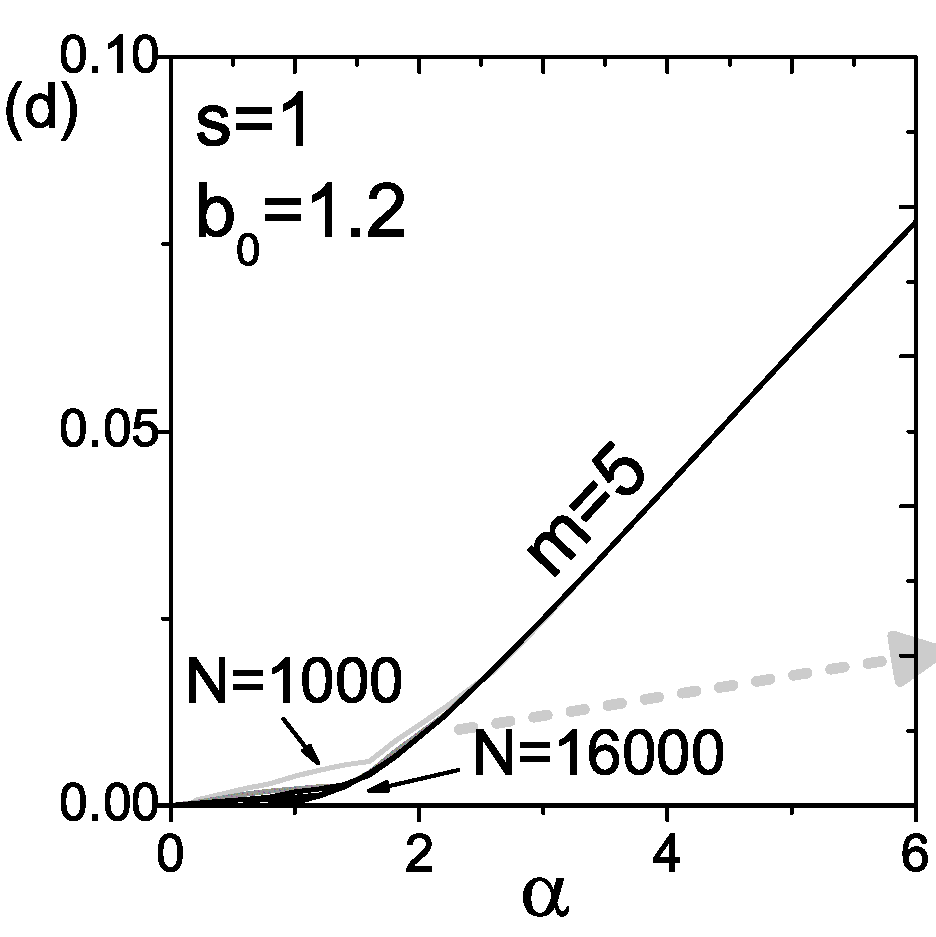}\includegraphics[width=3.9cm]{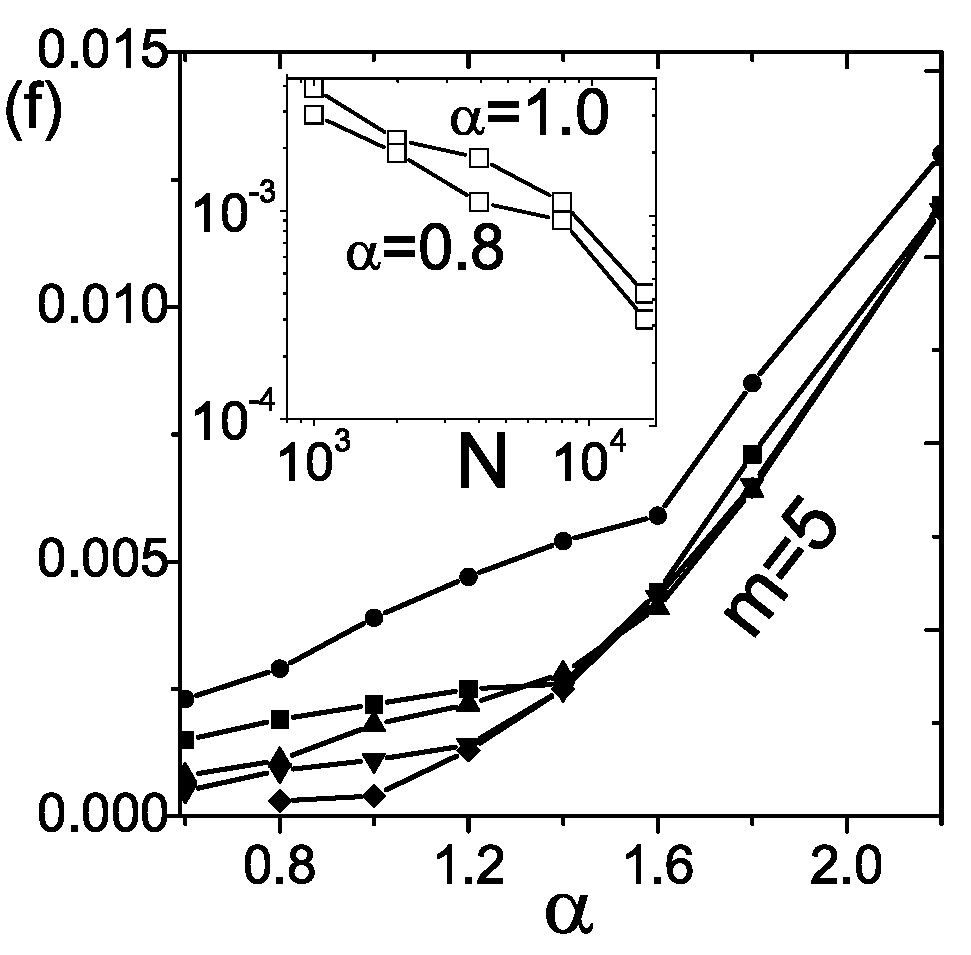}
\end{center}
\caption{(a) and (b) Growth rates of unstable perturbation modes of the indicated nonlinear BVBs. (c) and (d) The same but evaluated with $h=0.2$, $N=1000,2000,4000,8000$ and $16000$ from the lighter to the darker curves. (e) and (f) Zoom of the region where the growth rate depends appreciably on $N$. $N=1000$ (circles), $2000$ (squares), $4000$ (up triangles), $8000$ (down triangles), $16000$ (rhombuses). The insets additionally display the growth rates as functions of $N$ for particular values of $\alpha$, indicating a decay $\sim 1/N$.}
\label{Fig7}
\end{figure}
If there exists at least one eigenvalue $\kappa_I \equiv \mbox{Im}\{\kappa\} < 0$, then an instability will develop. In this case, it is expected to manifest itself in a weakly perturbed BVB through the breakup of the first and brightest ring, and possibly the secondary rings too, and the number of fragments resulting from this breakup is expected to be equal to the winding number $m$ of the mode with the largest growth rate. Also, several unstable modes with similar growth rate may compete, or induce different number of fragments in different rings. The analysis of Eqs. (\ref{PERT}) and (\ref{EIGEN}) for all integers $m=0,1,2\dots$ covers all type of weak perturbations, since these equations result from introducing a nonlinear BVB  plus a general perturbation, i. e., $\tilde A=\tilde A_s+ \varepsilon p(\rho,\varphi,\zeta)$, into the NLSE, linearizing, expanding $p$ in azimuthal harmonics, and solving the equations for each harmonic \cite{porras-desyatnikov2005}.

We solved the problem in Eq. (\ref{EIGEN}) numerically, as done previously for radial perturbations ($m=0$) to the fundamental nonlinear Bessel beam \cite{porras-porras2004,porras-polesana2007,porras-porras2015}. Weaker localization and the structured radial profile of the nonlinear BVBs make the numerical procedure used to solve Eq. \ref{EIGEN} more difficult when compared to other cases, such as the vortex solitons in the complex Ginzburg-Landau model \cite{porras-mihalache2006,porras-aranson2002,porras-akhmediev2007,porras-knobloch2015}.
We transform it into an algebraical eigenvalue problem of a $2N\times 2N$ matrix by introducing a discretization in a radial grid of step $h$ and $N$ points, so that the largest radius is $Nh$, with boundary conditions $u\sim \rho^{|s+m|}$, $v\sim \rho^{|s-m|}$ for $\rho\rightarrow 0$, and $u,v=0$ at $\rho=Nh$. Stability or instability is inferred from the spectrum of $2N$ eigenvalues
in the double limit $h\rightarrow 0$, $N\rightarrow\infty$. For fixed truncation radius $Nh$, no substantial difference between the eigenvalues is found provided that the nonlinear BVB profile is adequately sampled (e. g., $h=0.1, N=2000$ and $h=0.2$, N=1000). We then focus on increasing truncation radius $Nh$ by fixing $h$ and increasing the number of points $N$, since important differences are found. In case of instability with low growth rate $|\kappa_I|$, these differences arise because the associated eigenmode $u,v$ presents the slow exponential decay $\sim e^{\kappa_I\rho/2}$ \cite{porras-polesana2007}, so that it is only suitably reproduced by the truncated system if $Nh$ is much larger than $2/|\kappa_I|$. In other terms, the minimum reliable growth rate obtainable from the truncated system is $|\kappa_I|\gg 2/Nh$. We used special routines for sparse matrices to obtain eigenvalues of matrices up to $32000\times 32000$ in order to improve accuracy for the small growth rates involved, sampling our system with $N = 16000$ points. These technical details are explained to stress that proving stability with this method requires, strictly speaking, to infer the limit $N\rightarrow\infty$.

Figures \ref{Fig7}(a) and (b) show results for the growth rates of unstable modes of nonlinear BVBs with $s=1$, two choices of $b_0$ and varying $\alpha$, and Figs. \ref{Fig7}(c-f) illustrate how these results are obtained. For all $m$, the growth rates computed with finite $N$ behave as almost straight lines that tend to cut the horizontal axis at certain positive values of $\alpha$, but this behavior is interrupted when the growth rate is comparable to the minimum reliable value $2/Nh$, as seen in Figs. \ref{Fig7} (c) and (d) for the two modes with higher growth rate or leading to instability down to lower values of $\alpha$. For high values of $\alpha$, no difference is observed with increasing $N$, and these values are attributed to a genuine instability of the corresponding nonlinear BVB in the limit $N\rightarrow\infty$. In Figs. \ref{Fig7}(e) and (f) we plot the detailed behavior of the growth rate in the region of small $\alpha$ where relevant differences with $N$ are observed. The growth rate is seen to converge to positive values for some $\alpha$, but is observed to approach zero for the smallest values of $\alpha$. The insets of Figs. \ref{Fig7}(e) and (f) indicate an approximate decay $1/N$ for  the particular values of $\alpha$ where convergence to a positive value is not observed. Similar behavior is observed for all other relevant modes $m$. Up to the limit of our computational capability, we can then say that there exist nonlinear BVBs in Kerr media that are stable against all type of small perturbations, as shown in Figs. \ref{Fig7} (a) and (b); here, for $M =4$ and $s =1$ these are nonlinear BVBs with $\alpha \lesssim 1.1$ for $b_0 = 1.2$, and with $\alpha \lesssim 1$ for $b_0 = 1.6$.

\begin{figure}[tb]
\begin{center}
\includegraphics*[width=3.8cm]{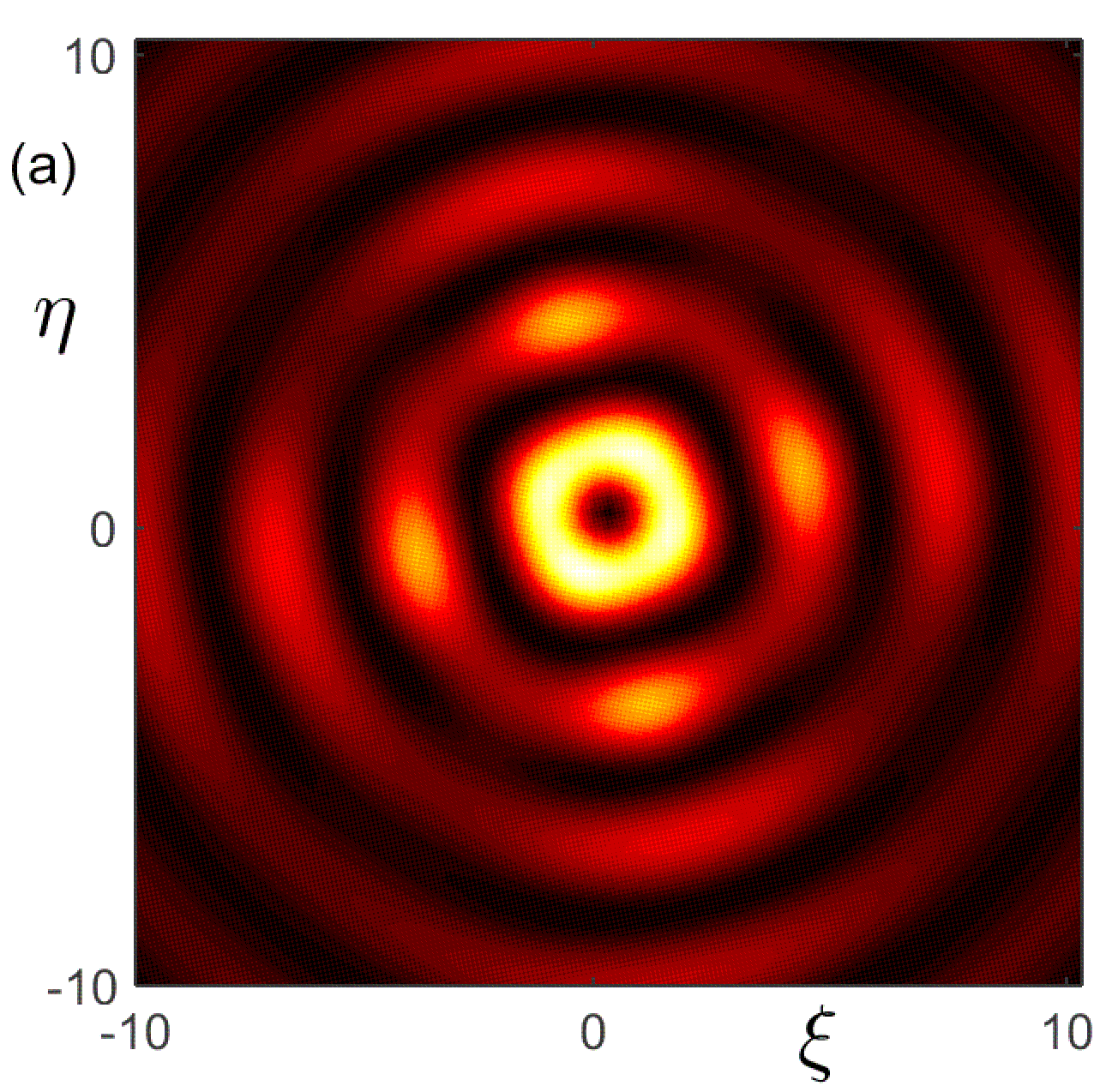}\includegraphics*[width=3.8cm]{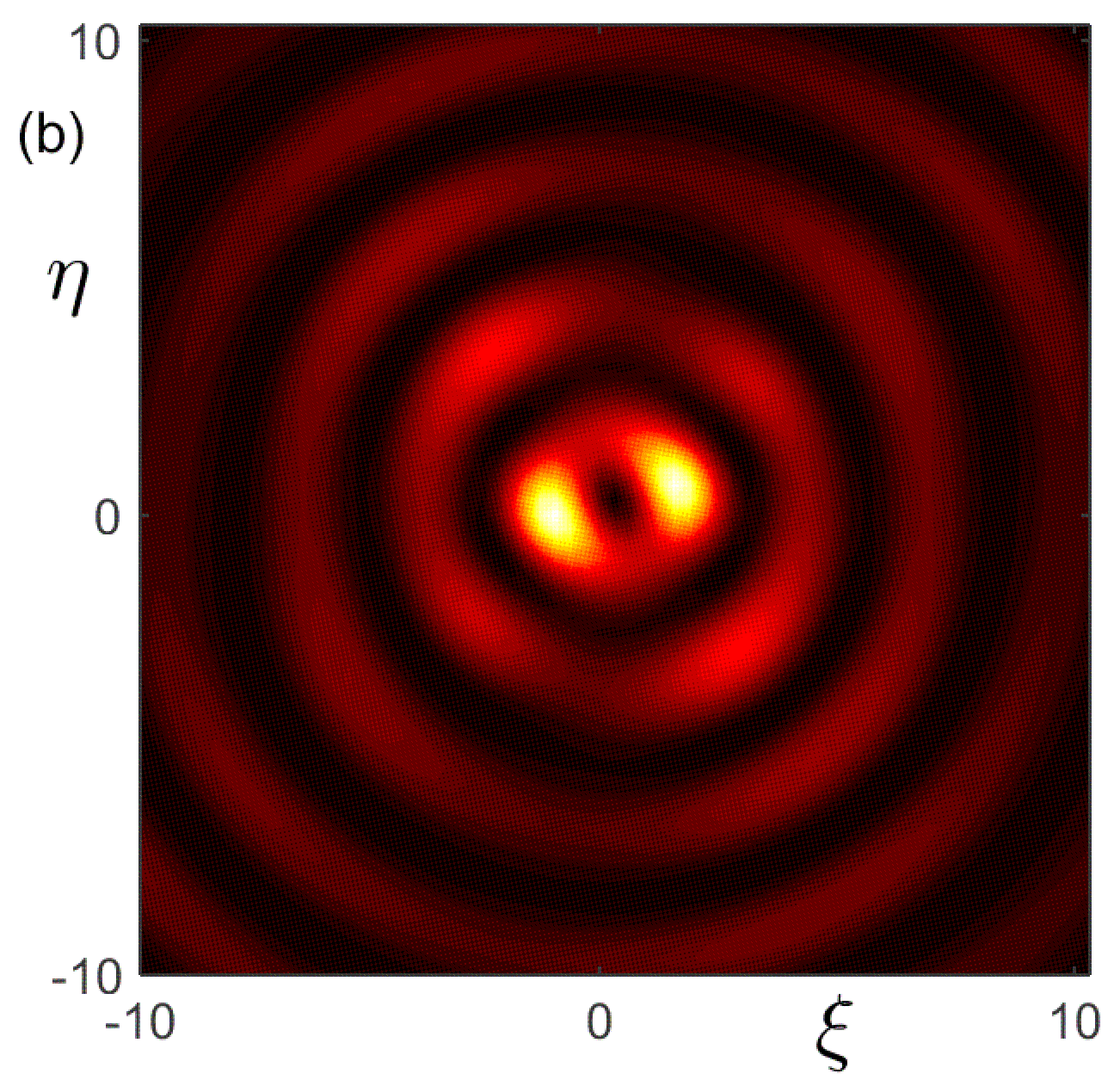}\includegraphics*[width=3.8cm]{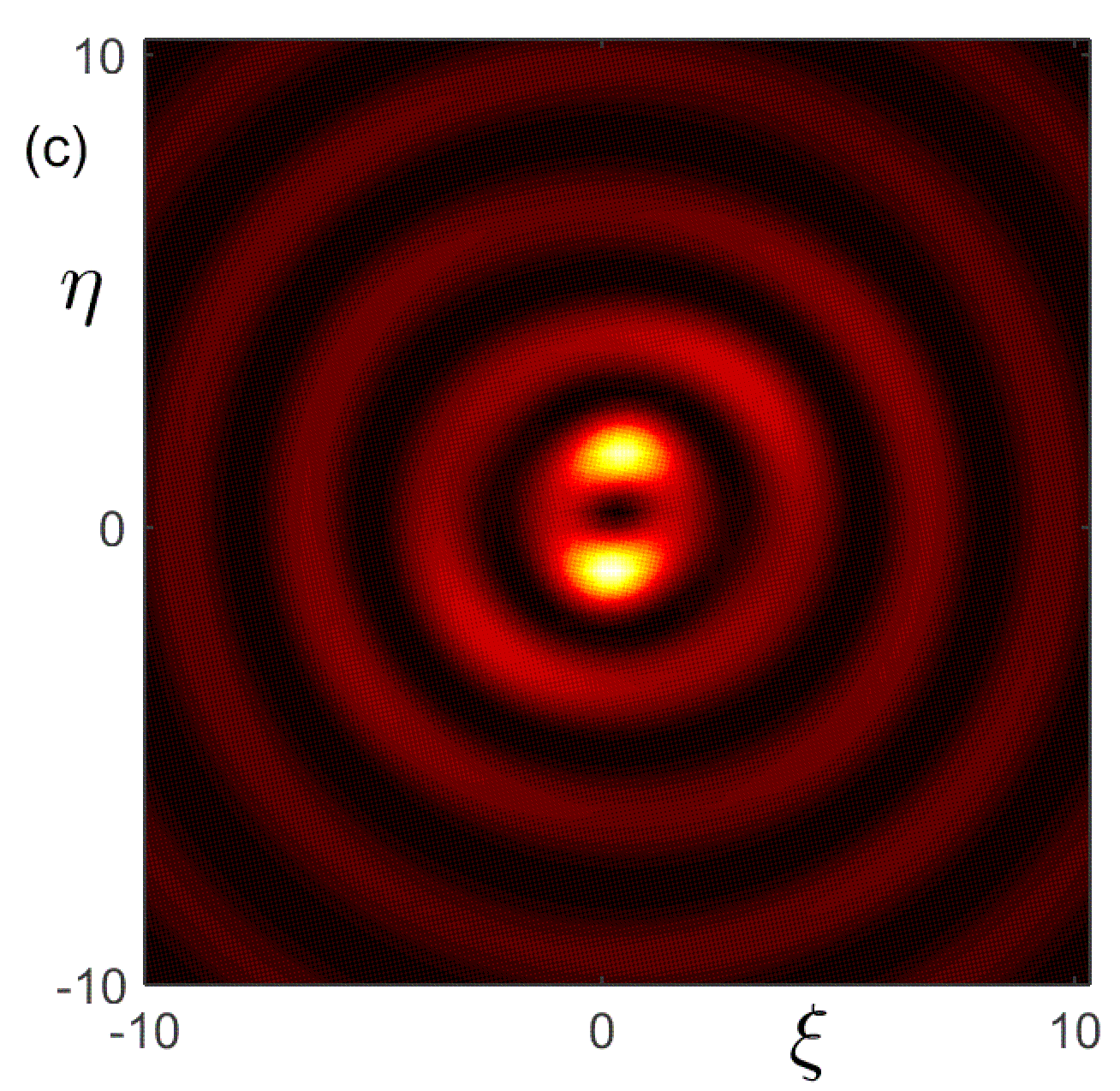}
\end{center}
\caption{For $M=4$, $s=1$ and $b_{0}=1.6$, the transverse intensity
distributions $|\tilde{A}|^{2}$ of the initially perturbed vortex with (a) $\alpha =2.2$ at $\protect\zeta =100$, (b) $\alpha =2.8$ at $\zeta =45$, and (c) $\alpha =4$ at $\zeta =15$. The numbers of fragments into which the unstable
vortices split are exactly predicted by the linear-stability analysis.}
\label{Fig8}
\end{figure}

Direct numerical simulations of the NLSE in Eq. \eqref{NLSE3} were also carried out using a split-step Fourier method. In all cases, nonlinear BVBs initially perturbed by random noise have their rings broken into fragments moving along circular trajectories if unstable. Comparing Fig. \ref{Fig8} with Fig. \ref{Fig7} (b), it can be verified that the number of fragments is exactly equal to the winding number $m$ of the unstable mode with the largest instability growth rate. The larger the instability growth rate, the sooner the BVB breaks up. On a special note, a mode competition can be seen in Fig. \ref{Fig8} (b), where the main ring splits into two fragments and the outer ones into four, in agreement with the fact that there are two perturbation eigenmodes which are nearly equally unstable.

\begin{figure}[tb]
\begin{center}
\includegraphics*[width=3.8cm]{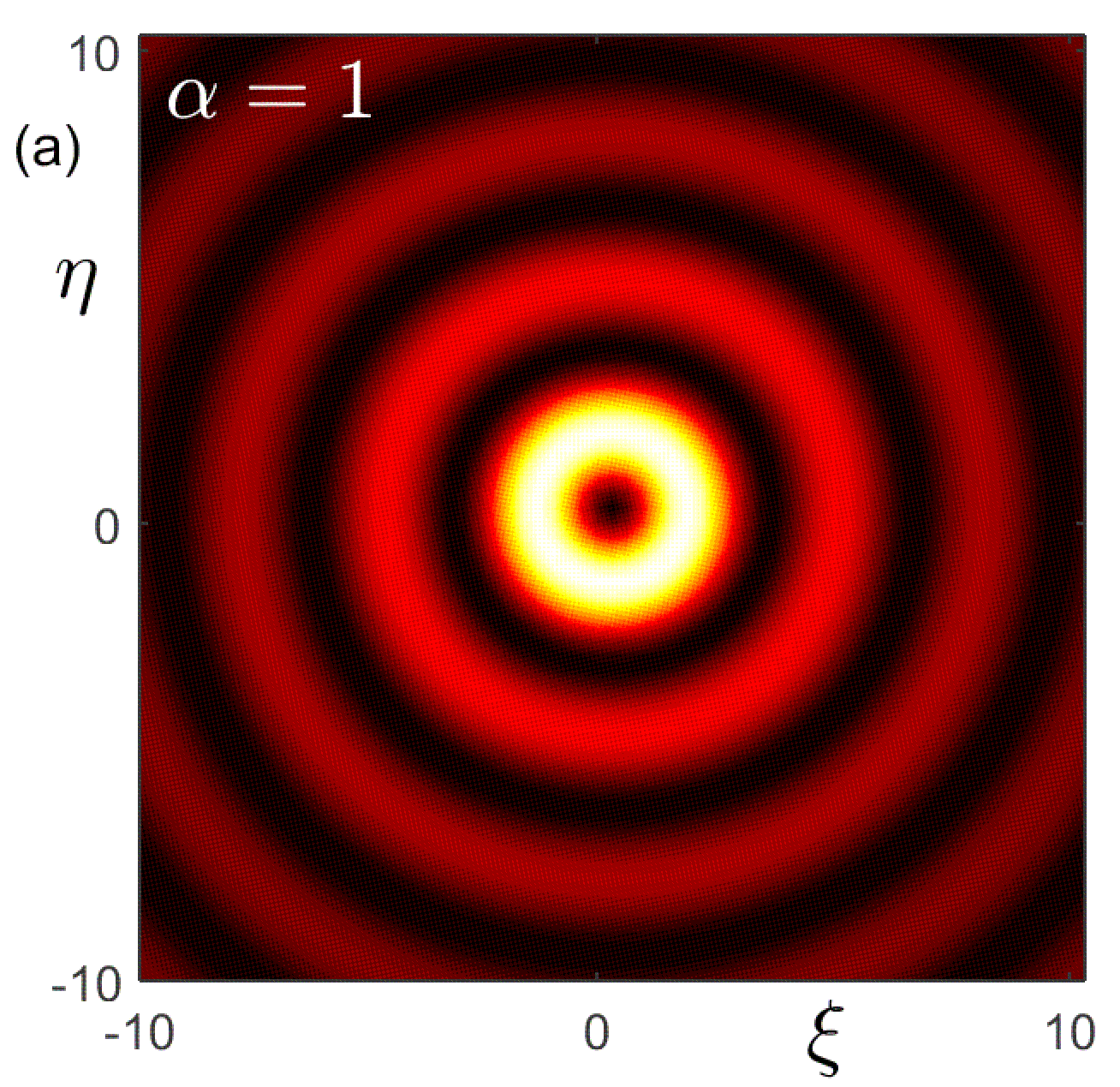}\includegraphics*[width=3.8cm]{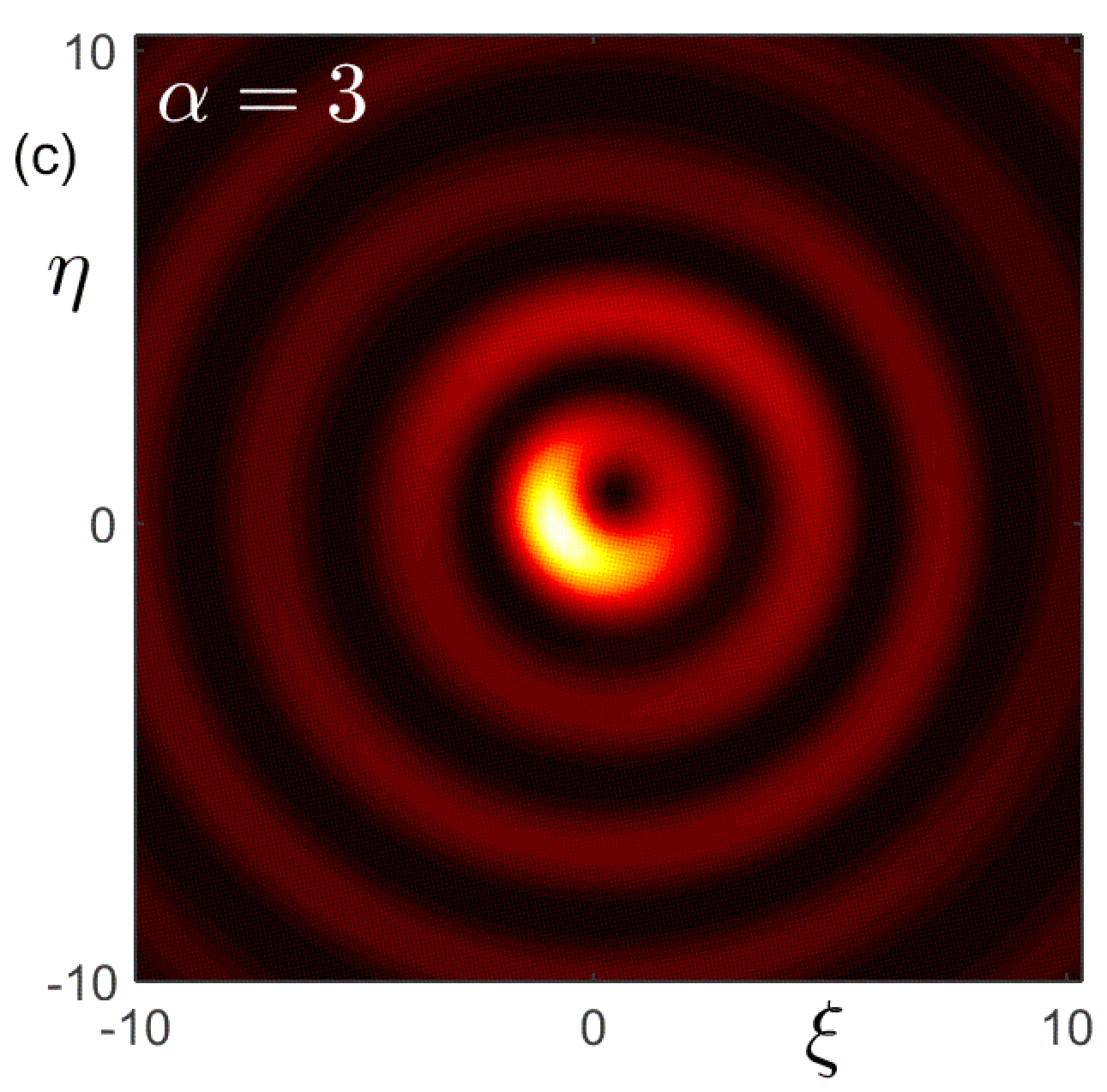}\includegraphics*[width=3.8cm]{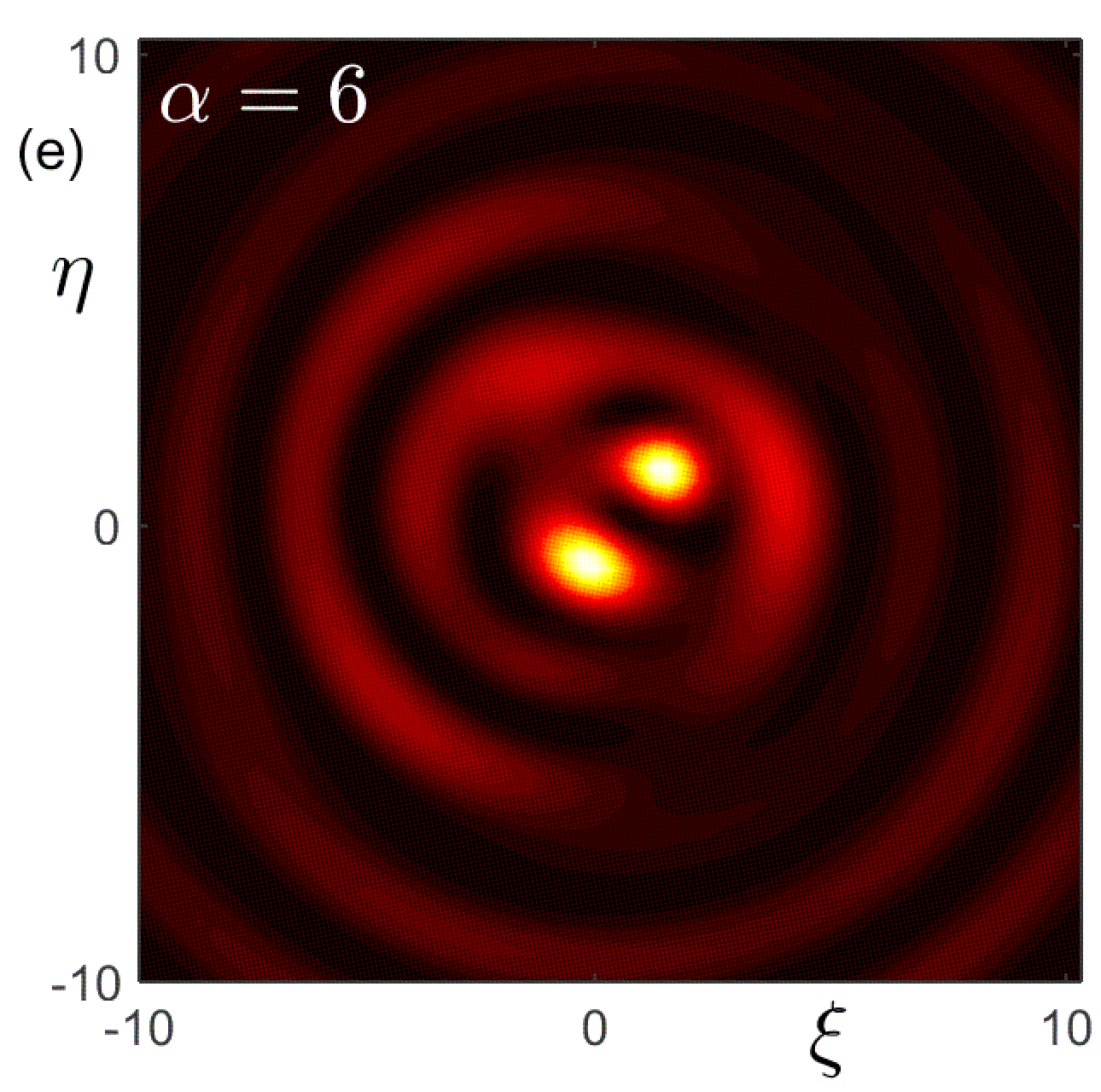}
\includegraphics*[width=3.8cm]{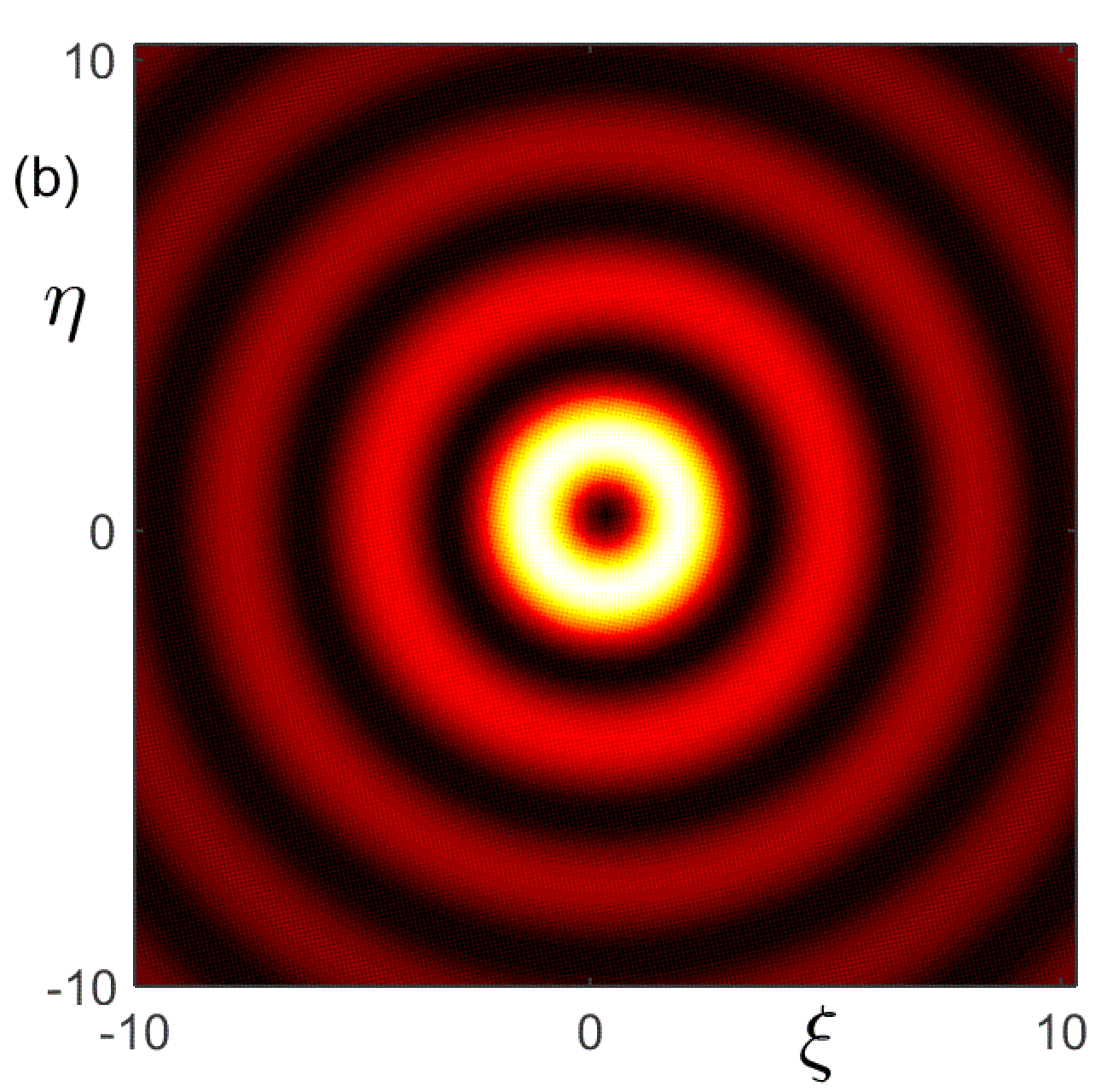}\includegraphics*[width=3.8cm]{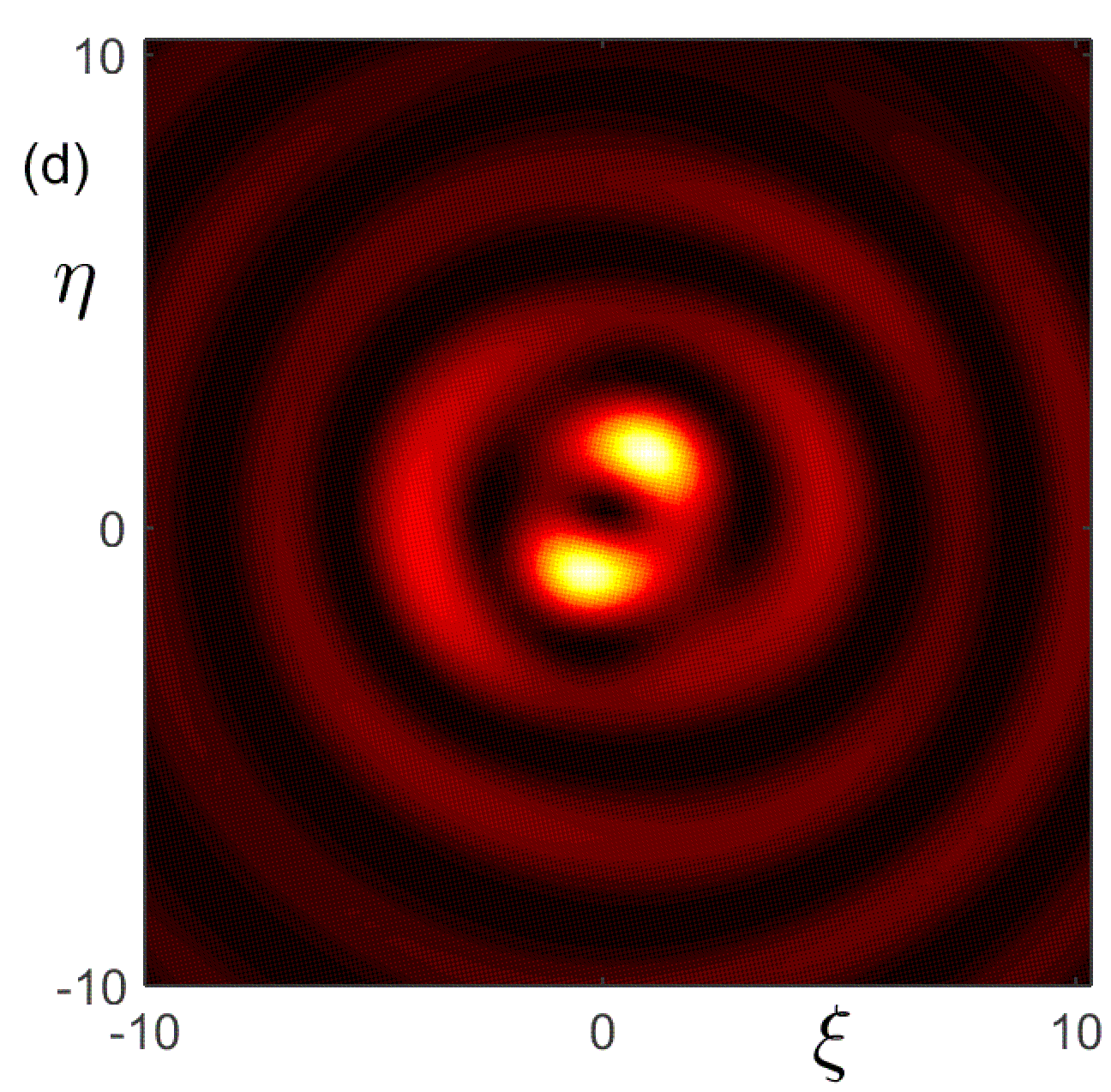}\includegraphics*[width=3.8cm]{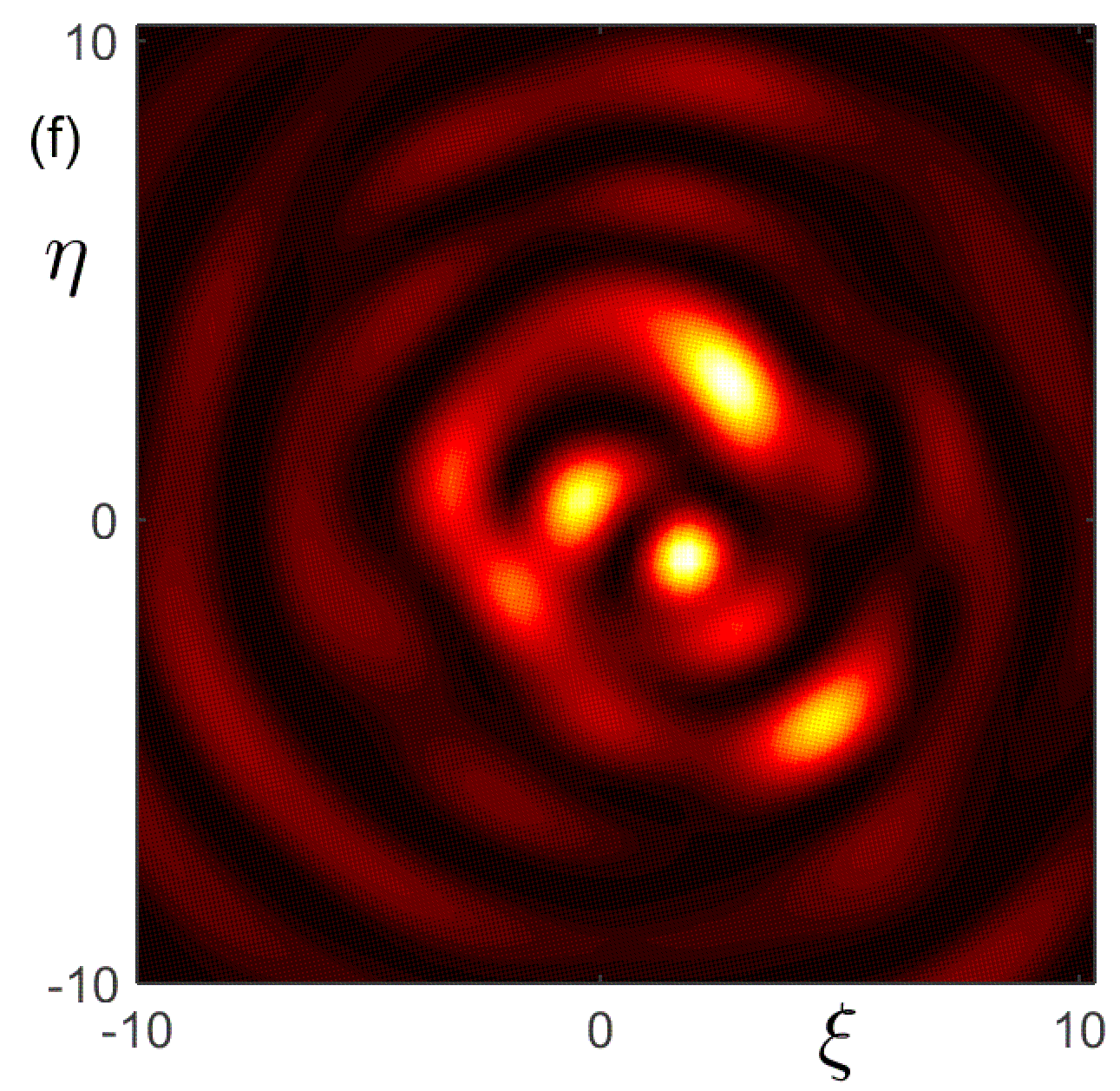}
\includegraphics*[width=3.8cm]{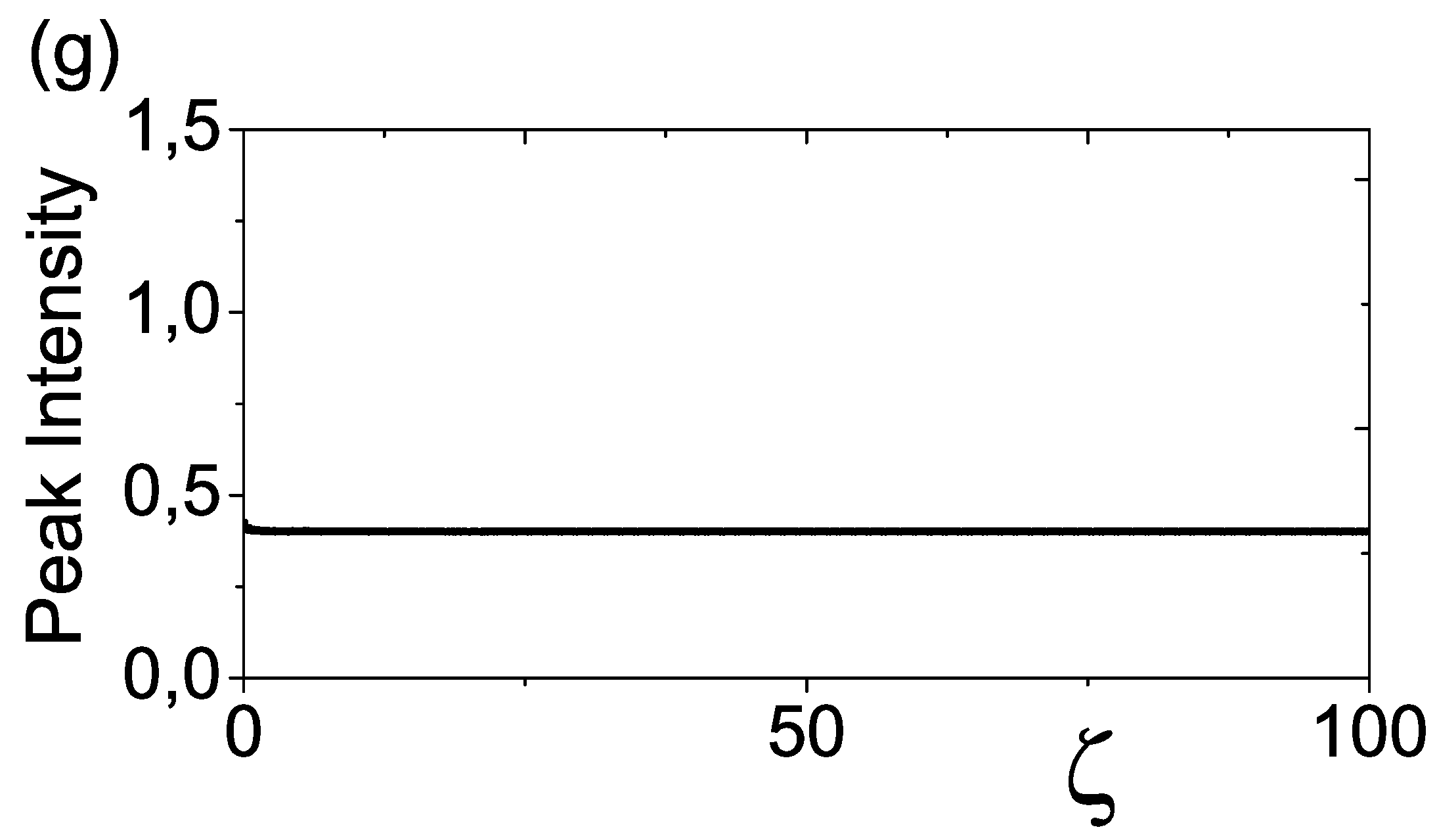}\includegraphics*[width=3.8cm]{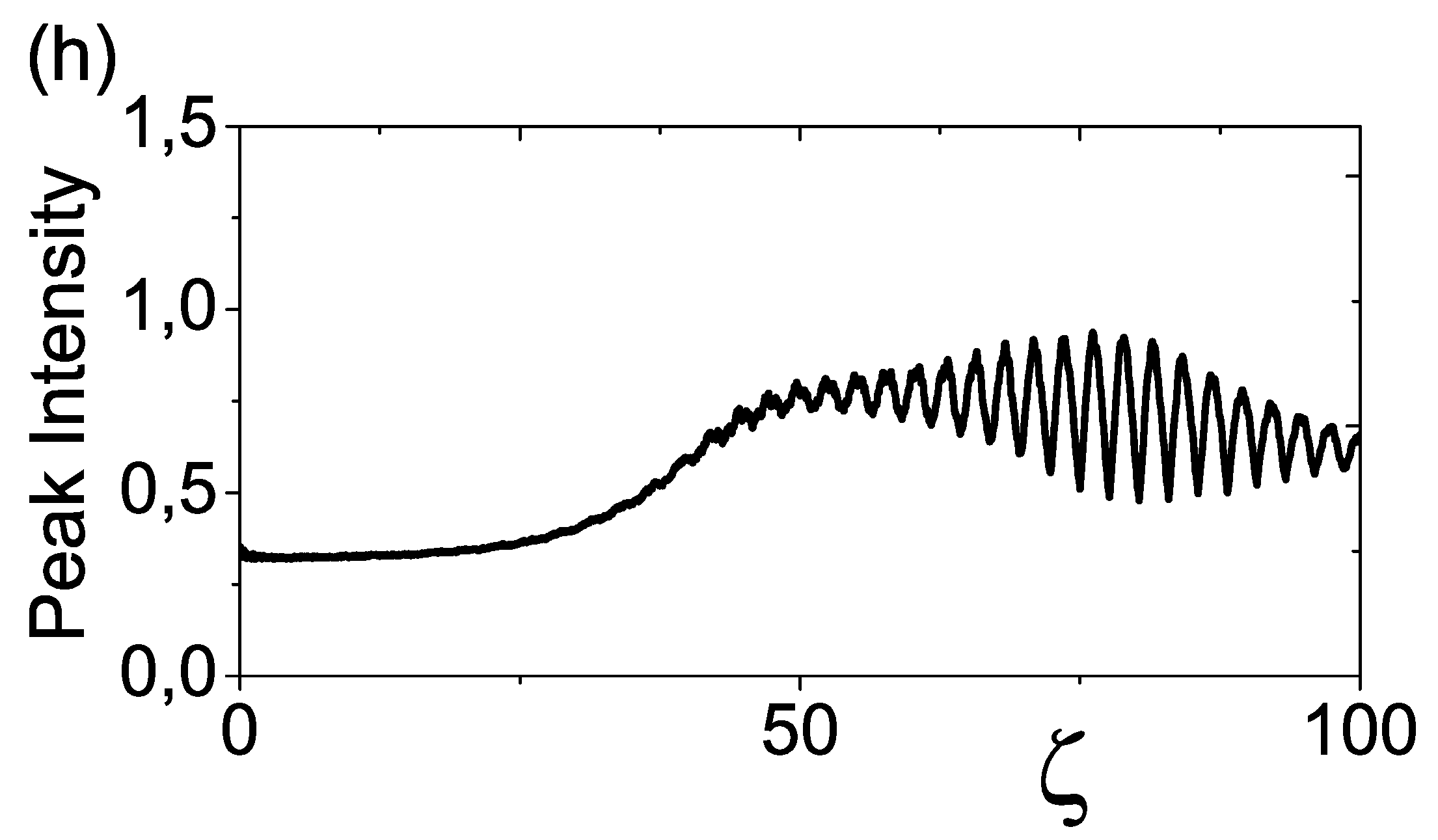}\includegraphics*[width=3.8cm]{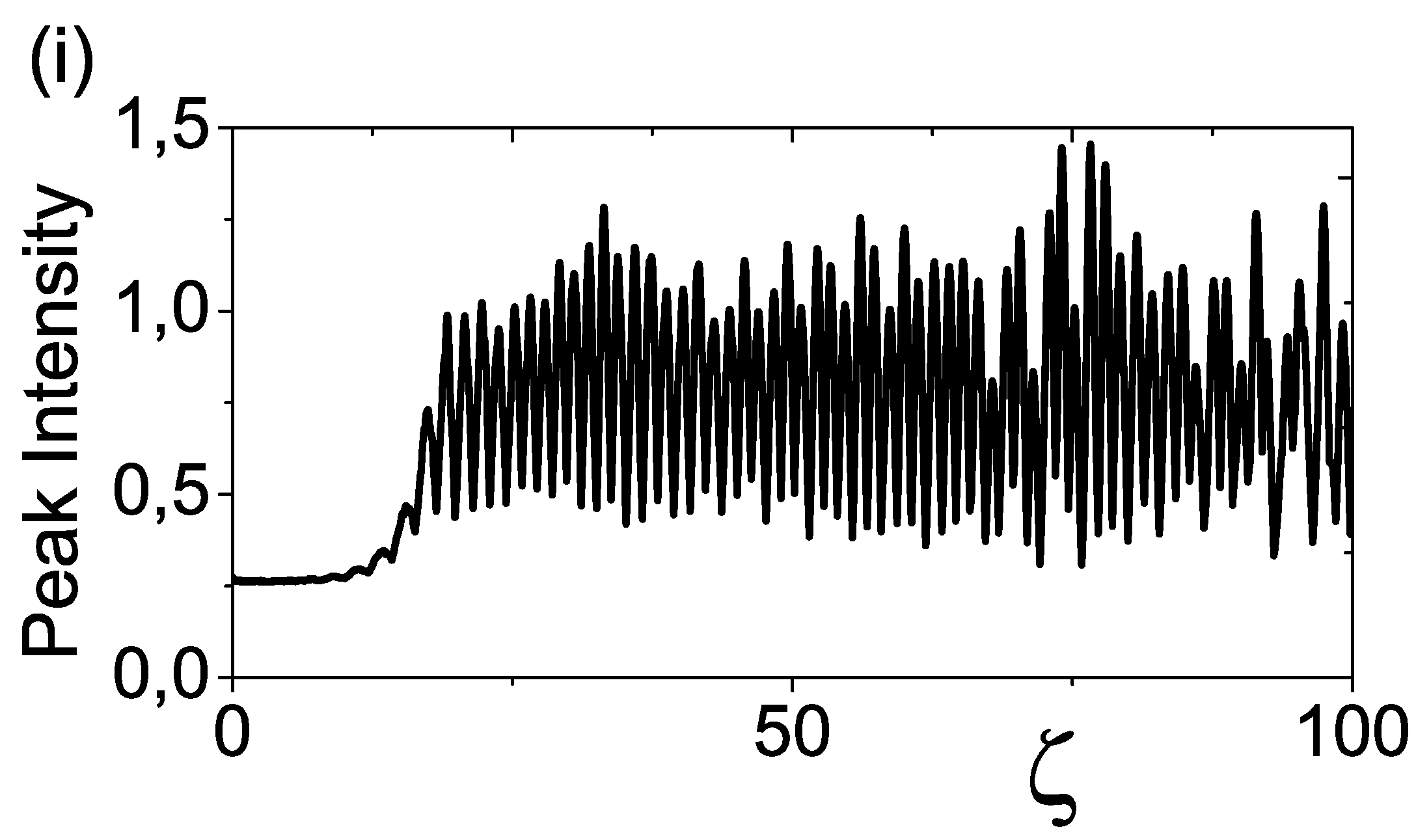}
\end{center}
\caption{For $M=4$, $s=1$ and $b_{0}=1.2$, the transverse intensity distributions, $|\tilde{A}|^{2}$, of the initially perturbed BVB with $\alpha=1$ at propagation distances (a) $\zeta=10$ and (b) $\zeta=300$, with $\alpha=3$ at propagation distances (c) $\zeta =40$ and (d) $\zeta =96$, and with $\alpha =6$ at (e) $\zeta =23$ and (f) $\zeta=93$. The results in (a,b) corroborate the stability, and the results in (c-f) demonstrate the secondary breakup of fragments produced by the primary instability. (g-i) Peak intensity in the three respective cases with $\alpha=1, 3$ and $6$ as functions of the propagation distance.}
\label{Fig9}
\end{figure}

If stability is predicted by our model, as for $\alpha=1$ in Fig. \ref{Fig7}(a), our simulations (ran up to $\zeta = 300$) show that the nonlinear BVB absorbs the random perturbation and keeps its shape throughout the whole propagation, as illustrated in Figs. \ref{Fig9} (a) and (b).
For unstable nonlinear BVBs, the development of the instability into large perturbations is not generally determined by the number of circulating fragments in the small-perturbation regime, as known for the case of usual vortex solitons \cite{porras-desyatnikov2005}. This can be readily seen in the following examples: Fig. \ref{Fig7} (a) for $\alpha=3$ predicts the single circulating fragment that indeed appears in Fig. \ref{Fig9} (c). This fragment breaks afterwards into two circulating fragments observed in Fig \ref{Fig9} (d). Also, Fig. \ref{Fig7} (a) for $\alpha=6$ predicts the two circulating fragments that shown in Fig. \ref{Fig9} (e). These two fragments decay at longer propagation distances into numerous randomly placed, non-rotating splinters that appear and disappear in the course of the propagation, as illustrated in Fig. \ref{Fig9} (f). For further purposes, we have also plotted in Figs. \ref{Fig9} (g), (h) and (i) the peak intensity in the course of propagation for the three perturbed nonlinear BVBs considered. For $\alpha=3$, instability leads to quasi periodic oscillations that are damped after the secondary splitting. For the more unstable case with $\alpha=6$, where the filaments are non-rotating and disordered, the behavior of the peak intensity indicates an endless chaotic regime continuously fed by the reservoir.

\begin{figure}[tb]
\begin{center}
\includegraphics[width=4.5cm]{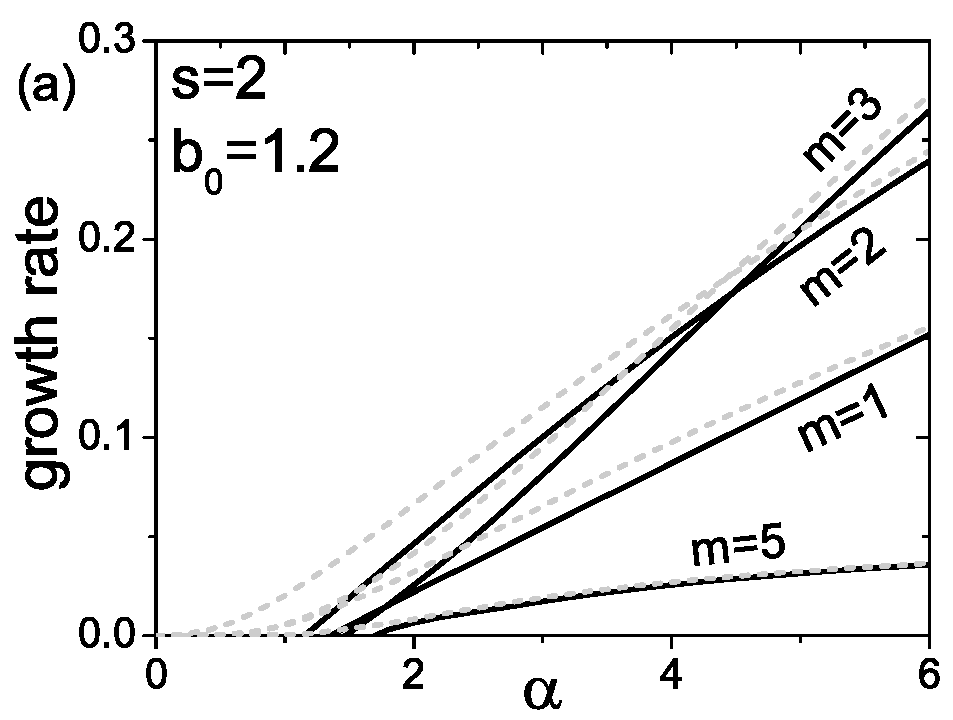} \includegraphics[width=4.5cm]{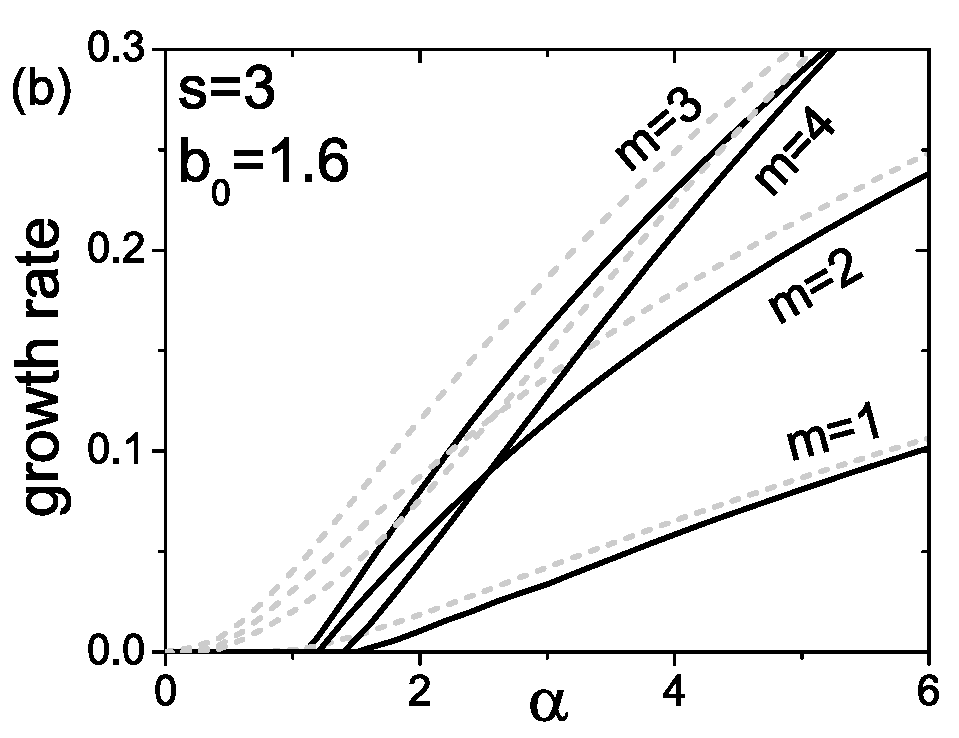}
\end{center}
\caption{For $M=4$, the growth rates of the most unstable azimuthal modes of
BVBs with $s=2$ and $s=3$ (solid curves), and of their counterparts with the same parameters in
the absence of absorption (dashed curves).}
\label{Fig10}
\end{figure}

The existence of stable nonlinear BVBs with multiple vorticity ($|s|>1$) is proven in Fig. \ref{Fig10}, which summarizes the results of the linear-stability analysis for cases $s=2$ and $s=3$. A similar outcome is expected to take place for $s>3$. Note that the stabilization of vortices with $s>1$ is achieved for the first time with this setup, as previously attempts failed \cite{porras-dodd1996,porras-adhikari2001,porras-saito2002,porras-alexander2002,porras-carr2006,porras-mihalachemazilu2006,porras-malomed2007,porras-sakaguchi2014,porras-zhang2015}.

The mechanism behind vortex stability in the present system can be understood from a deeper observation of Fig. \ref{Fig10}. In it, we also plot the instability growth rates for nonlinear BVBs in the fully transparent medium after setting all dissipative terms in the equations to zero in our model. As expected, all nonlinear BVBs are unstable in this case. The BVB intensity profiles obtained with and without dissipation in all cases shown in Fig. \ref{Fig10} are almost identical, aside that with dissipation the contrast of the radial oscillations is a bit smaller than unity. However, the beams in the medium without dissipation are unstable, and the beams in the medium with dissipation and with $\alpha \lesssim 1$ are stable. In this respect, we note that stabilization is reported to occur in Refs. \cite{porras-jukna2014,porras-xie2015} when the cone angle increases (i.e., $\alpha$ decreases), and qualitatively explained as inefficient nonlinear wave mixing or suppressed growth of the modulation instability at these cone angles. The results presented in Fig. \ref{Fig10} from the linear-stability analysis quantify the stabilizing effect of using large cone angles with and without dissipation. Besides the quantification of the effect, it also demonstrates that the stabilization is never complete without taking dissipation into consideration. This means, in other words, that the mere increase of the cone angle does not produce, by it self, completely stable nonlinear BVBs without dissipation.
\section{Tubular, rotating, and speckle-like filament regimes in axicon-generated nonlinear Bessel vortex beams}
\label{sec:porras-exp}

All preceding results regard to ideal, linear or nonlinear, conical beams in the sense that they transport infinite power.
We now consider the experimental implications of these results in experiments that generate actual Bessel beams with finite power and their nonlinear counterparts. In most of these arrangements, a Gaussian beam with embedded vorticity $s$ is passed through an axicon \cite{porras-polesana2008,porras-porras2015,porras-jukna2014}, or equivalent device to produce a conical beam \cite{porras-xie2015}. The nonlinear medium is usually placed close or immediately after the axicon, so that we can model the optical field at the entrance plane of the medium by  $A(r,\varphi,0)=\sqrt{I_G}\exp(-r^2/w_0^2)\exp(-i kr\theta)\exp(is\varphi)$ \cite{porras-jukna2014,porras-xie2015}, where $\sqrt{I_G}$ is the amplitude of the Gaussian beam, $w_0$ its width, and $\theta$ the cone angle imprinted by the setup. Expressed in our dimensionless variables, this field reads as $\tilde{A}_{G}(\rho,\varphi,0)=b_G\exp (-\rho^2/\rho _0^2)\exp(-i\rho)\exp(is\varphi)$, where amplitudes, transversal and axial lengths scale as specified in Eqs. \eqref{SCALING}. If we had linear propagation after the axicon, then a finite-power version of the linear BVB $\tilde{
A}_{B}\simeq b_{B}J_{s}(\rho)\exp (is\varphi)$ would form at the axicon focus, with intensity $b_{B}^{2}=\pi
\rho _{0}e^{-1/2}b_{G}^{2}$. The focus is placed at $\rho_0/4$ in our dimensionless axial coordinate, half of the Bessel distance $\rho_0/2$  ($w_0/2\theta$ and $w_0/\theta$, respectively, in physical units).

With this arrangement, quasi-stationary and rotary regimes of nonlinear propagation after the axicon have been described in Ref. \cite{porras-jukna2014}, and quasi-stationary, rotary and speckle-like (chaotic) regimes in Ref. \cite{porras-xie2015}. In those works, the quasi-stationary regime is associated with the formation of a nonlinear BVB, which is supposed to be a stable nonlinear BVB, and in Ref. \cite{porras-jukna2014} the rotary regime is conjectured to be associated with either an unstable nonlinear BVB or nonexistence of a stationary state in the specific experimental configuration. In neither case the respective nonlinear BVB was identified, so that these conjectures were not verified.

Recent investigation for the vortex-less case ($s=0$) \cite{porras-porras2015} has shown that the dynamics of the nonlinear propagation after the axicon is determined in the Bessel zone by the fundamental ($s=0$) nonlinear Bessel beam with amplitude $|b_{\rm in}|=b_B$. This means that, in the vortex-less situation, the law of conservation of the amplitude of the inward H\"ankel component, described in Sect. \ref{sec:porras-selec}, holds also in actual settings with finite power, but is reinterpreted here by stating that the nonlinear BVB that acts as an attractor in the Bessel zone is that with the same cone angle and with amplitude $|b_{\rm in}|$ equal to that of the linear Bessel beam that would be formed at the axicon focus in linear propagation. It has been further verified that if this fundamental nonlinear Bessel beam is stable, the system smoothly develops into it about the center of the Bessel zone and decays afterwards. On the other hand, if the nonlinear BVB is unstable, an unsteady regime characterized by the signatures of the development of the instability forms in the Bessel zone.

The same conclusion appears to be valid for the nonlinear BVB of arbitrary topological charge ($s\neq 0$), as supported by the extensive numerical simulations carried out. The dynamics in the Bessel zone is also determined by the nonlinear BVB with $b_0$ such that the amplitude of the inward H\"ankel component is $|b_{\rm in}|=b_B$, and its cone angle and vorticity $s$ are the same as those from its linear counterpart. The complete agreement between the prediction obtained from the stability analysis for a specific nonlinear BVB and the observed azimuthal-symmetry-breaking dynamics in the Bessel zone demonstrates it.

\begin{figure}[t]
\begin{center}
\includegraphics*[width=3.8cm]{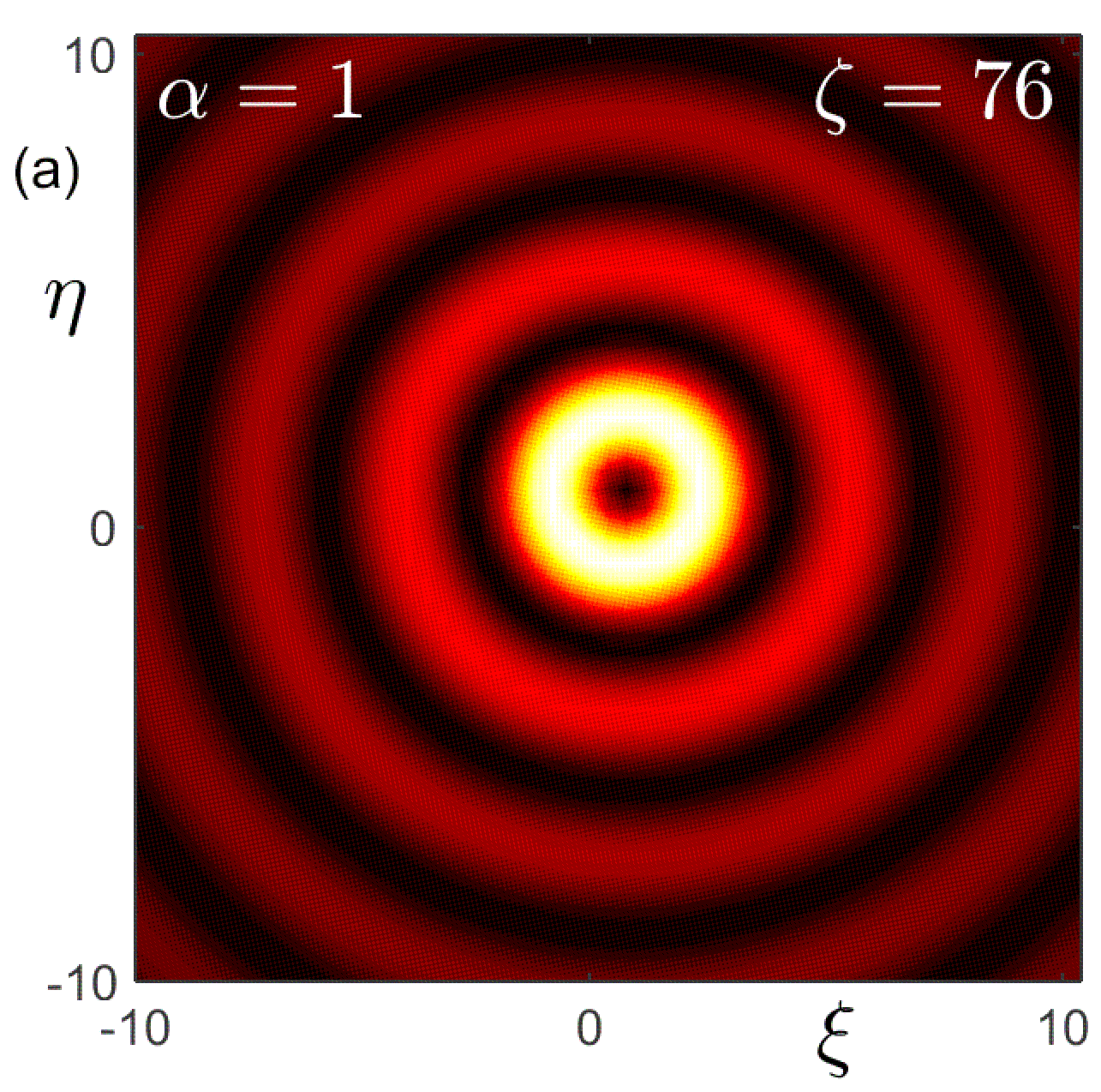}\includegraphics*[width=3.8cm]{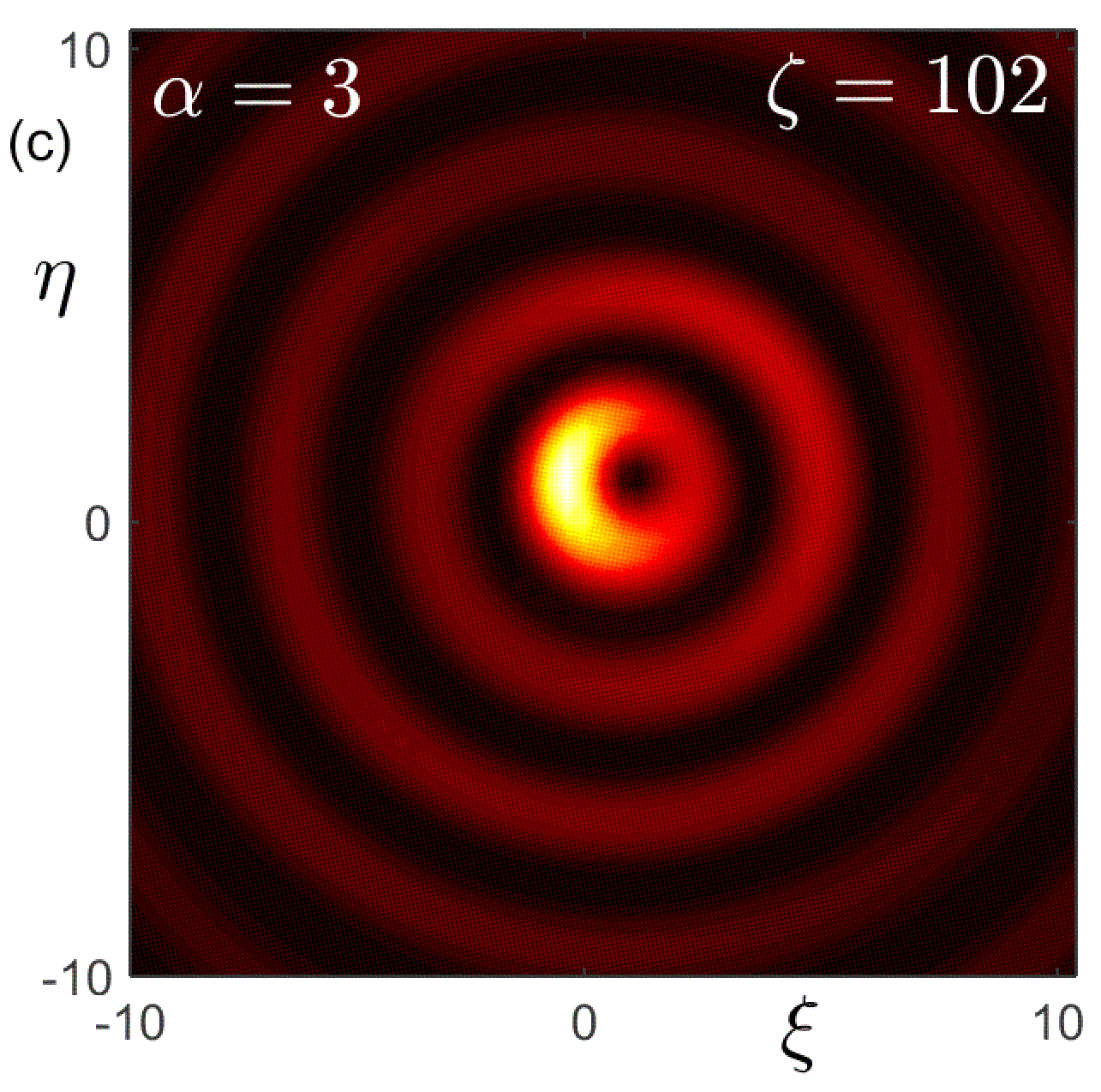}\includegraphics*[width=3.8cm]{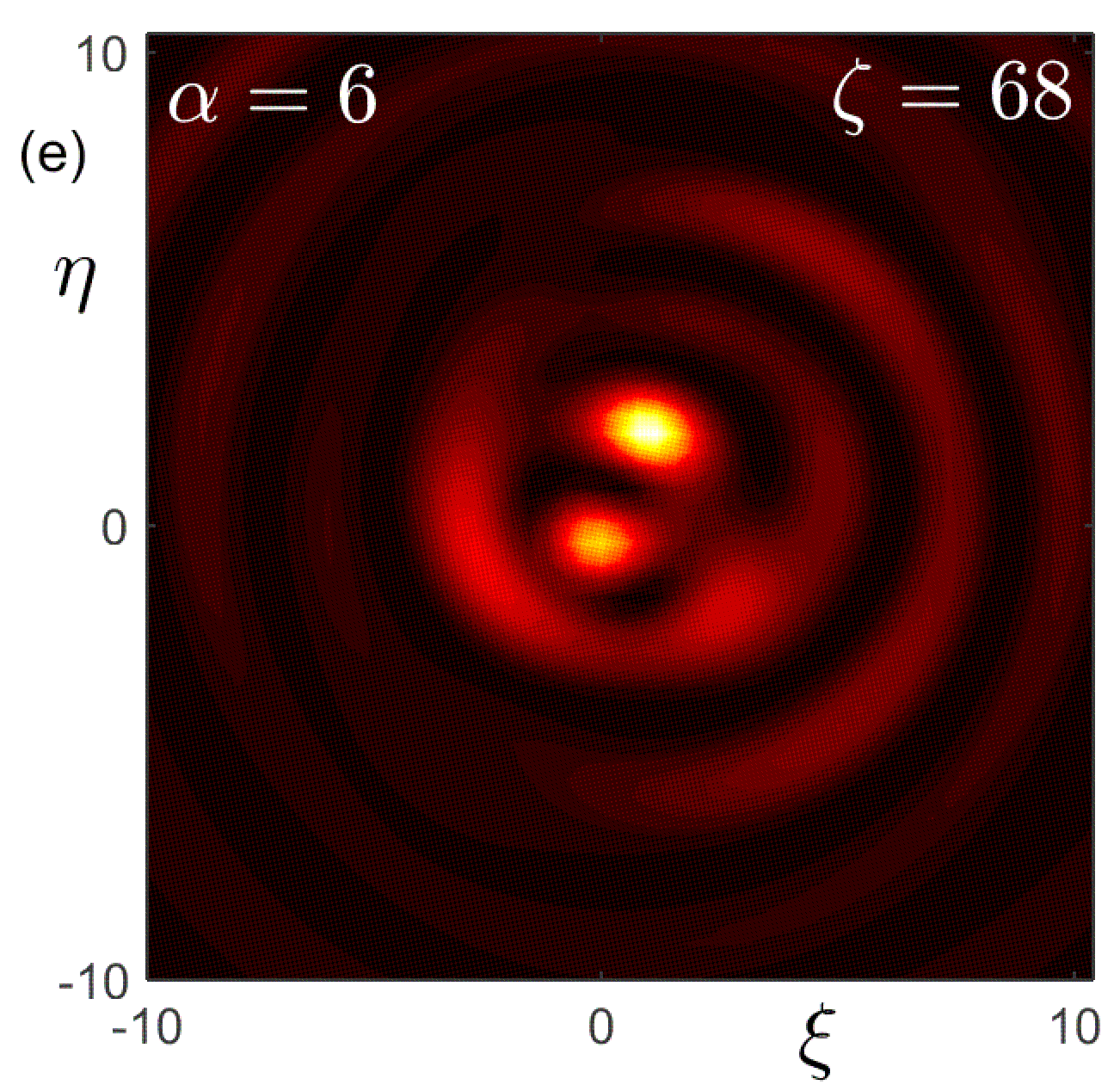}
\includegraphics*[width=3.8cm]{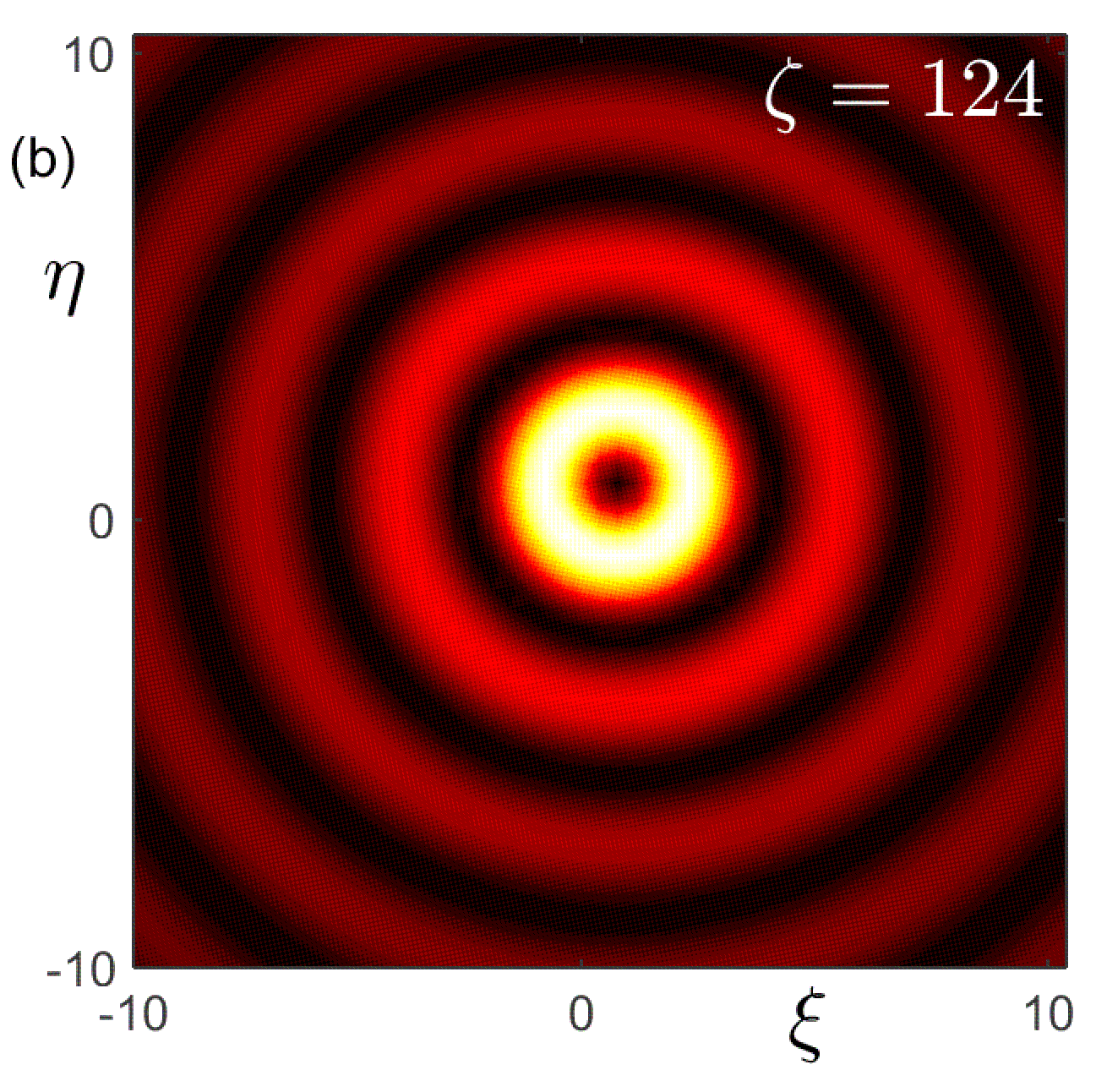}\includegraphics*[width=3.8cm]{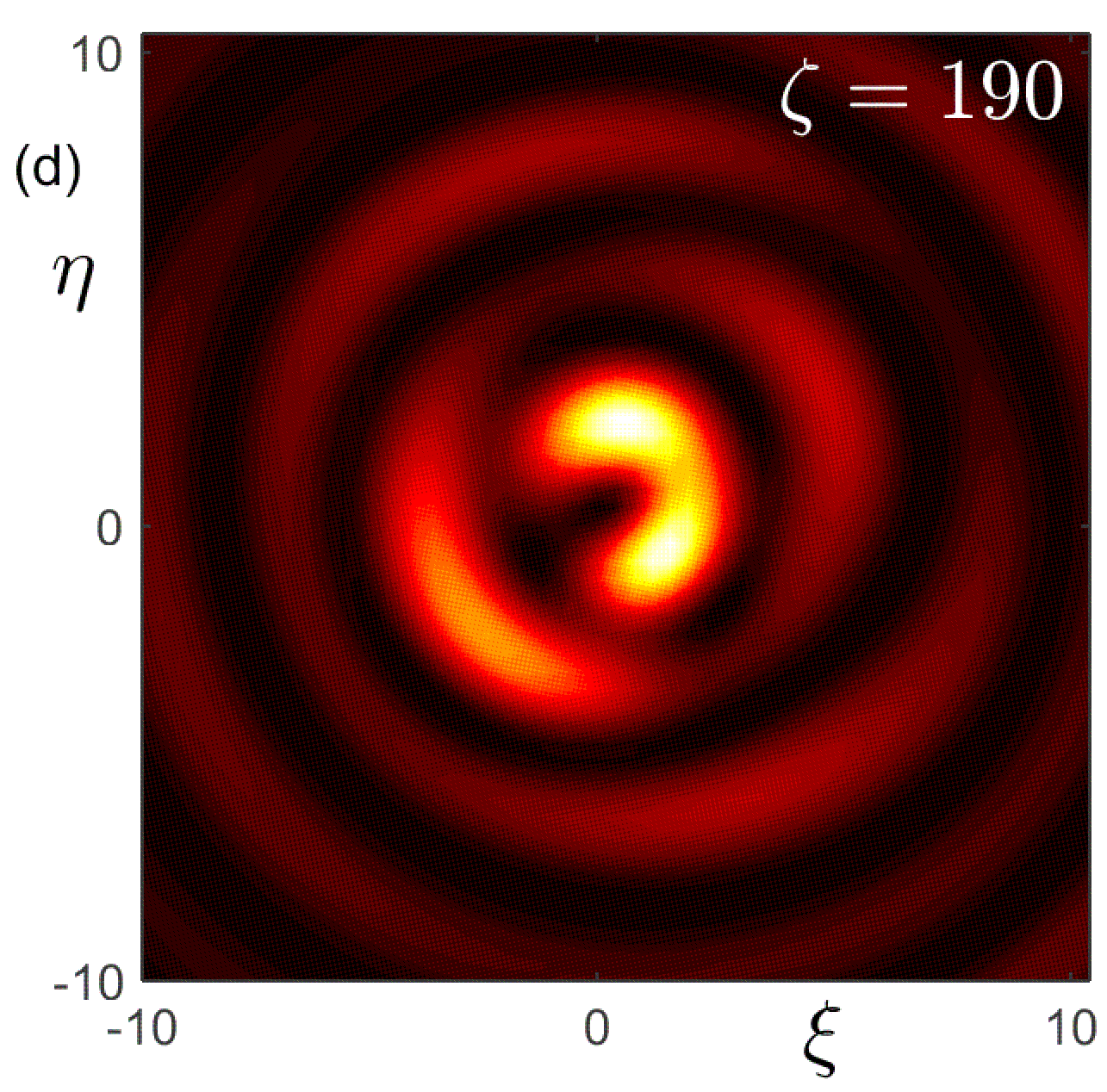}\includegraphics*[width=3.8cm]{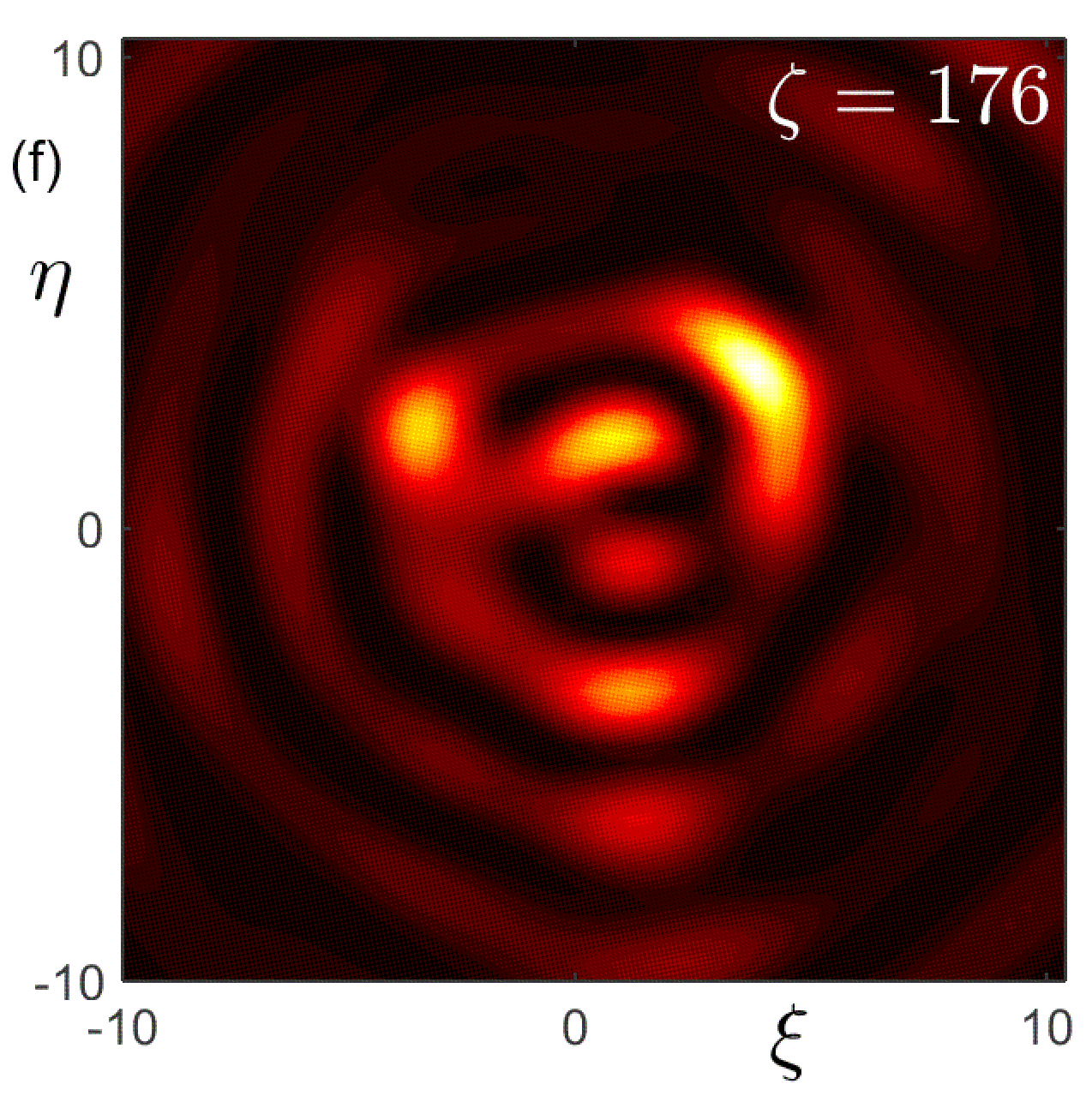}
\includegraphics*[width=3.8cm]{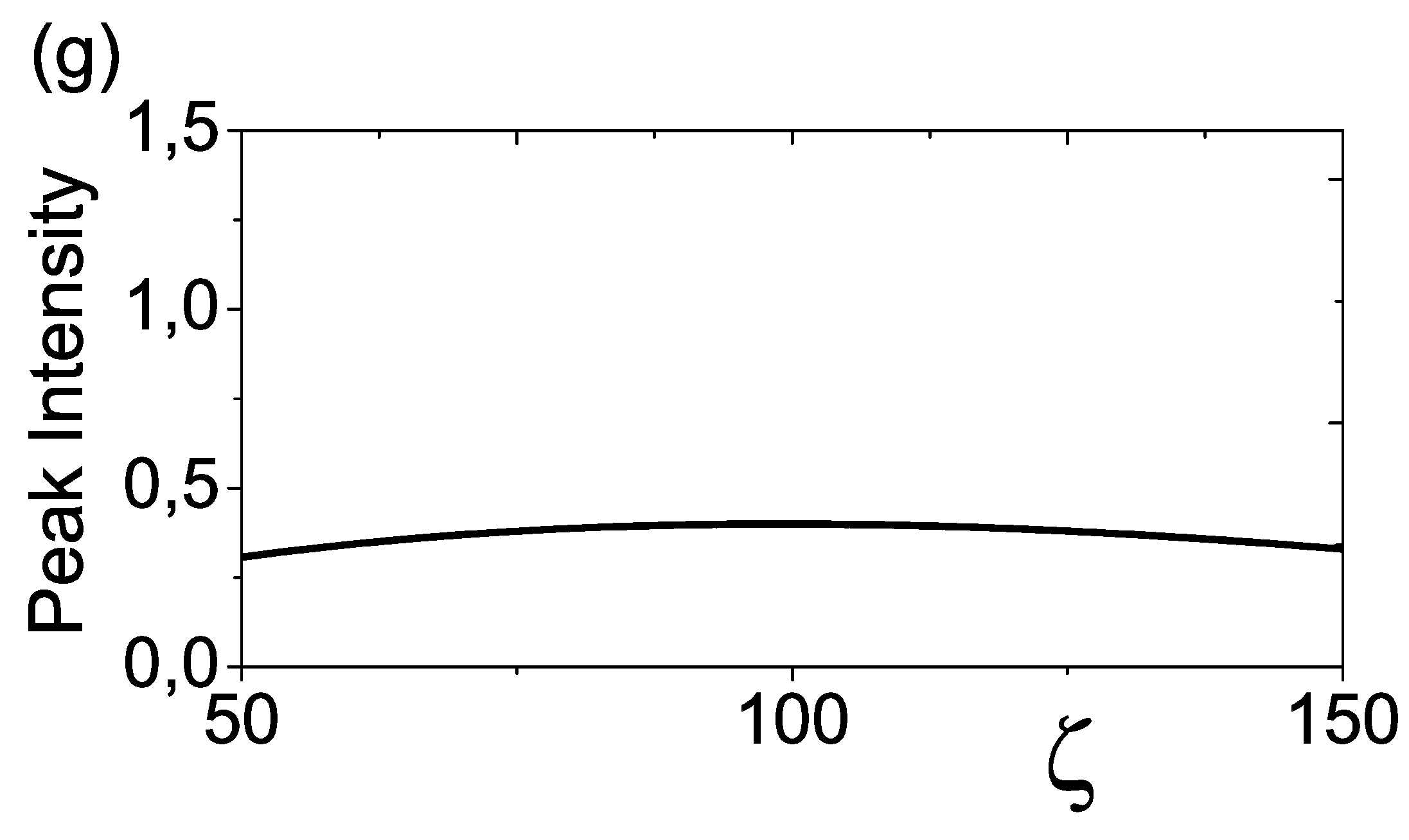}\includegraphics*[width=3.8cm]{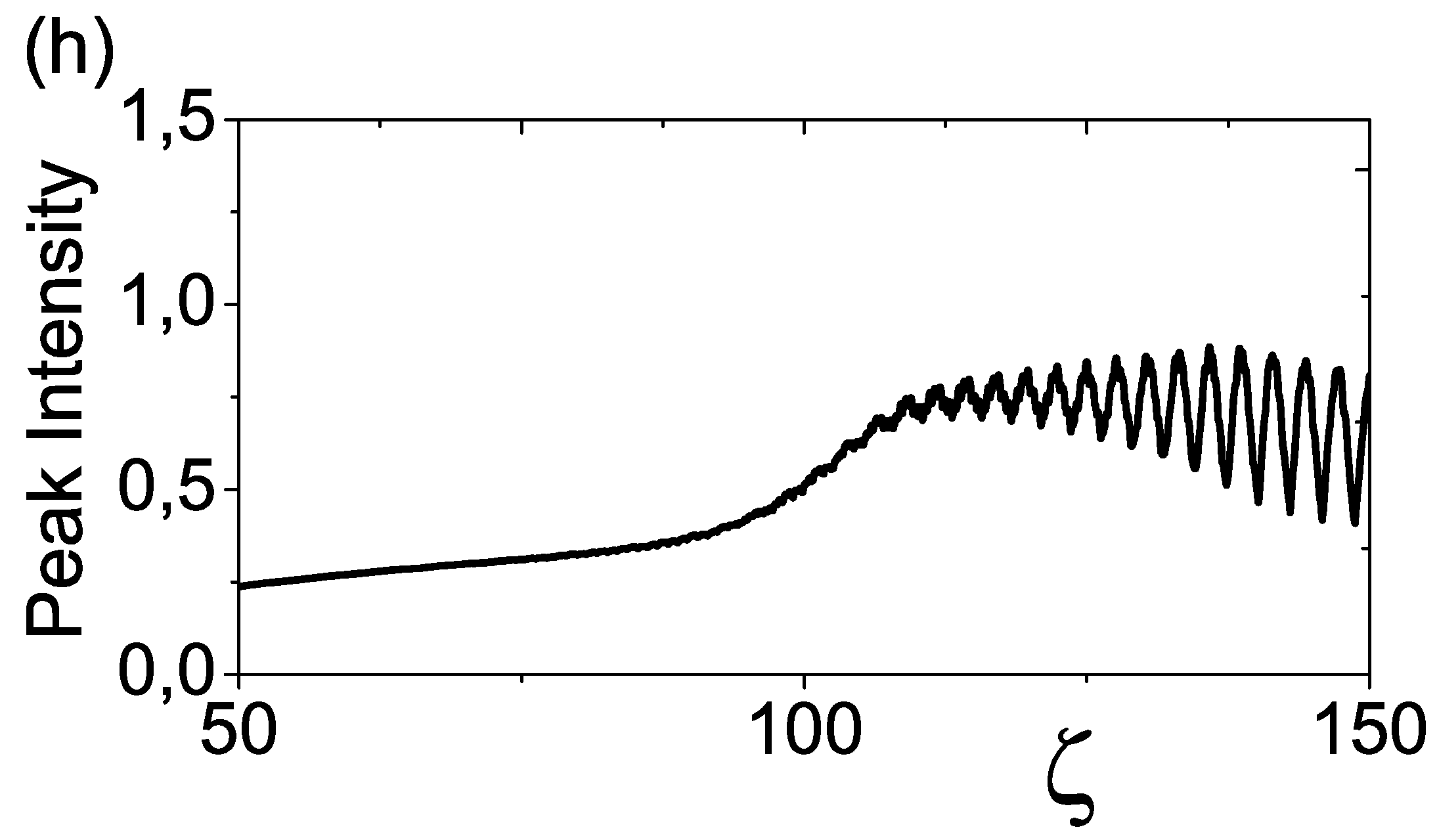}\includegraphics*[width=3.8cm]{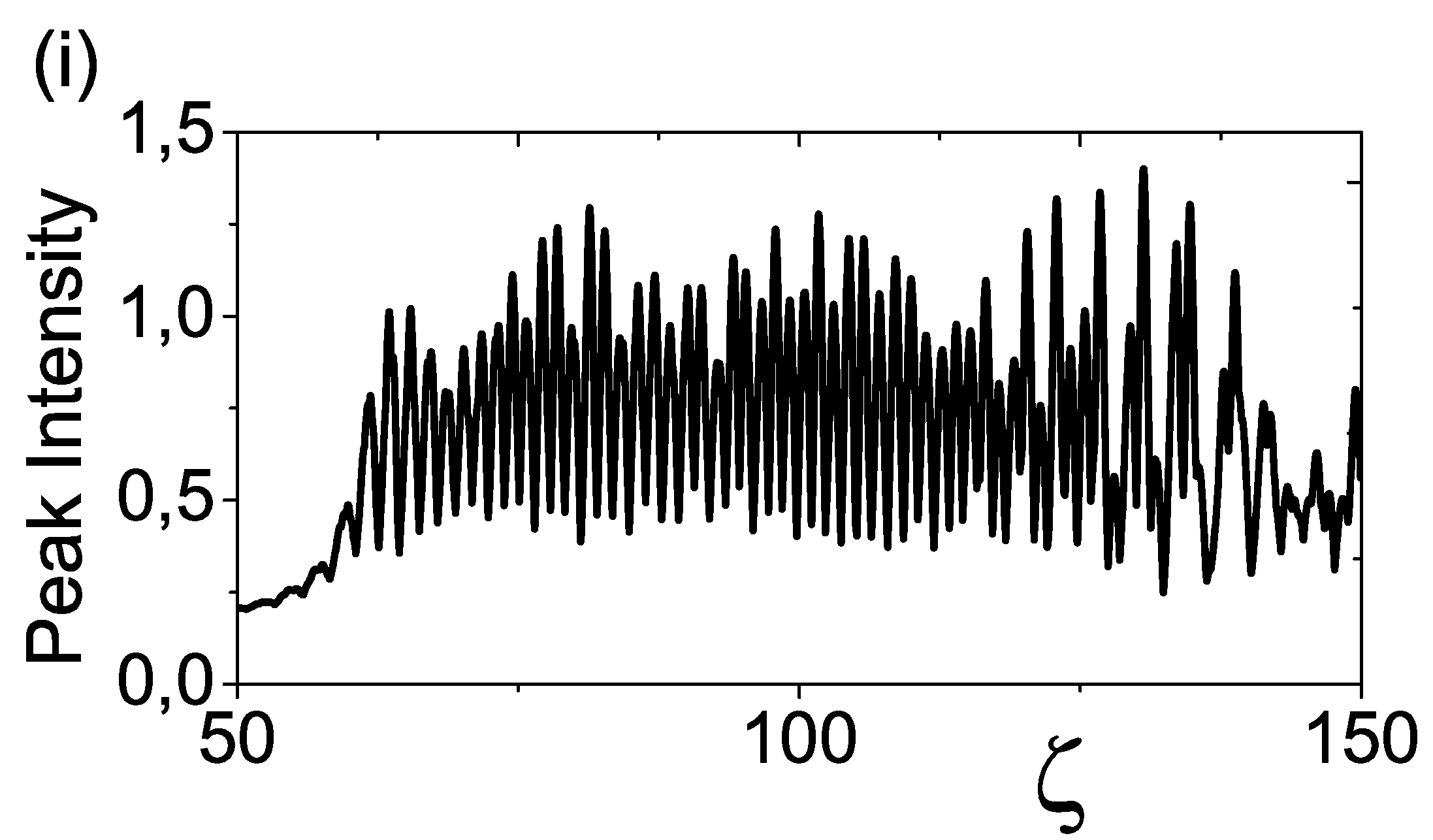}
\end{center}
\caption{Normalized intensity profiles at the propagation distances indicated in the panels for $\alpha =1$ (left), $\alpha =3$ (center), and $\alpha=6$ (right), in the presence of the four-photon absorption, produced by axicons illuminated by vortex-carrying Gaussian beams with $\rho_0=400$ and $s=1$, but the different amplitudes $b_G=0.0402$ (left), $0.0341$ (center) and $0.030$ (right), yielding respective linear Bessel amplitudes $b_B=1.11,0.94$ and $0.829$. These values are chosen such that the three nonlinear BVBs with $s=1$ and $|b_{\mathrm{in}}|=b_B$ in samples with the
corresponding values of $\alpha$ and $M$, have the same amplitude parameter $b_{0}=1.2$, which can be identified using the method explained in Sect. \ref{sec:porras-asym} or in Ref. \cite{porras-porras2014}. (g-i) Peak intensity as a function of the propagation distance in the central part of the Bessel zone for the three cases. This figure is to be compared with Fig. \ref{Fig9} and illustrates that the dynamics in the Bessel zone reproduces that in the development of instability, if any, of each nonlinear BVB.}
\label{Fig11}
\end{figure}

Three particularly representative examples are shown in Fig. \ref{Fig11}. These are three illuminating Gaussian beams, respectively with $b_G = 0.0402$ (left), $0.0341$ (center) and $0.030$ (right), all of them with an embedded single vortex $s = 1$ and $\rho_0 =400$. The Bessel amplitudes yielded under linear propagation would then be, in this order, $b_B = 1.11, 0.94$ and $0.829$. These illuminating beams are chosen such that the attracting nonlinear BVBs in respective nonlinear media with $M=4$ and $\alpha=1,3$ and $6$, and with $|b_{\rm in}|=b_B$, are characterized by $b_0 = 1.2$ in all three cases.

Under these conditions, Fig. \ref{Fig7}(a) obtained from our analysis, and the numerical simulations of the propagation of the perturbed respective nonlinear BVBs in Fig. \ref{Fig9}, predict instability for $\alpha =3$ and $6$, with inverse growth rates, or characteristic length of development of the instability, much shorter than the length of the Bessel zone, $\rho_0/2$. On the contrary, for $\alpha = 1$ stability is predicted in Figs. \ref{Fig7}(a) and \ref{Fig9}.

As seen in Figs. \ref{Fig11}(a) and (b), the stable nonlinear BVB with $\alpha=1$ is smoothly formed about the axicon focus ($\rho_0/4 = 100$), followed by a smooth decay towards the end of the Bessel zone. The propagation thus resembles the robust propagation of the stable nonlinear BVB in Figs. \ref{Fig9}(a) and (b), only limited by the finite power of the reservoir, which depleted at the end of the Bessel zone.
The evolution of the instability leading to the azimuthal breakup can be clearly observed in Figs. \ref{Fig11}(c) and (d) for $\alpha=3$ and Figs. \ref{Fig11}(e) and (f) for $\alpha=6$. The dynamics within the Bessel zone starts from the small-perturbation regime predicted by the linear-stability analysis in Fig. \ref{Fig7} (a), and then proceeds to the large-perturbation regime displayed in Figs. \ref{Fig9}(c) and (d) for $\alpha=3$ and Figs. \ref{Fig9}(e) and (f) for $\alpha=6$. At $\alpha =3$, a rotatory regime with initially one spot that later breaks into two is observed, as in the development of instability in Figs. \ref{Fig9} (c) and (d). At $\alpha =6$, a rotatory regime with initially two fragments later breaks into randomly placed non-rotating spots, indicating a stronger instability, also as in the development of instability in Figs. \ref{Fig9} (e) and (f).
The behavior of the peak intensity in the central part of the Bessel zone is plotted in Figs. \ref{Fig11} (g-i) in the three representative examples  studied. In all them, the peak intensity reproduces the dynamics also observed in the development of the instability, if any, triggered by noise of the respective nonlinear BVBs plotted in Figs. \ref{Fig9} (g-i).

When compared to the quasi-stationary and rotary regimes described numerically in \cite{porras-jukna2014}, the similarity is remarkable. A strong parallel can also be established with the quasi-stationary, rotary and speckle-like regimes described experimentally in \cite{porras-xie2015}. When subjected to the linear-stability analysis procedure, the previously mentioned representative examples reveal that all three regimes admit a common explanation in terms of stability or instability of a specific nonlinear BVB: the one preserving the cone angle, the topological charge and with amplitude $b_0$ satisfying the condition $|b_{\rm in}|=b_B$. The existence and uniqueness of a single nonlinear BVB satisfying the condition $|b_{\rm in}|=b_B$ is ensured considering that for a given $M$, $\alpha$ and $s$, $|b_{\rm in}|$ is a strictly growing function of $b_0$ taking all values between $|b_{\rm in}| = 0$ and $|b_{\rm in}| = \infty$, as seen in Figs. \ref{Fig5} (b) and (c).

It is possible, for example, to study more in depth the experimental observation of the quasi-stationary, rotating and speckle-like regimes reported in Fig. 4 of \cite{porras-xie2015}. Based on the values of the material constants, cone angles and three pulse energies, we can characterize the three attracting nonlinear BVBs as those defined by $M =5$, $s =3$, $\alpha = 14.89$, and $b_0 =0.368$, $0.822$ and $1.644$. Through the stability analysis we obtain, respectively, the largest dimensionless growth rates in each case as $0.026$, $0.245$ and $0.784$, or, multiplying by $|\delta|$, $2.48$ cm$^{-1}$, $23.4$ cm$^{-1}$ and $74.9$ cm$^{-1}$ in physical units. The different observed behaviors can be understood when we compare the length of the Bessel zone ($\sim 0.072$ cm), with the associated characteristic lengths of the instability development, or inverse growth rates: $0.402$ cm, $0.043$ cm and $0.013$ cm. In the first case the instability had not yet the opportunity to develop; in the second case the rotating filaments observed correspond to the azimuthal primary break in the nonlinear BVB instability; and in the third case the random filaments observed correspond to the full development of the nonlinear BVB instability.

Lastly, the existence of truly stable nonlinear BVBs demonstrated in this work guarantees the existence of a regime of tubular-beam propagation. The only limitation on this regime comes by the finite amount of power that can be stored in the reservoir, whose depletion will delimit the Bessel zone. Thus, the stable vortex tubules may be increased indefinitely by increasing the power supplied to the reservoir, for example, through the increase of $\rho_0$.

\section{Summary}
\label{sec:porras-sum}

In this Chapter we have reviewed the properties of nonlinear BVBs, nonlinear conical beams that propagate without any change, including any attenuation, in homogeneous Kerr media with nonlinear absorption. These beams form narrow and (ideally) infinitely long tubes of light where energy and orbital angular momentum can be transferred to matter, but the beam energy and angular momentum are continuously restored by spiral currents coming from a reservoir at large radial distances. Nonlinear BVBs arise naturally in the propagation of linear BVBs at intensities at which multi-photon absorption are significant.

Though a linear-stability analysis we have demonstrated that nonlinear BVBs may be stable against azimuthal symmetry breaking and collapse. These stable vortices can have multiple vorticities, and do not require materials with specific nonlinearities. They can propagate robustly in common dielectrics such as air, water, or optical glasses at high enough intensities, typically tens of TW/cm$^2$ in gases or a fraction TW/cm$^2$ in condensed matter. It is then not surprising that, contrary to previous settings, these vortices have been observed before their stability is demonstrated. The proof of stability has important implications in these experiments with powerful pulsed Bessel beams and their applications. The tubular filamentation regime, for instance, is only limited by the amount of power stored in the reservoir (and depleted at the end of the Bessel zone), and as such can be enlarged by simply increasing the amount of power of the illuminating beam.

From the stability analysis and diagnostic numerical simulations we have extracted simple underlying laws that govern the propagation of powerful Bessel beams in nonlinear media, and that apply both in ideal situations with infinite power and in actual experiments. We have shown that the propagation is always governed by a nonlinear BVB attractor. We we have identified it as the nonlinear BVB preserving the cone angle, the topological charge and the inward component of the power flux of the Bessel beam. We point out that attractors are not necessarily stable attractors, as is well-known in the field of nonlinear and chaotic dynamical systems. The tubular, rotating and speckle-like (chaotic) regimes observed in experiments are just manifestations of the stability/instability properties of the attracting nonlinear BVB. Understanding these laws is of fundamental importance for improving the applications of these powerful nonlinear Bessel beams.

\section*{Acknowledgments}
M.A.P. acknowledges support from Projects of the Spanish Ministerio de Econom\'{\i}a y Competitividad No. MTM2015-63914-P, and No. FIS2013-41709-P.

\bibliographystyle{spmpsci}
\bibliography{bibporras}\printindex\end{document}